\documentstyle[aps,prb,multicol,graphics,epsfig]{revtex}
\begin{document}
\title
{\bf
The {\em GW} method
}
\author
{
F. Aryasetiawan$^{1}$ and O. Gunnarsson$^2$\\[12pt]
$^1${\em Department of Theoretical Physics}, \\
{\em University of Lund,}\\
{\em S\"olvegatan 14A, }\\
{\em S-223 62 Lund, Sweden}\\[12pt] 
$^2${\em Max-Planck-Institut f\"{u}r Festk\"{o}rperforschung,}\\
{\em Heisenbergstrasse 1,}\\
{\em 70569 Stuttgart,}\\ 
{\em Federal Republic of Germany}
}
\maketitle

\begin{abstract}

Calculations of
ground-state and excited-state properties of materials have been
one of the major goals of condensed matter physics. Ground-state properties
of solids
have been extensively investigated for several decades
within the standard density functional theory. 
Excited state properties, on the other hand,
were relatively unexplored in {\it ab initio} calculations until a decade
ago.
The most suitable approach up to now for studying excited-state properties
of extended systems is the Green function method. 
To calculate the Green function one requires the self-energy operator which
is non-local and energy dependent. In this article we describe the $GW$
approximation which has turned out to be
a fruitful approximation to the self-energy.
The Green function theory,
numerical methods for carrying out the self-energy calculations, simplified
schemes, and
applications to various systems are described.
Self-consistency issue and new developments beyond the
$GW$ approximation are also discussed as well as
the success and shortcomings of the $GW$ approximation.
\end{abstract}

\tableofcontents
\section{Introduction}

The Hamiltonian for a many-electron system is given by (in atomic units $%
\hbar =m=e=1$) 
\begin{equation}
H=\sum_i[-\frac 12\nabla ^2({\bf r}_i)+V({\bf r}_i)]+\frac 12\sum_{i\ne j}%
\frac 1{|{\bf r}_i-{\bf r}_j|}
\end{equation}
where $V$ is a local external potential such as the field from the nuclei.
Solving the above Hamiltonian has been a major problem in molecular and
condensed matter physics. The Coulomb interaction in the last term makes the
Hamiltonian difficult to solve. For a small system, such as an atom or a
small molecule, it is possible to obtain the many-particle ground-state
wavefunction by means of the configuration interaction (CI) method (Boys 1950,
Dykstra 1988). 
In this
method the wavefunction is expanded as a sum of Slater determinants 
(Slater 1930) whose
orbitals and coefficients are determined by minimizing the total energy. A
very accurate ground state wavefunction and energy can be obtained. The
computational effort, however, scales exponentially with the system size so
that applications to large molecules or solids are not feasible. For excited
states, the computational effort becomes very large already for a small
system. Fortunately, in practice we are interested in quantities which do
not require the full knowledge of the wavefunctions. For example, we are
interested in the total energy, excitation spectra and expectation values of
single-particle operators which can be obtained from the Green function
described in a later section.

Approximate theories are 
usually concerned with finding a good single-particle
approximation for the Coulomb term. The earliest of these theories is the
Hartree approximation (1928)
where the non-local Coulomb term is replaced by an
average local Coulomb potential (Hartree potential) from all the electrons.
Although it gives reasonable results, due to a cancellation between exchange
and correlation, the Hartree theory is not accurate enough in many cases. An
extension of the Hartree theory which takes into account the fermionic
nature of the electrons leads to the Hartree-Fock approximation (HFA) 
(Fock 1930) where
in addition to the average local Coulomb potential there is a non-local
exchange potential which reflects the Pauli exclusion principle. For systems
with an energy gap in their excitation spectra, the HFA gives a
qualitatively reasonable result. In fact the HFA works quite well for atoms
but for insulating solids, the energy gap is in most cases too large. The
reason for this can be traced back to the neglect of correlations or
screening which is not too important in atoms but crucial in solids. The HFA
already takes into account to a large extent correlation between electrons
of the same spin since the Pauli exclusion principle (exchange) prevents
them to get close together. Two electrons of opposite spin on the other hand
are allowed to occupy the same single-particle state at the cost of a large
Coulomb energy. Correlations keep electrons away from each other, creating a
screening hole around each electron which reduces the interaction with the
other electrons and thereby the Coulomb energy. The energy cost of
transferring an electron from one site to a neighbouring site is substantially
reduced by screening.
In the tight-binding limit, i.e. for localized states, the energy gap is
approximately given by $U$
and this is essentially the effective
Coulomb energy of the states that form the gap. Thus, correlation or
screening reduces the gap from its Hartree-Fock value. In metals, the
absence of correlations in the HFA leads to qualitatively wrong results such
as zero density of states at the Fermi level due to a logarithmic
singularity in the derivative of the single-particle spectra with respect to
$k$ at $k_F$ (see, e.g., Ashcroft and Mermin 1976).

To simulate the effect of correlations, Slater introduced the 
$X \alpha$ method where the exchange potential is modelled by a local
potential of the form $V^x=\alpha n^{1/3}$, derived from the electron gas
and scaled by a constant $\alpha $ to simulate correlations. $n$ is the
local electron density (Slater 1951a, 1951b, 1953, 1974). 
This method has been quite successful in calculating
ground-state properties and excitation spectra but it is semiempirical. 
The $X\alpha$ theory may be regarded as a precursor of the modern density
functional theory (DFT) (Hohenberg and Kohn 1964, Kohn and Sham 1965)
which has become a standard method for calculating
ground-state properties of molecules and solids. Recent reviews on DFT
may be found in Jones and Gunnarsson (1989) and Dreizler and Gross (1990).
In DFT, the ground-state
energy can be shown to be a functional of the ground-state density and
satisfies the variational principle with respect to the density. The
explicit form of the functional in terms of the density is not known and
such an explicit functional may not exist. Using the variational property of
the energy functional, one arrives at a set of single-particle equations,
the Kohn-Sham (KS) equations (Kohn and Sham 1965), 
to be solved self-consistently: 
\begin{equation}
\lbrack -\frac 12\nabla ^2+V^H+V^{\rm xc}]\phi _i=\varepsilon _i\phi _i
\end{equation}
\begin{equation}
n=\sum_i^{\rm occ}|\phi _i|^2
\end{equation}
where $V^H$ and $V^{\rm xc}$ are the Hartree and exchange-correlation potential
respectively. In practical applications, the functional containing the
effects of exchange and correlations is approximated by the local
density approximation (LDA) where the density in the exchange-correlation
potential of the electron gas is replaced by the local
density of the real system (Kohn and Sham 1965). The KS eigenvalues $%
\varepsilon _i$ have no clear physical meaning except for the highest
occupied which corresponds to the ionization energy (Almbladh and von Barth
1985a). Although there is no
theoretical justification, they are often interpreted as single-particle
excitation energies corresponding to excitation spectra of the system upon
a removal or addition of an electron. In many cases, in particular in sp
systems, the agreement with photoemission spectra is quite good but as will
be described in the next section, there are also serious discrepancies.

A proper way of calculating single-particle excitation energies or
quasiparticle energies (Landau 1956, 1957)
is provided by the Green function theory
(Galitskii and Migdal 1958, Galitskii 1958). It can be
shown that the quasiparticle energies $E_i$ can be obtained from the
quasiparticle equation: 
\begin{equation}
\lbrack -\frac 12\nabla ^2({\bf r})+V^H({\bf r})]\Psi _i({\bf r})
+\int
d^3r^{\prime }\;\Sigma({\bf r},{\bf r}^{\prime };E_i)\Psi _i({\bf r}%
^{\prime })=E_i\Psi _i({\bf r})  \label{qpenergyeqn}
\end{equation}
The non-local and energy dependent potential $\Sigma$, or the
self-energy, contains the effects of exchange and correlations. It is in
general complex with the imaginary part describing the damping of the
quasiparticle. It can be seen that the different single-particle
theories amount to approximating the self-energy operator $\Sigma$.
Calculations of $\Sigma$ are unfortunately very difficult even for the
electron gas. We must resort to approximations and this review describes the 
$GW$ approximation (GWA) (Hedin 1965a)
which is the simplest working approximation beyond
the HFA that takes screening into account.

\subsection{Problems with the LDA}

The LDA has been very successful for describing ground-state properties such
as total energies and structural properties. There is no clear
theoretical justification why the LDA KS eigenvalues should give excitation
energies, and even the exact $V^{\rm xc}$ is not supposed to give the exact
quasiparticle energies. Nevertheless, 
the LDA KS eigenvalues are often found to be in
good agreement with the quasiparticle energies measured in photoemission
experiments. 
Despite of its success, there are serious
discrepancies already in sp-systems and they become worse in d- and f-systems:

\begin{itemize}
\item  The band width in Na is 10-20 \% too large, 3.2 eV in LDA $vs$ 
2.65 eV experimentally (see, however, section V-A for the experimental band
width).

\item  The band gaps in semiconductors Si, GaAs, Ge, etc. are systematically
underestimated, as much as 100 \% in Ge.

\item  The band width in Ni is $\sim $ 30 \% too large, 4.5 eV in LDA $vs$
3.3 eV experimentally (H\"ufner {\em et al} 1972, Himpsel, Knapp, and
Eastman 1979)

\item  In f-systems, the LDA density of states is in strong disagreement
with experiment.

\item  In the Mott-Hubbard insulators of transition metal oxides
the LDA band gap is much too small (Powell and Spicer 1970, H\"ufner {\em et
al} 1984, Sawatzky and Allen 1984) and in some cases the LDA gives $%
qualitatively$ wrong results. For example, the Mott-Hubbard insulators CoO
and the undoped parent compound of the high $T_{\mbox{c}}$ 
material La$_2$CuO$_4$ are
predicted to be metals (see e.g. Pickett 1989)

\item  The magnetic moments in the transition metal oxides are
systematically underestimated (Alperin 1962, Fender {\em et al} 1968,
Cheetham and Hope 1983)
\item
In alkali-metal clusters, 
the ionization energies calculated within the LDA are too low compared
to experiment (Ishii, Ohnishi, and Sugano 1986, Saito and Cohen 1988).
\item
In the LDA, the image potential seen by an electron in the vacuum far from a
surface decays exponentially instead of the expected
$-1/4(z-z_0)$ decay where $z$ is the coordinate normal to the surface and
$z_0$
is the position of the image plane (Lang and Kohn 1973).
\end{itemize}

Strictly speaking, one should not blame the LDA for all of
these discrepancies 
since many of them are related to excited state properties
which are outside the domain of DFT. However, excited state properties of
finite systems can be calculated within the time-dependent extension of DFT
(Runge and Gross 1984, Gross, Dobson, and Petersilka 1996).

\subsection{Theories beyond the LDA}

When discussing theories beyond the LDA, a distinction should be made
between theories which attempt to find better energy functionals but which
lie within DFT and those theories which attempt to mimic the self-energy in
order to obtain better quasiparticle energies but which are then usually
outside DFT. The GWA belongs to the latter.

\subsubsection{Gradient corrections}

Since the LDA is based on the homogeneous electron gas, it is natural to
take into account the inhomogeneity in the charge density of real systems by
including gradient corrections in the energy functional. There are two main
approaches. The first is a semiempirical approach where the
exchange-correlation functional is modelled by a functional containing
parameters which are adjusted to give the best fit to the cohesive energies
of a number of ''standard molecules''. The most successful of these models
are due to Becke (1988, 1992, 1996). 
The other approach attempts to calculate the coefficients in
the gradient expansion from first principles 
(Langreth and Mehl 1983,
Svendsen and von Barth 1996, Springer, Svendsen, and von Barth 1996,
for a recent development see Perdew, Burke, and
Ernzerhof 1997 and references therein). While in general gradient
corrections give a significant improvement in the total energy 
(Causa and Zupan 1994, Philipsen and Baerends 1996, Dal Corso 1996)
there is almost no major improvement for quasiparticle energies
(Dufek {\em et al} 1994).

\subsubsection{LDA+U}

One of the problems with the LDA is the absence of orbital dependence in the
exchange-correlation potential. Since the potential does not distinguish
between orbitals with different $m$-quantum numbers,
for systems containing a partially filled d-
or f-shell one obtains a corresponding partially filled band with a
metallic type electronic structure unless the exchange and
crystal field splitting create a gap between the up and down channel. Thus,
the late transition metal oxides, which are insulators, are predicted to be
metals by the (nonspin-polarized) LDA. In the LDA+U method 
(Anisimov, Zaanen, and Andersen 1991, Anisimov {\em et al} 1993,
Lichtenstein, Zaanen, and Anisimov 1995) and its generalization
(Solovyev, Dederichs, and Anisimov 1994,
Solovyev, Hamada, and Terakura 1996), an orbital
dependent potential U acting only on localized d- or f-states is
introduced on top of the LDA potential. For Mott-Hubbard insulators or rare
earth metal compounds where the partially filled 3d or 4f bands are
split by the Coulomb interaction, forming the upper and lower
Hubbard band, the LDA+U works reasonably well (Anisimov {\em et al} 1997). 
The bandstructure, however,
is unsatisfactory. Another problem with the method arises for systems with
partially filled $3d$ shells which are metallic, like the transition metals.
In this case, the LDA+U would produce unphysical results since it would
split the partially filled band.

\subsubsection{Self-interaction correction (SIC)}

Apart from the problem with orbital dependence in the LDA, there is another
problem associated with an unphysical interaction of an electron with
itself. In DFT, only the highest occupied state is free from
self-interaction but in LDA, there is in general self-interaction for all
states. This self-interaction is explicitly subtracted out in the SIC
formalism resulting in an orbital-dependent potential (Cowan 1967, Lindgren
1971, Zunger {\em et al} 1980, Perdew and Zunger 1981). Self-interaction is
significant for localized states and it tends to zero for extended states
since in the latter case the charge is spread over the crystal and therefore
the Coulomb interaction of an electron with itself is of order $1/N$. Since
self-interaction is usually positive, one would expect
the LDA eigenvalues for localized
states to be too high, as is indeed the case. For atoms, SIC therefore
lowers the LDA eigenvalues giving better agreement with experiment. The
orbital-dependent potential in SIC can describe Mott-Hubbard insulators
although the bandstructure is not likely to be satisfactory. SIC predicts
all the 3d monoxides to be insulators except VO which is correctly predicted
to be a metal (Svane and Gunnarsson 1990, Szotek, Temmerman, and Winter 1993,
Arai and Fujiwara 1995). 
SIC however fails to give a delocalized paramagnetic solution
for the doped high ${\rm T}_{{\rm C}}$ compounds and a similar problem is
anticipated in LDA+U. For more itinerant systems, SIC does not give
localized solutions, and it then reduces to the LDA. Accordingly,
application to semiconductors, e.g. Si, does not increase the LDA gap.

\subsubsection{A generalized KS scheme}

A recent attempt to improve the LDA description of quasiparticle energies is
to choose a non-interacting reference system having the same ground-state
density as the real system, like in the conventional KS scheme, but with a
non-local potential (Seidl {\em et al} 1996). 
Since the potential is non-local, the choice is not
unique, different non-local potentials may generate the same ground-state
density. A particular choice is a non-local screened exchange potential
minus its local form. By definition, the functional from which this
potential is derived, is zero at the correct density. The KS equations
consist of the usual LDA exchange-correlation potential plus the chosen
non-local potential. For semiconductors Si, GaAs, Ge, InP and InSb, this
method improves the values of the band gap. Applications to other systems
have not been performed so far.

Another recent scheme proposed by Engel and Pickett (1996)
incorporates part of the correlation energy into the kinetic energy functional
which may be thought of as mass renormalization.
This scheme is shown to improve the description of
band gaps in silicon and germanium 
but it gives negligible correction for diamond and carbon.

\subsubsection{The optimized effective potential method}

A natural way of improving the LDA would be to calculate the exchange energy
exactly and to generate a local exchange potential by taking a functional
derivative of the exchange energy functional with respect to the density
(Kotani 1995, Kotani and Akai 1996, Bylander and Kleinman 1995a,b). 
The correlation energy can be approximated by the LDA value.
The original idea of this method is due to Talman and Shadwick (1976)
in their work on atoms 
where the Hartree-Fock total energy is minimized with
orbitals restricted to be 
solutions to single-particle Hamiltonians with local potentials.
The scheme is also known as the optimized effective potential method.
Applications of this approach to several
semiconductors and insulators (C, Si, Ge, MgO, CaO, and MnO) yields
encouraging results regarding the band gaps which in most cases 
are improved from the corresponding LDA values. 
However,
the scheme is still within the density functional formalism and it is intended
to improve the energy functional rather than the self-energy.
One advantage of this scheme is the possibility to systematically improve
the energy functional. With regard to {\em GW} calculations, the scheme may
provide better starting points than the LDA.
The scheme may be extended to include correlations by using a screened
interaction potential.
\subsection{Motivations for the GWA}

The theories described above have drawbacks when applied to calculating
quasiparticle energies. Gradient corrections attempt to improve total
energies but do not address the problem of improving quasiparticle energies.
Indeed, applications to transition metal oxides, where the gap is zero or
grossly underestimated by the LDA, do not give any significant improvement.
SIC theory only applies to localized occupied states and numerical
calculations show that the resulting eigenvalues are too low. The LDA+U is
designed for systems with localized states split by the Coulomb correlation,
forming the upper and lower Hubbard band. Applications to transition metals,
however, would lead to unreasonable results. The generalized KS scheme using
screened exchange, like the other theories discussed above, has no energy
dependence which can be important in some cases. Moreover, the choice of the
non-local potential is rather arbitrary. Since the theory has not been
applied extensively, it is difficult to judge its usefulness.
The exact exchange approach should in principle improve
total energies when correlations are also taken into account
but it does not address the LDA problems with quasiparticles. 

The GWA is derived systematically from many-body perturbation theory. The
form of the self-energy in the GWA is the same as in the HFA but the Coulomb
interaction is dynamically screened, remedying the most serious deficiency
of the HFA. The self-energy in the GWA is therefore non-local and energy
dependent. The GWA is physically sound because it is qualitatively correct
in some limiting cases (Hedin 1995)
which allows applications to a large class of
materials, metals or insulators. 
\begin{itemize}
\item
In atoms, screening is small and the GWA
approaches the HFA which is known to work well for atoms. 
\item
In the electron
gas, screening is very important which is taken into account in the GWA and
for semiconductors it can be shown that the GWA reduces the Hartree-Fock
gaps. 
\item
For a core electron excitation, the GWA corresponds to the classical
Coulomb relaxation energy of the other electrons due to the adiabatic
switching-on of the core hole potential, which is just what is 
to be expected
physically. 
\item
For a Rydberg electron in an atom, the GWA gives the classical
Coulomb energy of the Rydberg electron due to the adiabatic switching-on of
an induced dipole in the ion core. 
\item
For the decay rate and the energy loss per unit time
of a fast electron in an electron gas,
the GWA gives the correct formula.
\end{itemize}
The GWA has been applied with success to
a wide class of systems ranging from simple metals to transition metals and
their compounds.

So far, the GWA has been applied mainly to calculate single-particle
excitation spectra, but it is also possible to calculate the total energy
(Lundqvist and Samathiyakanit 1969,
von Barth and Holm 1997, Farid, Godby, and Needs 1990)
and the expectation value of any single-particle operator in the
ground state. 

\subsection{A short historical survey}

The earliest attempt to include correlations beyond the HFA in the form of
$GW$ theory was probably the work of Quinn and Ferrell (1958)
for the electron gas.
Their calculations, however, were limited to states around the Fermi energy
and several approximations were made. DuBois (1959a, 1959b)
also calculated the self-energy
of the electron gas within a $GW$ type theory but his calculations were only
for small values of the electron gas parameter $r_s<1$ or for high densities
since $(4\pi/3)r^3_s=\rho$, where $\rho$ is the electron density. His
results have therefore received less attention
because they are not directly relevant to real metals
which have $r_s\sim 2-5$ (Al $r_s\sim$ 2, Cs $r_s\sim$ 5).
The first full calculation of the self-energy within
the GWA for the electron gas was performed by Hedin (1965a). 
He also showed in a
systematic and rigorous way how the self-energy can be expanded in powers of
the dynamically
screened Coulomb interaction, with the GWA as the first term in this
expansion. Later on, Lundqvist (1967a, 1967b, 1968)
did extensive calculations of the
self-energy of the electron gas for various densities and studied the
spectral functions. Rice (1965)
used a different version of what is conventionally
known as the GWA, including vertex corrections (corrections beyond the GWA).
His results are similar to those of Hedin. Later on, 
Mahan and his group (Mahan and Sernelius 1989, Frota and Mahan 1992)
performed extensive self-energy calculations for the electron gas using
various forms of the $GW$-type approximations, studying the importance of
vertex corrections.

Due to computational difficulties, the GWA was not applied to real materials
until the mid eighties starting with the work of Hybertsen and Louie 
(1985a,b, 1986, 1987b) on
semiconductors with encouraging results. At about the same time, Godby, Sham
and Schl\"{u}ter (1986, 1987a,b, 1988) 
did the same calculations and their results are in good
agreement with those of Hybertsen and Louie (1986). 
We should also mention an
earlier calculation for diamond using the tight-binding approach by
Strinati, Mattausch, and Hanke (1982) although it is
superseded by later calculations. The good results for semiconductors
encouraged further applications to more complicated systems, transition
metals and their compounds, jellium surface, sodium clusters and to 
f-systems. By now the GWA has become a standard method for including
correlations beyond the HFA.

\section{Theory}

In this section we describe a brief summary of the Green function theory and
derive the GWA. More details on the Green function theory may be found in
standard text books on many-body theory (e.g. Nozi\`eres 1964,
Fetter and Walecka 1971,
Inkson 1984, Mahan 1990) and in the review article by Hedin and
Lundqvist (1969).

\subsection{The Green function and the self-energy}

To study the electronic excitation spectrum of a solid, one performs
photoemission experiments where photons with a certain energy $\omega$ are
used as projectiles to knock out electrons. By measuring the kinetic energy
of the photoemitted electrons along a certain direction ${\bf k}$ and using
the conservation of energy and momentum, the excitation spectrum $E({\bf k})$
of the solid can be obtained: 
\begin{equation}
\omega = K.E. + E({\bf k})
\end{equation}
A photoemission experiment then measures the excitation spectrum of a solid
with the presence of a hole (occupied density of states). An inverse
photoemission experiment or BIS uses electrons as probes to measure the
unoccupied density of states or the excitation spectrum with an additional
electron.

In the limit of large kinetic energy (''sudden approximation'') (Hedin and
Lundqvist 1969) of the photoemitted or probing electron, the spectrum is
directly related to the one-particle Green function which is
defined as 
\begin{eqnarray}
iG(x,x^{\prime }) &=&
\langle N|T[\hat{\psi}(x)\hat{\psi}^{\dagger
}(x^{\prime })]|N\rangle
\\ \nonumber\\ 
&=&\left\{ 
\begin{array}{rcll}
\langle N|\hat{\psi}(x)\hat{\psi}^{\dagger }(x^{\prime })|N\rangle & 
{\rm for} & 
t>t^{\prime } & ({\rm electron}) \\ 
-\langle N|\hat{\psi}^{\dagger }(x^{\prime })\hat{\psi}(x)|N\rangle & 
{\rm for} & 
t<t^{\prime } & ({\rm hole})
\end{array}
\right.
\end{eqnarray}
$|N\rangle $ is the exact N-electron ground state, 
$\hat{\psi}(x)$ is a field
operator in the Heisenberg representation
which annihilates an electron at $x=({\bf r},t)$, and $T$ is the
time-ordering operator which arises naturally from the time development
operator defined later in equation (\ref{timedevelopment}). 
The physical interpretation of the Green function is
that for $t^{\prime }>t$ it is is the probability amplitude that a hole
created at $x$ will propagate to $x^{\prime }$ and for $t>t^{\prime }$, the
probability amplitude that an electron added at $x^{\prime }$ will propagate
to $x$. Thus, the Green function describes the photoemission and inverse
photoemission processes.

From the Green function we can obtain

\begin{itemize}
\item  The expectation value of any single-particle operator in the
ground state.

\item  The ground-state energy.

\item  The one-electron excitation spectrum.
\end{itemize}

In this review, we are mainly interested in the excitation spectra. The
first and the second properties have not been explored for real systems.

From the Heisenberg equation of motion for the field operator 
\begin{equation}
i\frac{\partial \hat{\psi}(x)}{\partial t}=
[\hat{\psi}(x),\hat{H}]
\end{equation}
where the Hamiltonian is given by 
\begin{eqnarray}
\hat{H} &=&\int d^3r\hat{\psi}^{\dagger }(x)h_0(x)
\hat{\psi}(x) 
+\frac 12\int d^3rd^3r^{\prime }\hat{\psi}^{\dagger }(x)
\hat{\psi}%
^{\dagger }(x^{\prime })v({\bf r}-{\bf r}^{\prime })
\hat{\psi}(x^{\prime })%
\hat{\psi}(x)
\end{eqnarray}
we obtain the equation of motion for the Green function: 
\begin{equation}\label{eqnofmotion}
{\left[ i\frac \partial {\partial t}-h_0(x)\right] G(x,x^{\prime} )} 
-\int dx''M (x,x'')G(x'',x^{\prime })=\delta (x-x^{\prime })
\end{equation}
where the mass operator (Hartree potential + self-energy)
$M$ is defined to be such that 
\begin{eqnarray}
\int dx_1M (x,x_1)G(x_1,x^\prime ) &=&-i\int d^3r_1v\left( {\bf r-r}%
_1\right)  \nonumber \\
&&\times \langle N|T\left[ \hat{\psi}^{\dagger }({\bf r}_1,t)
\hat{\psi}({\bf %
r}_1,t)\hat{\psi}({\bf r},t)\hat{\psi}^{\dagger }({\bf r}^{\prime
},t^{\prime })\right] |N\rangle  \label{defsigma}
\end{eqnarray}
$h_0$ is the kinetic energy operator plus a local external potential. The
quantity on the right hand side is a special case of a two-particle Green
function: 
\begin{equation}
G_2(1,2,3,4)=(i)^2\langle N|T\left[ \hat{\psi }(1)\hat{\psi }(3)%
\hat{\psi }^{\dagger }(4)\hat{\psi }^{\dagger }(2)\right] |N\rangle
\label{G2}
\end{equation}
where $1 \equiv x_1=({\bf r}_1 ,t_1)$ etc.

The self-energy may be evaluated in at least two ways, either by using
Wick's theorem (Wick 1950, Fetter and Walecka 1971)
or by Schwinger's functional derivative method
(Schwinger 1951, Martin and Schwinger 1959). We follow the latter.
This is done by introducing a time varying field $\phi ({\bf r,}t)$ which
is used as a mathematical tool for evaluating the self-energy and it will be
set to zero once the self-energy is obtained.
Working in the interaction (Dirac) picture we have
\begin{equation}
|\psi _D({\bf r},t)\rangle =\hat{U}(t,t_0)|\psi _D({\bf r},t_0)\rangle
\label{diracstate}
\end{equation}
The time development operator $\hat{U}$ is given by 
\begin{equation}
\hat{U}(t,t_0)=T\exp \left[ -i\int_{t_0}^t
 d\tau \hat{\phi }(\tau )\right]
\label{timedevelopment}
\end{equation}
\begin{equation}
\hat{\phi }(\tau )=\int d^3r\phi ({\bf r},\tau )\hat{\psi }%
_D^{\dagger }({\bf r},\tau )\hat{\psi }_D({\bf r},\tau )
\end{equation}
The relationship between operators in the Heisenberg and Dirac
representations is 
\begin{equation}
\hat{\psi }({\bf r},t)=\hat{U}^{\dagger }(t,0)\hat{\psi }_D(%
{\bf r},t)\hat{U}(t,0)  \label{heisenbergdirac}
\end{equation}
The field operator $\hat{\psi }_D$ satisfies 
\begin{equation}
i\frac \partial {\partial t}\hat{\psi }_D=\left[ \hat{\psi }_D,%
\hat{H}(\phi =0)\right]
\end{equation}
so it is the same as the unperturbed ($\phi =0$) Heisenberg operator. The
Green function can now be written as
\begin{equation}
iG(1,2)=\frac{\langle N^0|T\left[ \hat{%
U}(\infty ,-\infty )\hat{\psi }_D(1)\hat{\psi }_D^{\dagger }(%
2)\right] |N^0\rangle }{\langle N^0|\hat{U}%
(\infty ,-\infty )|N^0\rangle }  \label{defgengreenfn}
\end{equation}
By taking functional derivative of $G$ with respect to $\phi $ we get 
\begin{equation}
\frac{\delta G(1,2)}{\delta \phi (3)}=G(1,2)G(3,3^{+})-G_2(1,2,3,3^{+})
\label{dgdphi}
\end{equation}
Using the above result for $G_2$ in the 
definition of $M $ in equation (\ref{defsigma}) the term $GG$ gives the
Hartree potential $V^H$ and we define 
\begin{equation}\label{sigma} 
\Sigma=M-V^H.
\end{equation} 
The equation of motion for the Green function becomes
\begin{equation}\label{eqnofmotionxc}
{\left[ i\frac \partial {\partial t}-H_0(x)\right] G(x,x^{\prime} )} 
-\int dx''\Sigma (x,x'')G(x'',x^{\prime })=\delta (x-x^{\prime })
\end{equation}
where
\begin{equation}\label{H0} 
H_0= h_0 + V^H + \phi
\end{equation} 
Using the identity
\begin{equation}\label{ginvg} 
\frac \delta {\delta \phi }(G^{-1}G)=G^{-1}\frac{\delta G}{\delta \phi }+%
\frac{\delta G^{-1}}{\delta \phi }G=0 
\;\;\;\;\;\;\rightarrow \;\;\;\;\;\;\;\frac{\delta G}{\delta \phi }
= -G\frac{\delta G^{-1}}{\delta \phi }G
\end{equation} 
and evaluating $\delta G^{-1}/\delta\phi$, where from equation 
(\ref{eqnofmotionxc}) 
\begin{equation}\label{Ginv} 
G^{-1} = i\frac{\partial}{\partial t} - H_0-\Sigma,
\end{equation} 
we get 
\begin{equation}
\Sigma
 (1,2)=i\int d3d4G(1,3^+)W(1,4)\Lambda (3,2,4)  \label{sigmageneral}
\end{equation}
$W$ is the screened Coulomb potential 
\begin{equation}
W(1,2)=\int d3\epsilon ^{-1}(1,3)v(3-2),  \label{Wgeneral}
\end{equation}
\begin{equation}
\epsilon ^{-1}(1,2)=\frac{\delta V(1)}{\delta \phi (2)}
\end{equation}
where $V$ is the sum of the Hartree and the external potential: 
\begin{equation}
V=V^H+\phi  \label{totalpot}
\end{equation}
$\Lambda $ is the vertex function 
\begin{eqnarray}
\Lambda (1,2,3)\; &=&-\frac{\delta G^{-1}(1,2)}{\delta V(3)}  \nonumber
\\
&=&\delta (1-2)\delta (2-3)+\frac{\delta \Sigma (1,2)}{\delta V(3)}
\nonumber \\
&=&\delta (1-2)\delta (2-3)+ \int d(4567) 
\frac{\delta \Sigma (1,2)}{\delta G(4,5)} G(4,6) G(7,5) \Lambda(6,7,3)
\label{vertex}
\end{eqnarray}
The second line is obtained from equation (\ref{Ginv})
and the last line by using the 
chain rule $\delta \Sigma/\delta V=
(\delta \Sigma/\delta G)( \delta G/\delta V)$ and by
using the identity in equation
(\ref{ginvg}) and the definition of $\Lambda$.

Fourier transformation of equation (\ref{eqnofmotionxc}) gives
(with $\phi$ now set to zero)
\begin{equation}\label{ftdyson} 
\left[ \omega -H_0\left( {\bf r}\right) \right] G\left( {\bf r,r}^{\prime
},\omega \right) -\int d^3r''\Sigma ({\bf r},{\bf r}'',\omega )
G({\bf r}''%
,{\bf r}^{\prime },\omega )=\delta ({\bf r}-{\bf r}^{\prime }) 
\end{equation} 
If $G_0$ is the Green function corresponding to $\Sigma=0$, then we
have the Dyson equation 
\begin{equation}\label{dyson} 
G=G_0+G_0\Sigma G 
\end{equation} 
The first term $G_0(1,2)$ is a direct propagation from $1$ to $2$ without
exchange-correlation interaction and $\Sigma      $ contains all possible
exchange-correlation interactions with the system that an electron can have
in its propagation from $1$ to $2$.

In practical application, $G_0$ corresponds to $H_0=H^{\rm Hartree}+V^{\rm xc}$
where $V^{\rm xc}$ is some local and energy-independent exchange-correlation
potential, e.g. $V_{\rm LDA}^{\rm xc}$. In this case, the Dyson equation becomes 
\begin{equation}
G=G_0+G_0\Delta \Sigma      G  \label{deltadyson}
\end{equation}
where $\Delta \Sigma      =\Sigma      -V^{\rm xc}.$

\subsection{The polarization and response function}

The response function is an important quantity in the evaluation of the
self-energy. It is related to the inverse dielectric function $\epsilon
^{-1} $ as follows: 
\begin{eqnarray}
\epsilon ^{-1} &=&\frac{\delta V}{\delta \phi }  \nonumber \\
&=&1+v\frac{\delta \rho }{\delta \phi }  \nonumber \\
&=&1+v\frac{\delta \rho }{\delta V}\frac{\delta V}{\delta \phi }
\label{definvepsi}
\end{eqnarray}
The response function is defined as 
\begin{equation}
R(1,2)=\frac{\delta \rho (1)}{\delta \phi (2)}  \label{defresponse}
\end{equation}
which gives the change in the charge density upon a change in the 
{\em external} field.
We note that
the above response function is a time-ordered one which is related to the
physical (causal) response function $R^R$ by (Fetter and Walecka 1971)
\begin{equation}
{\rm Re} R(\omega) = {\rm Re} R^R(\omega),\;\;\; 
{\rm Im} R(\omega) {\rm sgn}\omega={\rm Im} R^R(\omega)
\end{equation} 
The polarization function is defined as
\begin{equation}
P(1,2)=\frac{\delta \rho (1)}{\delta V(2)}  \label{defpolarization}
\end{equation}
which gives the change in the charge density upon a change in the 
{\em total} (external + induced) field.
Noting that 
\begin{equation}
\rho (1)=-iG(1,1^{+})  \label{rho}
\end{equation}
we can write 
\begin{equation}
P(1,2)=-i\int d3d4G(1,3)\Lambda (3,4,2)G(4,1^{+})  \label{polarizationfunc}
\end{equation}
In summary we have 
\begin{equation}
\epsilon ^{-1}=1+vR  \label{invepsi}
\end{equation}
\begin{equation}
\epsilon =1-vP  \label{epsi}
\end{equation}
\begin{equation}
R=P+PvR  \label{eqnresponse}
\end{equation}
\begin{eqnarray}
W &=&v+vPW  \nonumber \\
&=&v+vRv  \label{eqnW}
\end{eqnarray}
\subsection{The Hedin equations}
Summarizing the results in the previous sections, we arrive at the well-known
set of coupled integral equations (Hedin 1965a, Hedin and Lundqvist 1969).
From equations (\ref{sigmageneral}), (\ref{dyson}),
(\ref{vertex}), and
(\ref{eqnW})
we have
\begin{eqnarray}\label{hedin1} 
\Sigma(1,2)&=&i\int d(34) G(1,3^+)W(1,4)\Lambda(3,2,4)
\\
G(1,2)&=& G_0(1,2) + \int d(34) G_0(1,3)\Sigma(3,4)G(4,2)
\label{hedin2}
\\
\Lambda(1,2,3)
&=&\delta (1-2)\delta (2-3)+ \int d(4567) 
\frac{\delta \Sigma (1,2)}{\delta G(4,5)} G(4,6) G(7,5) \Lambda(6,7,3)
\label{hedin3}
\\
W(1,2)&=&v(1,2) + \int d(34) v(1,3)P(3,4)W(4,2)
\label{hedin4}
\end{eqnarray} 
where $P$ is given in equation (\ref{polarizationfunc}).
Like $G$, $\Lambda$ and $W$ satisfy Dyson-like equations.
Starting from a given approximation for $\Sigma$ the above set of
equations can be used to generate higher order approximations.
Although the equations are exact, a straightforward expansion for the
self-energy in powers of the screened interaction may yield unphysical
results such as negative spectral functions (Minnhagen 1974,
Schindlmayr and Godby 1997). In fact, the expansion itself
is only conditionally convergent due to the long-range nature of the
Coulomb potential.
So far there 
is no systematic way of choosing which diagrams to sum.
The choice is usually dictated by physical intuition.

\subsection{Quasiparticles}

From the classical theory of Green functions the solution to equation (\ref
{ftdyson}) can be written in a spectral representation: 
\begin{eqnarray}\label{Gclassic}
G({\bf r},{\bf r}^{\prime },\omega )=\sum_i\frac{\Psi _i({\bf r},\omega
)\Psi _i^{\dagger }({\bf r}^{\prime },\omega )}{\omega -E_i(\omega )}
\end{eqnarray}
where $\Psi _i$ are solutions to the quasiparticle equation: 
\begin{equation}
H_0\left( {\bf r}\right) \Psi _i\left( {\bf r},\omega \right) +\int
d^3r\;\Sigma      ({\bf r},{\bf r}',\omega )\Psi _i({\bf r}',\omega
)=E_i(\omega )\Psi _i({\bf r},\omega )  \label{qpeqn}
\end{equation}
In a crystal, the index $i$ may be associated with the Bloch wave vector 
and band index.
The eigenvalues $E_i$ are in general complex and the quasiparticle
wavefunctions are not in general orthogonal because $\Sigma      $ is not
Hermitian but both the real and imaginary part of $\Sigma      $ are
symmetric. Suppose at some $\omega =\omega _i$ we find that $\omega _i={\rm 
\mathop{\rm Re}
}E_i(\omega _i)$. If ${\rm Im} E_i(\omega_i)$ is small,
then the imaginary part of $G$ is expected to have a peak
at this energy (quasiparticle peak) with a life-time given by $1/{\rm 
\mathop{\rm Im}
}E_i(\omega _i)$. 
It may happen that $\omega-{\rm Re}E_i(\omega)$ 
is zero or close to zero at some other
energies and if the corresponding Im$E_i(\omega)$ are small, then we get
satellites.
For a non-interacting
system, $\Sigma      $ is Hermitian and therefore $E_i$ is real so that the
quasiparticle has an infinite life-time.

The spectral representation can also be obtained directly from the
definition of $G$ by inserting a complete set of $(N\pm 1)$-electron states
in between the field operators and performing Fourier transformation,
keeping in mind that the field operators are in the Heisenberg representation,
i.e. $\hat{\psi}(t)=\rm{exp}(i\hat{H}t)\hat{\psi}(0)\rm{exp}
(-i\hat{H}t)$. 
\begin{equation}\label{Gspectral}
G({\bf r},{\bf r}^{\prime },\omega )=\int_{-\infty }^\mu d\omega ^{\prime }%
\frac{A({\bf r},{\bf r}^{\prime },\omega ^{\prime })}{\omega -\omega
^{\prime }-i\delta }+\int_\mu ^\infty d\omega ^{\prime }\frac{A({\bf r},{\bf %
r}^{\prime },\omega ^{\prime })}{\omega -\omega ^{\prime }+i\delta }
\end{equation}
The spectral function or density of states $A$ is given by 
\begin{eqnarray}
A({\bf r},{\bf r}^{\prime },\omega ) &=&-\frac 1\pi {\rm 
\mathop{\rm Im}
}\;G({\bf r},{\bf r}^{\prime },\omega ){\rm sgn}(\omega -\mu )  \nonumber \\
&=&\sum_ih_i({\bf r})h_i^{*}({\bf r}^{\prime })\delta [\omega -\mu +e(N-1,i)]
\\
&&+\sum_ip_i^{*}({\bf r})p_i({\bf r}^{\prime })\delta [\omega -\mu -e(N+1,i)]
\end{eqnarray}
where 
\begin{eqnarray}
h_i({\bf r}) &=&\langle N-1,i|\hat{\psi}({\bf r},0)|N\rangle \\
p_i({\bf r}) &=&\langle N+1,i|\hat{\psi}^{\dagger }({\bf r},0)|N\rangle
\end{eqnarray}
$|N\pm 1,i\rangle $ is the $i$th eigenstate of the $N\pm 1$ electrons with
an excitation energy 
\begin{eqnarray}
e(N\pm 1,i)=E(N\pm 1,i)-E(N\pm 1)
\end{eqnarray}
which is positive and $E(N\pm 1)$ is the ground-state energy of the $N\pm 1$
electrons. The quantity $\mu $ is the chemical potential 
\begin{eqnarray}
\mu &=&E(N+1)-E(N)  \nonumber \\
&=&E(N)-E(N-1)+O(1/N)
\end{eqnarray}
The physical meaning of the poles of $G$ is therefore the exact excitation
energies of the $N\pm 1$ electrons. 
Since the poles of $G$ in equations
(~\ref {Gclassic}) and (~\ref{Gspectral}) must be the same,
it follows that the real part of $E_i(\omega _i)$ are also the
excitation energies of the $N\pm 1$ electrons.
For a very large system, the poles are usually so close together
that it is meaningless to talk about the individual poles,
and in an infinitely large system, the poles form a branch cut.
In this case, it is more meaningful to 
interpret the excitation spectrum in terms of quasiparticles with energies
${\rm Re} E_i(\omega_i)$ and life-times $1/{\rm Im} E_i(\omega_i)$.

From equation (\ref{dyson}), the spectral function $A$ is schematically
given by 
\begin{eqnarray}
A(\omega ) &=&\frac 1\pi \sum_i|{\rm 
\mathop{\rm Im}
}\;G_i(\omega )|  \nonumber \\
&=&\frac 1\pi \sum_i\frac{|{\rm 
\mathop{\rm Im}
}\;\Sigma _i     (\omega )|}{|\omega -\varepsilon _i-{\rm 
\mathop{\rm Re}
}\;\Delta \Sigma _i     (\omega )|^2+|{\rm 
\mathop{\rm Im}
}\;\Sigma _i     (\omega )|^2}
\end{eqnarray}
where $G_i$ is the matrix element of $G$ in an eigenstate $\psi _i$ of the
non-interacting system $H_0$. $A$ is usually peaked at each energy $%
E_i=\varepsilon _i+{\rm 
\mathop{\rm Re}
}\Delta \Sigma _i     (E_i)$ (quasiparticle peak) with a life-time given by $%
1/|{\rm 
\mathop{\rm Im}
}\;\Sigma _i     (E_i)|$ and renormalization factor 
(weight of the Lorentzian) 
\begin{equation}
Z_i=\left[ 1-\frac{\partial {\rm 
\mathop{\rm Re}
}\;\Delta \Sigma _i     (E_i)}{\partial \omega }\right] ^{-1}<1
\label{zfactor}
\end{equation}
At some other energies $\omega _p$, the denominator may be small and $%
A(\omega _p)$ could also show peaks or satellite structure which can be due
to plasmon excitations or other collective phenomena.

If we start with a single-particle Hamiltonian which is in some sense
close to the true
interacting Hamiltonian, the quasiparticles of the former Hamiltonian
are just a set of $\delta $%
-functions centred at the single-particle eigenvalues. If the interaction is
switched on, typically the delta functions become broadened since the
single-particle states can now decay to other excitations and lose some
weight which might appear as collective excitations or satellite structures.
The term quasiparticle is arbitrary but we usually refer to quasiparticle as
an excitation originating from a single-particle state and to satellite as
an excitation not contained in the approximate non-interacting system.

The quasiparticle energy can also be calculated to first order in
$E_i-\varepsilon_i$ as follows: 
\begin{eqnarray}
E_i &=&\varepsilon _i+{\rm 
\mathop{\rm Re}
\Delta }\Sigma _i     (E_i)  \nonumber \\
&=&\varepsilon _i+{\rm 
\mathop{\rm Re}
\Delta }\Sigma _i     (\varepsilon _i)+(E_i-\varepsilon _i)\frac{\partial 
{\rm 
\mathop{\rm Re}
\Delta }\Sigma _i     (\varepsilon _i)}{\partial \omega }  \nonumber \\
&=&\varepsilon _i+Z_i{\rm 
\mathop{\rm Re}
\Delta }\Sigma _i     (\varepsilon _i)  \label{qpenergy}
\end{eqnarray}

\subsection{The $GW$ approximation}

The GWA may be regarded as a generalization of the Hartree-Fock
approximation (HFA) but with a dynamically
screened Coulomb interaction. The non-local
exchange potential in the HFA is given by 
\begin{equation}
\Sigma ^x({\bf r,r}^{\prime })=-\sum_{{\bf k}n}^{\rm occ}
\psi _{{\bf k}n}({\bf r}%
)\psi _{{\bf k}n}^{*}({\bf r}^{\prime })v({\bf r-r}^{\prime })
\label{sigmaexch}
\end{equation}
In the Green function theory, the exchange potential is written as 
\[
\Sigma ^x({\bf r,r}^{\prime },t-t^{\prime })=iG({\bf r,r}^{\prime
},t-t^{\prime })v({\bf r-r}^{\prime })\delta (t-t^{\prime }) 
\]
which when Fourier transformed yields equation (\ref{sigmaexch}). The GWA
corresponds to replacing the bare Coulomb interaction $v$ by a screened
interaction $W$ :

\begin{equation}
\Sigma (1,2)=iG(1,2)W(1,2)  \label{defgwspacetime}
\end{equation}
This is physically well motivated especially in metals where the HFA leads
to unphysical results such as zero density of states at the Fermi level, due
to the lack of screening. Formally, the GWA is obtained by neglecting the
second term in the vertex function in equation (\ref{vertex}), i.e. setting
$\Lambda(1,2,3)= \delta(1-2)\delta(2-3)$. Fourier
transforming equation (\ref{defgwspacetime}) we get 
\begin{equation}\label{sigmagw} 
\Sigma 
({\bf r,r}^{\prime },\omega )=\frac i{2\pi }\int d\omega ^{\prime }G(%
{\bf r,r}^{\prime },\omega +\omega ^{\prime })W({\bf r,r}^{\prime },\omega
^{\prime }) 
\end{equation} 
For a non-interacting $G_0$ the imaginary part of the
correlation part of the self-energy can be evaluated explicitly: 
\begin{eqnarray}
\mathop{\rm Im}
\Sigma^{c} ({\bf r,r}^{\prime },\omega \leq \mu )=\pi \sum_{{\bf k}%
n}^{\rm occ}
\psi _{{\bf k}n}({\bf r})\psi _{{\bf k}n}^{*}({\bf r}^{\prime }) 
\mathop{\rm Im}
W^c({\bf r,r}^{\prime },\varepsilon _{{\bf k}n}-\omega )\theta (\varepsilon _{%
{\bf k}n}-\omega )  \label{imsignegw}
\end{eqnarray}
\begin{eqnarray}
\mathop{\rm Im}
\Sigma^{c} 
({\bf r,r}^{\prime },\omega >\mu )=-\pi \sum_{{\bf k}n}^{\rm unocc}\psi
_{{\bf k}n}({\bf r})\psi _{{\bf k}n}^{*}({\bf r}^{\prime }) 
\mathop{\rm Im}
W^c({\bf r,r}^{\prime },\omega -\varepsilon _{{\bf k}n})\theta (\omega
-\varepsilon _{{\bf k}n})  \label{imsigposw}
\end{eqnarray}
where 
\begin{equation}\label{Wcdef} 
W^c = W - v
\end{equation} 
is the frequency-dependent part of $W$.
The above results are obtained by expressing $G$ and $W^c$ in their
spectral representations. The spectral representation of $G$ is given in
equation (\ref{Gspectral}). For $W^c$ it is given by
\begin{equation}\label{Wspectral} 
W^c({\bf r}{,{\bf r}^{\prime }},\omega )=\int_{-\infty }^0d\omega ^{\prime
}\;\frac{D({\bf r}{,{\bf r}^{\prime }},\omega ^{\prime })}{\omega -\omega
^{\prime }-i\delta }+\int_0^\infty d\omega ^{\prime }\;\frac{D({\bf r}{,{\bf
r}^{\prime }},\omega ^{\prime })}{\omega -\omega ^{\prime }+i\delta }
\label{whilbert}
\end{equation}
$D$ is proportional to the imaginary part of $W$ and defined to be
anti-symmetric in $\omega $: 
\begin{eqnarray}
D({\bf r}{,{\bf r}^{\prime }},\omega ) &=&-\frac 1\pi 
{\rm Im}\;W^c({\bf r}{,{\bf
r}
^{\prime }},\omega ){\rm sgn}(\omega ) \\
D({\bf r}{,{\bf r}^{\prime }},-\omega ) &=&-D({\bf r}{,{\bf r}^{\prime }}
,\omega )
\end{eqnarray}

The spectral representation of the correlation part of the self-energy is
\begin{equation}
\Sigma^c ({\bf r,r}^{\prime },\omega )=
\int_{-\infty }^\mu d\omega ^{\prime }%
\frac{\Gamma({\bf r},{\bf r}^{\prime },\omega ^{\prime })}{\omega -\omega
^{\prime }-i\delta }+\int_\mu ^\infty d\omega ^{\prime }
\frac{\Gamma({\bf r},{\bf %
r}^{\prime },\omega ^{\prime })}{\omega -\omega ^{\prime }+i\delta }
\label{realsig}
\end{equation}
where
\begin{equation}
\Gamma({\bf r},{\bf r}^{\prime },\omega ) =-\frac 1\pi {\rm 
\mathop{\rm Im}
}\;\Sigma^{c}
({\bf r},{\bf r}^{\prime },\omega ){\rm sgn}(\omega -\mu ) 
\end{equation} 
The real part of $\Sigma^c$ can be obtained
by performing the principal value integration 
(Kramers-Kronig
relation or Hilbert transform). 

A physically appealing way of expressing the self-energy is by dividing it
into a screened-exchange term $\Sigma _{{\rm SEX}}$ and a Coulomb-hole term $%
\Sigma _{{\rm COH}}$ (COHSEX) (Hedin 1965a, Hedin and Lundqvist 1969). 
It is straightforward to verify that the real
part of the self-energy can be written as 
\begin{equation}
{\rm 
\mathop{\rm Re}
}\Sigma _{{\rm SEX}}\left( {\bf r,r}^{\prime },\omega\right) 
=-\sum_i^{{\rm occ}%
}\psi _i({\bf r})\psi _i^{*}({{\bf r}^{\prime }})
{\rm Re}W\left( {\bf r,r}^{\prime
},\omega-\varepsilon _i\right)  \label{SEX}
\end{equation}
\begin{equation}
{\rm 
\mathop{\rm Re}
}\Sigma _{{\rm COH}}\left( {\bf r,r}^{\prime },\omega\right) 
=\sum_i\psi _i({\bf r%
})\psi _i^{*}({{\bf r}^{\prime }}){\rm P}\int_0^\infty d\omega'\frac{D\left( 
{\bf r,r}^{\prime },\omega' \right) }
{\omega-\varepsilon _i-\omega'}  \label{COH}
\end{equation}
The physical interpretation of $\Sigma _{{\rm COH}}$ becomes clear in the
static approximation due to Hedin (1965a). 
If we are interested in a state with energy $E$ close to
the Fermi level, the matrix element $\langle \psi |{\rm 
\mathop{\rm Re}
}\Sigma _{{\rm COH}}(E)|\psi \rangle $ 
picks up most of its weight from states with energies $\varepsilon_i$
close to $E$ in energy. We may then assume that $E-\varepsilon_i$ is
small compared to the main excitation energy of $D,$ which is at the
plasmon energy. If we set $E-\varepsilon _i=0,$ we get 
\begin{equation}
{\rm 
\mathop{\rm Re}
}\Sigma _{{\rm COH}}\left( {\bf r,r}^{\prime }\right) =\frac 12\delta \left( 
{\bf r-r}^{\prime }\right) W^c\left( {\bf r,r}^{\prime },0\right) 
\end{equation} 
This is simply the interaction energy of the quasiparticle with the induced
potential due to the screening of the electrons around the quasiparticle.
The factor of 1/2 arises from the adiabatic growth of the interaction. In
this static COHSEX approximation, $\Sigma _{{\rm COH}}$ becomes local.

The polarization function needed to evaluate $W$ is calculated within the
random phase approximation (RPA) (Pines and Bohm 1952, Bohm and Pines 1953,
Lindhard 1954, Pines 1961, Gell-Mann and Brueckner 1957) 
which corresponds to neglecting the second
term in the vertex function and using a non-interacting $G_0$ : 
\begin{eqnarray}
P({\bf r,r}^{\prime },\omega ) &=&\sum_{{\rm spin}}\sum_{{\bf k}n}^{{\rm occ}%
}\sum_{{\bf k}^{\prime }n^{\prime }}^{{\rm unocc}}\psi _{{\bf k}n}^{*}({\bf r%
})\psi _{{\bf k}^{\prime }n^{\prime }}({\bf r})\psi _{{\bf k}^{\prime
}n^{\prime }}^{*}({\bf r}^{\prime })\psi _{{\bf k}n}({\bf r}^{\prime }) 
\nonumber \\
&&\times \left\{ \frac 1{\omega -\varepsilon _{{\bf k}^{\prime }n^{\prime
}}+\varepsilon _{{\bf k}n}+i\delta }-\frac 1{\omega +\varepsilon _{{\bf k}%
^{\prime }n^{\prime }}-\varepsilon _{{\bf k}n}-i\delta }\right\}
\label{rpapol}
\end{eqnarray}
We have used the fact that for every $\psi_{{\bf k}n}$ there is
$\psi_{{\bf -k}n}^*$ with the same eigenvalue due to the 
time-reversal symmetry.
The wavefunction $\psi _{{\bf k}n}$ is normalized to unity in the entire
space. 
The physical meaning of the RPA is that the electrons respond to the
total field (external + induced field) as if they were non-interacting.

\section{Numerical methods}

One of the main computational problems is to calculate the polarization in
equation (\ref{rpapol}). Since $P$ can be long-ranged for crystals, the
conventional method of calculating it is to use Bloch basis functions.
Noting that $P({\bf r+T,r}^{\prime }{\bf +T)}=P({\bf r,r}^{\prime })$ , $P$
can be expanded in general as follows 
\begin{equation}
P({\bf r,r}^{\prime },\omega )=\sum_{{\bf q}ij}B_{{\bf q}i}({\bf r})P_{ij}(%
{\bf q},\omega )B_{{\bf q}j}^*({\bf r}^{\prime })  \label{polexpansion}
\end{equation}
where $\left\{ B_{{\bf q}i}\right\} $ is a set of Bloch basis functions
large enough to describe $P$ and the sum over ${\bf q}$ is restricted to the
first Brillouin zone. The matrix elements of $P$ in the basis are given by 
\begin{eqnarray*}
P_{ij}({\bf q},\omega ) &=&\sum_{{\rm spin},{\bf k}}\sum_n^{{\rm occ}%
}\sum_{n^{\prime }}^{{\rm unocc}}\langle B_{{\bf q}i}\psi _{{\bf k}n}|\psi _{%
{\bf k}+{\bf q\,}n^{\prime }}\rangle \langle \psi _{{\bf k}+{\bf q\,}%
n^{\prime }}|\psi _{{\bf k}n}B_{{\bf q}j}\rangle \\
&&\times \left\{ \frac 1{\omega -\varepsilon _{{\bf k}+{\bf q\,}n^{\prime
}}+\varepsilon _{{\bf k}n}+i\delta }-\frac 1{\omega +\varepsilon _{{\bf k}+%
{\bf q\,}n^{\prime }}-\varepsilon _{{\bf k}n}-i\delta }\right\}
\end{eqnarray*}
Similarly, the Coulomb potential $v$ can be expanded as in equation (\ref
{polexpansion}). The screened potential $W$ can then be calculated from equation (%
\ref{eqnW}). 

The matrix element of the imaginary part of the self-energy 
in a state $\psi_{{\bf q}n}$ is given by
\begin{eqnarray}\label{imsig} 
{\rm Im}\Sigma^c_{{\bf q}n}(\omega) 
&=&\pi
\sum_{{\bf k}}\sum_{n^{\prime}\leq\mu} \sum_{ij} \langle \psi_{{\bf
q}n}
\psi_{{{\bf k}-{\bf q}},n^{\prime}} |B_{{\bf k}i} \rangle\;
{\rm Im}W^c_{ij}({\bf k},\omega-\varepsilon_{{{\bf k}-{\bf q}},
n^{\prime}})
\nonumber \\ 
&&\;\;\;\;\;\;\;\;\;\;\;\;\;\;\;\;\;\;\;
\times \; \langle B_{{\bf k}j}
|\psi_{{
{\bf k}-{\bf q}},n^{\prime}} \psi_{{\bf q}n} \rangle \;
\theta(\omega-\varepsilon_{{{\bf k}-{\bf q}},n^{\prime}})
\;\;\;\;\;\;\;\;\;\;{\rm for}\;\;\omega\leq\mu 
\end{eqnarray}
If we are only interested in quasiparticle energies, it is more
favourable to perform the frequency
integration in the expression for the self-energy in equation 
(\ref{sigmagw}) along the imaginary axis (Godby, Schl\"uter, and
Sham 1988). In this case, $W$ must also be
calculated along the imaginary axis which is advantageous since
the pole structure along the real axis is avoided.
For states away from the Fermi level, there is in addition
a contribution to the self-energy
from the poles of the Green function.

The choice of basis functions depends on the type of materials
we are interested in. For sp systems, plane waves are appropriate
especially when used in conjunction with pseudopotentials.
For systems containing rather localized states such as the 3d
and 4f systems, a large number of plane waves would be needed and
therefore localized basis functions are more suitable.

\subsection{Plane-wave basis}

In this case $B_{{\bf k}j}\rightarrow \exp \left[ i({\bf k+G)\cdot r}\right]
/\sqrt{\Omega }$ where the index $j$ is represented by the reciprocal
lattice vector {\bf G}
 and $\Omega $ is the unit cell volume. This is probably
the simplest basis with the following advantages:

\begin{itemize}
\item  Programming ease: The matrix elements can be calculated easily
particularly when the wave-functions are also expanded in plane waves.

\item  The Coulomb potential is diagonal with matrix elements given by $4\pi
/|{\bf k+G|}^2$.

\item  Good control over convergence.
\end{itemize}

The disadvantages are:

\begin{itemize}
\item  It is not feasible to do all-electron calculations. In many cases,
it is essential to include core electrons. For example, the exchange of a 
3d valence state with the 3s and 4p core states in the
late 3d transition metals is overestimated by the LDA by as much as 1
eV which would lead to an error of the same order in the pseudopotential
method.

\item  The size of the response matrix becomes prohibitively large for
narrow-band systems due to a large number of plane waves.

\item  No direct physical interpretation.
\end{itemize}

\subsection{Localized basis}

For systems containing 3d or 4f electrons, plane-wave basis
becomes very costly. Methods based on the
linear-muffin-tin-orbital (LMTO) basis
or Gaussian basis are more appropriate. In the LMTO method (Andersen 1975), 
the wave
functions are expanded as follows 
\begin{equation}
\psi _{{\bf k}n}({\bf r})=\sum_{{\bf R}L}\chi _{{\bf R}L}({\bf r,k})b_{{\bf k%
}n}({\bf R}L) 
\end{equation} 
where $\chi $ is the LMTO basis which in the atomic sphere approximation
(ASA) for ${\bf r}$ in the central cell is given by 
\begin{equation}
\chi _{{\bf R}L}({\bf r,k})=\phi _{{\bf R}L}({\bf r})+\sum_{{\bf R}^{\prime
}L^{\prime }}\stackrel{.}{\phi }_{{\bf R}^{\prime }L^{\prime }}({\bf r})h_{%
{\bf R}^{\prime }L^{\prime },{\bf R}L}({\bf k}) 
\end{equation} 
$\phi _{{\bf R}L}({\bf r})=\varphi _{{\bf R}l}(r)Y_L(\Omega )$ is a solution
to the
Schr\"{o}dinger equation inside a sphere centred on an atom at site ${\bf %
R}$ for a certain energy $\epsilon _\nu ,$ normally chosen at the centre of
the band. $\stackrel{.}{\phi }_{{\bf R}L}$ is the energy derivative of $\phi
_{{\bf R}L}$ at $\epsilon _\nu .$ One advantage of the LMTO method is that $%
\phi _{{\bf R}L}$ is independent of ${\bf k}.$ When forming the polarization
function in equation (\ref{rpapol}) we have products of wave functions for $%
{\bf r}$ or ${\bf r}^{\prime }$ of the form 
\begin{equation}
\psi \psi =[\phi \phi +(\phi \stackrel{.}{\phi }+\stackrel{.}{\phi }\phi )h+%
\stackrel{.}{\phi }\stackrel{.}{\phi }h^2]b^2 
\end{equation} 
It is then clear that the sets of products $\phi \phi ,$ $\phi \stackrel{.}{%
\phi },$ $\stackrel{.}{\phi }\stackrel{.}{\phi }$ form a complete basis for
the polarization function and the response function (Aryasetiawan and
Gunnarsson 1994a) since the latter can be
written as 
\begin{equation}
R=P+PvP+PvPvP+\ldots 
\end{equation} 
so that the basis for $R$ is completely determined by that of $P.$ Although
this product basis is not complete for the Coulomb potential, it is of no
consequence since $v$ is always sandwiched between two $P$'s. It can be
easily shown that the product basis is also complete for the self-energy.
Thus schematically 
\begin{eqnarray}
\Sigma ({\bf k}n,\omega ) &=&\langle \psi _{{\bf k}n}|iGW|\psi _{{\bf k}%
n}\rangle \nonumber\\
&=&\ \langle \psi _{{\bf k}n}\psi |v|\psi \psi _{{\bf k}n}\rangle +\langle
\psi _{{\bf k}n}\psi |vRv|\psi \psi _{{\bf k}n}\rangle
\end{eqnarray}
which shows that it is sufficient to expand $v$ in the product basis and the
latter is therefore a complete basis for the self-energy.

\noindent
\begin{figure}[t]
\unitlength1cm
\begin{minipage}[t]{15.0cm}
\rotatebox{0.0}
{\centerline{\epsfxsize=4.5in \epsffile{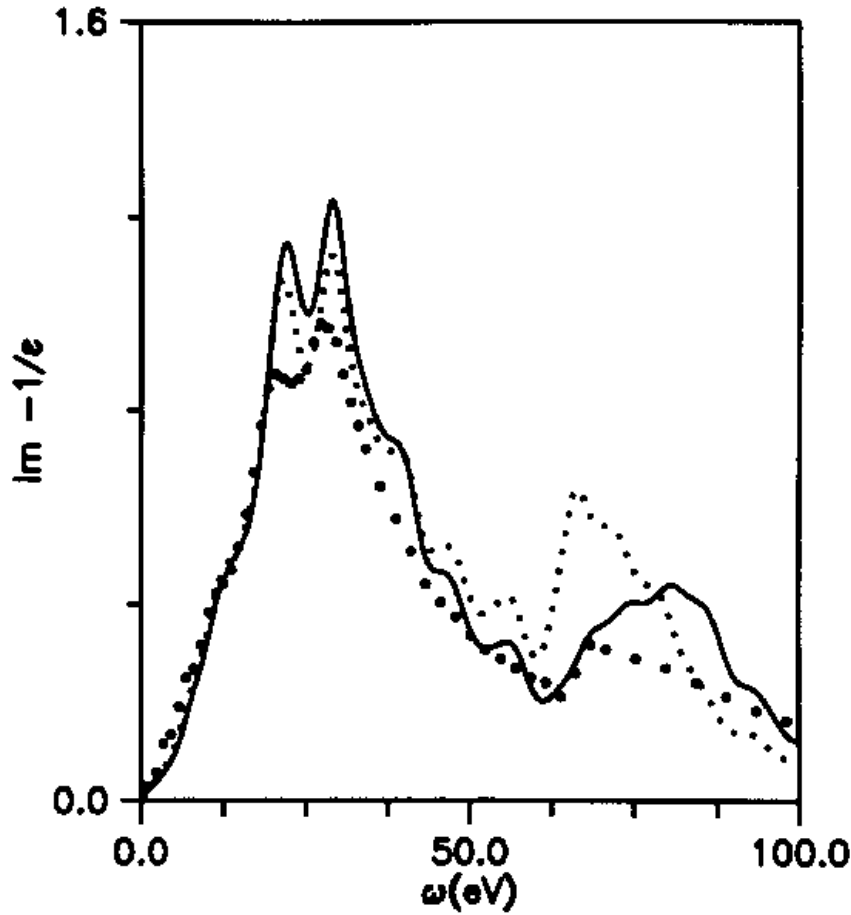}}}
\vskip0.3cm
\caption[]{\label{ag3}
The energy loss spectra of Ni for ${\bf q}$ = (0.25, 0 0) $2\pi/a$, 
$a=6.654 a_0$.
The large dots correspond to the experimental spectrum taken from
Feldcamp, Stearns, and Shinozaki (1979).
The solid line and small dots are respectively
the loss spectra with and without local field corrections due to the
inhomogeneity in the charge density. Both spectra are calculated with 4s, 4p,
3d, 4d, 4f, and 5g orbitals, including an empty sphere at (0.5, 0.5, 0.5)$a$
and core excitations.
After Aryasetiawan and Gunnarsson (1994).
}
\end{minipage}
\hfill
\end{figure}

\noindent

The number of product functions is still large, with nine spd orbitals 
we have $2[9(9+1)/2]$ products of $\phi\phi$ and $\dot\phi\dot\phi$
and $9\times 9$ products of $\phi\dot\phi$ 
giving in total 161 product functions.
With spdf orbitals the number of product functions is 528.
It can be reduced
considerably without too much loss of accuracy by neglecting the $\stackrel{.%
}{\phi }$ terms since they are small. Moreover, in the polarization function
there are no products between conduction states. Therefore in sp and d
systems products of $\phi _d\phi _d$ and $\phi _f\phi _f$ respectively can
be neglected. After these eliminations, the 
remaining product functions turn
out to have a significant number of linear dependencies, typically 30-50 \%,
which can be eliminated further giving $\sim 100$ optimized product
functions per atom for spdf orbitals. A product function is given by 
\begin{eqnarray}
\widetilde{B}_\alpha ({\bf r)} &=&\phi _{{\bf R}L}({\bf r})\phi _{{\bf R}%
L^{\prime }}({\bf r}) \nonumber\\
&=&\varphi _{{\bf R}l}(r)\varphi _{{\bf R}l^{\prime }}(r)Y_L(\Omega
)Y_{L^{\prime }}(\Omega)
\end{eqnarray}
where $\alpha \equiv ({\bf R,}LL^{\prime }).$ This function is non-zero only
inside an atomic sphere centred on atom ${\bf R.}$ 
There are no products
between orbitals centred on different spheres. The optimization is performed
by calculating the eigenvalues of the overlap matrix $O_{\alpha \beta
}=\langle \widetilde{B}_\alpha |\widetilde{B}_\beta \rangle $ and
subsequently neglecting eigenvectors with eigenvalues less than a certain
tolerance. The optimized product functions are then linear combinations 
\begin{equation}
B_\alpha =\sum_\gamma \widetilde{B}_\gamma z_{\gamma \alpha } 
\end{equation} 
where $z$ are the eigenvectors of the overlap matrix: $Oz=ez.$ 
Due to the localized property of the basis, the calculation of the dielectric
matrix scales as N$^3$. This approach
has been used with success to calculate loss spectra (Aryasetiawan and
Gunnarsson 1994b) and self-energy for 3d
systems (Aryasetiawan and Gunnarsson 1995).
The loss spectra of Ni calculated using the product basis is shown in figure
\ref{ag3}.

Another approach using a localized basis set is based on Gaussian functions
(Rohlfing, Kr\"uger, and Pollmann 1993).
The wave functions and the dielectric matrix are expanded in this basis. For
Si, for example, the number of Gaussian orbitals needed to perform {\em GW}
calculation is 40-60 whereas 350 plane waves are needed in the conventional
approach. This approach has been applied with success to a number of
semiconductors and insulators and to semicore states in Si, Ge,
and CdS as well as to the Si surface.

\subsection{The plasmon-pole approximation}

One of the major computational efforts in self-energy calculations is the
calculation of the screened interaction $W.$ The physical features of $W$
are well known; the imaginary part of $W$ is characterized by a strong peak
corresponding to a plasmon excitation at the plasmon frequency. This is
particularly evident in the case of the electron gas or the alkalis such as
Na and Al. The plasmon-pole approximation assumes that all the weight in Im $%
W$ resides in the plasmon excitation (Lundqvist 1967, Overhauser 1971,
Hedin and Lundqvist 1969, Hybertsen and Louie
1986). In the case of the electron gas, this is strictly true in the limit
of long wave length ${\bf q}\rightarrow 0.$ For finite ${\bf q}$, the
spectrum contains also particle-hole excitations at lower energies.
The particle-hole spectrum eventually merges with the plasmon excitation as $%
{\bf q}$ gets larger. Thus, in the simplest form, the plasmon-pole
approximation is given by (Lundqvist 1967, Hedin and Lundqvist 1969)
${\rm  \mathop{\rm Im}
\,}\epsilon ^{-1}\left( {\bf q,}\omega \right) =A_{{\bf q}}\delta \left(
\omega -\omega _{{\bf q}}\right) $. The two parameters $A_{{\bf q}}$ and $%
\omega _{{\bf q}}$ are determined from the static- and {\em f}-sum rules. In
the generalized plasmon-pole approximation due to Hybertsen and Louie (1986),
each
matrix component of the inverse dielectric function is written (for positive
frequency) 
\begin{equation}
{\rm 
\mathop{\rm Im}
\,}\epsilon _{{\bf GG}^{\prime }}^{-1}\left( {\bf q,}\omega \right) =A_{{\bf %
GG}^{\prime }}\left( {\bf q}\right) \delta \left( \omega -\omega _{{\bf GG}%
^{\prime }}\left( {\bf q}\right) \right) 
\end{equation} 
The corresponding real part is given by 
\begin{equation}
{\rm 
\mathop{\rm Re}
\,}\epsilon _{{\bf GG}^{\prime }}^{-1}\left( {\bf q},\omega \right) =\delta
_{{\bf GG}^{\prime }}+\frac{\Omega _{{\bf GG}^{\prime }}^2\left( {\bf q}%
\right) }{\omega ^2-\omega _{{\bf GG}^{\prime }}^2\left( {\bf q}\right) } 
\end{equation} 

\noindent
\begin{figure}[bt]
\unitlength1cm
\begin{minipage}[t]{15.0cm}
\rotatebox{0.0}
{\centerline{\epsfxsize=3.0in \epsffile{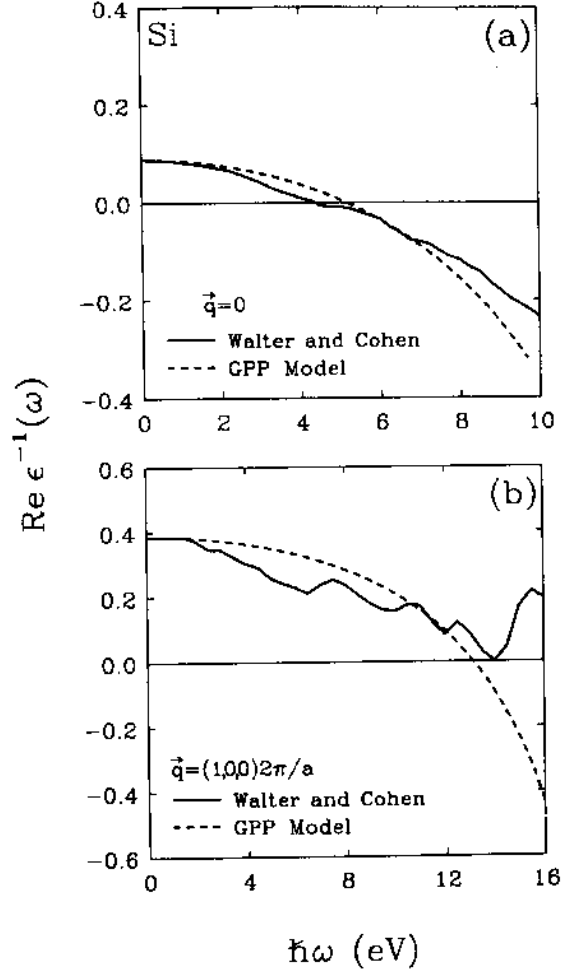}}}
\vskip0.3cm
\caption[]{\label{hl2}
Comparison between the numerically calculated inverse 
dielectric function (Walter and Cohen 1972) and 
the corresponding plasmon-pole results for Si. 
After Hybertsen and Louie (1988a).
}
\end{minipage}
\hfill
\end{figure}
\noindent
The effective bare plasma frequency $\Omega _{{\bf GG}^{\prime }}$ is
defined below. The unknown parameters $A_{{\bf GG}^{\prime }}\left( {\bf q}%
\right) $ and $\omega _{{\bf GG}^{\prime }}\left( {\bf q}\right) $ are
determined from the static limit of $\epsilon^{-1}$: 
\begin{equation}
{\rm 
\mathop{\rm Re}
\,}\epsilon _{{\bf GG}^{\prime }}^{-1}\left( {\bf q},0\right)
=\delta _{{\bf GG}^{\prime }}+\frac 2\pi {\rm P}\int_0^\infty d\omega \frac 1%
\omega {\rm 
\mathop{\rm Im}
\,}\epsilon _{{\bf GG}^{\prime }}^{-1}\left( {\bf q},\omega \right) 
\end{equation} 
and the {\em f}-sum rule: 
\begin{equation}
\int_0^\infty d\omega \,\omega \,{\rm 
\mathop{\rm Im}
\,}\epsilon _{{\bf GG}^{\prime }}^{-1}\left( {\bf q},\omega \right) =-\frac 
\pi 2\omega _p^2\frac{\left( {\bf q+G}\right) \cdot \left( {\bf q+G}^{\prime
}\right) }{\left| {\bf q+G}\right| ^2}\frac{\rho \left( {\bf G-G}^{\prime
}\right) }{\rho \left( 0\right) } 
\equiv -\frac \pi 2 \Omega^2_{\bf GG'}
\end{equation} 
Thus, there are no adjustable parameters.
The ${\em f}$-sum rule is true for the exact response function
since it is obtained from the
double commutator $\left[ \left[ H,\hat{\rho }_{{\bf q+G}}\right] ,%
\hat{\rho }_{{\bf q+G}^{\prime }}\right] $ which yields 
\begin{equation}
\frac 12\sum_s\left( E_s-E_0\right) \left\{ \langle 0|\hat{\rho }_{{\bf %
q+G}}|s\rangle \langle s|\hat{\rho }_{{\bf q+G}^{\prime }}^{\dagger
}|0\rangle +{\rm c.c.}\right\} =\left( {\bf q+G}\right) \cdot \left( {\bf q+G%
}^{\prime }\right) \rho \left( {\bf G-G}^{\prime }\right) 
\end{equation} 
The states $|s\rangle $ are the exact many-body excited states and $\rho
\left( {\bf G}\right) $ is the Fourier component of the electron density. It
can also be shown from the definition of the linear response dielectric
function in terms of the ground-state matrix element of the commutator of
the density operators that 
\begin{eqnarray}
\int_0^\infty d\omega \,\omega \,{\rm 
\mathop{\rm Im}
\,}\epsilon _{{\bf GG}^{\prime }}^{-1}\left( {\bf q,}\omega \right) &=&-%
\frac \pi 2v\left( {\bf q+G}\right) \nonumber\\
&&\times \frac 12\sum_s\left( E_s-E_0\right) \left\{ \langle 0|\hat{\rho 
}_{{\bf q+G}}|s\rangle \langle s|\hat{\rho }_{{\bf q+G}^{\prime
}}^{\dagger }|0\rangle +{\rm c.c.}\right\}
\end{eqnarray}
The {\em f}-sum rule follows immediately from the last two equations
(Hybertsen and Louie 1986) recalling that $\omega_p=4\pi \rho(0)$.
The quality of the plasmon-pole approximation is illustrated in figure 
\ref{hl2}.

One drawback of the Hybertsen and Louie model is that the plasmon frequency
may become complex, which is somewhat unphysical. A different model proposed
by von der Linden and Horsch (1988)
circumvents this problem by using the dielectric-bandstructure approach. Here one defines a Hermitian dielectric
matrix 
\begin{equation}
\epsilon _{{\bf GG}^{\prime }}\left( {\bf q},\omega \right) =\delta _{{\bf GG%
}^{\prime }}+\frac{4\pi }{\left| {\bf q+G}\right| \left| {\bf q+G}^{\prime
}\right| }\alpha_{{\bf GG}^{\prime }}^0\left( {\bf q},\omega \right) 
\end{equation} 
\begin{equation}
\alpha_{{\bf GG}^{\prime }}^0\left( {\bf q},\omega \right) =
\sum_{\rm spin}\sum_{{\bf k}%
}\sum_n^{\rm occ}\sum_m^{\rm unocc}
\frac{\langle {\bf k,}n|e^{i({\bf q+G)}\cdot {\bf %
r}}|{\bf k+q},m\rangle \langle {\bf k+q},m|e^{-i({\bf q+G}^{\prime })\cdot 
{\bf r}}|{\bf k},n\rangle }
{\omega -\varepsilon _{{\bf k}n}+\varepsilon _{{\bf k+q},m}} 
\end{equation} 
The inverse
static dielectric matrix is expressed in its eigen representation 
\begin{equation}
\epsilon _{{\bf GG}^{\prime }}^{-1}({\bf q},0)=\delta _{{\bf GG}^{\prime
}}+\sum_{i=1}^\infty U_{{\bf q},i}({\bf G})\left[ d _i^{-1}({\bf q}%
)-1\right] U_{{\bf q},i}^{*}({\bf G}^{\prime }) 
\end{equation}
The matrix $U$ is formed by the eigenvectors of the inverse dielectric
matrix. The plasmon-pole approximation is then obtained by introducing the
frequency dependence in the {\em eigenvalues}: 
\begin{equation}
d _i^{-1}({\bf q,}\omega )-1=\frac{z_i({\bf q})}{\omega ^2-\left[
\omega _i({\bf q})-i\delta \right] ^2} 
\end{equation} 
The eigenfunctions for $\omega\neq 0$ are approximated by the static
eigenfunctions (von der Linden and Horsch 1988).
As in the Hybertsen and Louie approach, the unknown quantities $z_i$ and $%
\omega _i$ are determined from the static limit
of $\epsilon^{-1}$ and the {\em f}-sum rule. These give 
\begin{equation}
z_i({\bf q})=\frac{\omega _p^2}{\rho (0)}\sum_{{\bf GG}^{\prime }}U_{{\bf q}%
,i}^{*}({\bf G})\frac{({\bf q+G)}\cdot ({\bf q+G}^{\prime })}{\left| {\bf q+G%
}\right| \left| {\bf q+G}^{\prime }\right| }\rho ({\bf G-G}^{\prime })U_{%
{\bf q},i}({\bf G}^{\prime }) 
\end{equation} 
and 
\begin{equation}
\omega _i^2({\bf q})=\frac{z_i({\bf q})}{1-d _i^{-1}({\bf q})} 
\end{equation} 
It can be shown that the pole strength $z_i$ can be expressed in terms of
the real-space eigenpotentials 
\begin{equation}
z_i({\bf q})=\frac{\omega _p^2}{\rho (0)}\int d^3r\rho ({\bf r})\left|
\nabla \Psi _{{\bf q}i}({\bf r})\right| ^2 
\end{equation}
where 
\begin{equation}
\Psi _{{\bf q}i}({\bf r})=\sum_{{\bf G}}\frac{U_{{\bf q}i}({\bf G})e^{-i(%
{\bf q+G})\cdot {\bf r}}}{\left| {\bf q+G}\right| } 
\end{equation} 
so that the pole strength is positive definite. The eigenvalues 
$d_i^{-1}$ of the static dielectric matrix lies between (0,1) which implies
that the plasmon frequencies $\omega _i$ are all real
(von der Linden and Horsch 1988).
A generalization of the plasmon-pole model has been proposed by
Engel {\em et al} (1991) and Engel and Farid (1992, 1993).

One drawback of the plasmon-pole approximation is that the imaginary part of
the self-energy is zero except at the plasmon-poles. As a consequence,
the life-time of the quasiparticles cannot be calculated.
Another drawback is its limited applicability to other than
sp systems. For more complex systems
where the plasmon excitations merge with the particle-hole excitations, 
it is not
clear anymore if the plasmon-pole approximation is appropriate.

\subsection{The space-time method}

Conventional ways of performing self-energy calculations express
all quantities in frequency and reciprocal space. This is natural because in
solids the Bloch momentum is a good quantum number and by working in
reciprocal space, one takes advantage of the lattice translational symmetry.
In the extreme case of the electron gas, the self-energy becomes diagonal in 
${\bf k}.$ In fact, it has been found numerically that in the Bloch-state
representation, the self-energy is almost diagonal even in systems that bear
little resemblance to the electron gas (Hybertsen and Louie 1986,
Aryasetiawan 1992a). The reason for this has been clarified recently by Hedin
(1995). The frequency representation is also
natural because most experiments generally focus on energy-dependent
measurements. For example, photoemission spectra are measured as functions
of momentum and energy. Theoretically, however, the self-energy
becomes a multidimensional convolution
when expressed in momentum-energy space
which is computationally very expensive as may be seen in equation
(\ref{imsig}). In
space-time representation, on the other hand, the self-energy in the GWA
takes a simple multiplicative form 
\begin{equation}
\Sigma \left( {\bf r,r}^{\prime },t\right) =iG\left( {\bf r,r}^{\prime
},t\right) W\left( {\bf r,r}^{\prime },t\right) 
\end{equation}
The response function also has a simple form 
\begin{equation}
P^0\left( {\bf r,r}^{\prime },t\right) =-iG^0\left( {\bf r,r}^{\prime
},t\right) G^0\left( {\bf r}^{\prime },{\bf r},-t\right) 
\end{equation} 
and the non-interacting Green function $G^0$ is given by 
\begin{equation}
G^0\left( {\bf r,r}^{\prime },t\right) =\left\{ 
\begin{array}{c}
\;\;i\sum_{{\bf k}n}^{\rm occ}
\psi _{{\bf k}n}\left( {\bf r}\right) \psi _{{\bf k%
}n}^{*}\left( {\bf r}^{\prime }\right) e^{i\varepsilon _{{\bf k}%
n}t},\;\;\quad t<0, \\ \\ 
-i\sum_{{\bf k}n}^{\rm unocc}
\psi _{{\bf k}n}\left( {\bf r}\right) \psi _{{\bf k}%
n}^{*}\left( {\bf r}^{\prime }\right) e^{-i\varepsilon _{{\bf k}n}t},\quad
t>0
\end{array}
\right. 
\end{equation} 
Evaluation of $G^0$ in real time is not advantageous due to the oscillatory
exponential term. In imaginary time this exponential term decays rapidly and 
the summation can be performed easily.
$G^0$ becomes (Rojas, Godby, and Needs 1995)
\begin{equation}
G^0\left( {\bf r,r}^{\prime },i\tau \right) =\left\{ 
\begin{array}{c}
\;\;i\sum_{{\bf k}n}^{\rm occ}
\psi _{{\bf k}n}\left( {\bf r}\right) \psi _{{\bf k%
}n}^{*}\left( {\bf r}^{\prime }\right) e^{-\varepsilon _{{\bf k}n}\tau
},\;\;\quad \tau <0, \\ \\
-i\sum_{{\bf k}n}^{\rm unocc}
\psi _{{\bf k}n}\left( {\bf r}\right) \psi _{{\bf k}%
n}^{*}\left( {\bf r}^{\prime }\right) e^{-\varepsilon _{{\bf k}n}\tau
},\quad \tau >0
\end{array}
\right. 
\end{equation} 
This is obtained by analytically
continuing the expression for $G^0$ in equation (\ref{Gspectral})
 from real to imaginary
energy and taking Fourier transform from imaginary energy to imaginary time.
From translational and lattice symmetry, ${\bf r}$ can be restricted to lie
in the irreducible wedge but ${\bf r}^{\prime }$ must run over all grid
points which are located inside a large sphere of radius $R_{\max }$ centred
on ${\bf r}.$ The method was tested for jellium and Si and typical values
for the grids are $\Delta r={\rm 0.5}$ a.u., $\Delta \tau ={\rm 0.3}$ a.u., $%
R_{\max }={\rm 18}$ a.u., and $\tau _{\max }={\rm 10}$ a.u., which give
convergence in the quasiparticle energy differences to 0.05 eV and absolute
values to 0.1 eV (Rojas, Godby, and Needs 1995).

Although the polarization function $P^0$ takes a particularly simple form in
space-time representation, calculating $\epsilon ^{-1}$ in real space is
still computationally prohibitive because of the large size of matrices in
inverting the dielectric function. In practice, one performs six-dimensional
fast Fourier transform from $\left( {\bf r,r}^{\prime }\right) $ to $\left( 
{\bf k,G,G}^{\prime }\right) $ and one-dimensional Fourier transform from
imaginary time to imaginary energy and solves for $\epsilon ^{-1}$ as a
matrix equation in ${\bf G,G}^{\prime }$ for each ${\bf k}$ and $i\omega .$
After forming $W_{{\bf GG}^{\prime }}({\bf k},i\omega )$ one Fourier
transforms back to real space and to imaginary time representation. The
advantages of this method are that
the computational effort is much reduced
compared to the conventional techniques since the double summation over $%
{\bf k}$ points and bands are avoided and there is no need for
a plasmon-pole approximation.

\begin{figure}[t]
\unitlength1cm
\rotatebox{-0.5}
{\centerline{\epsfxsize=4.5in \epsffile{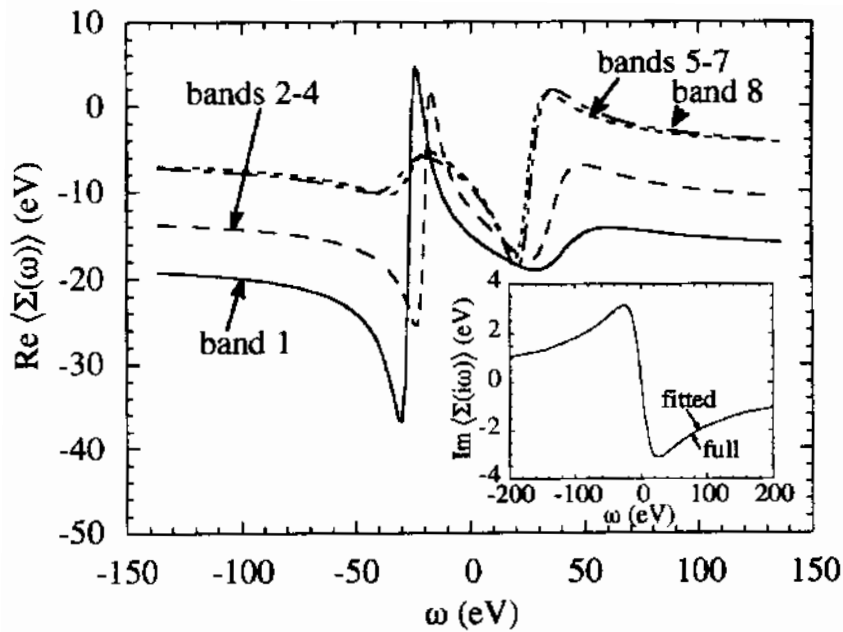}}}
\vskip0.3cm
\caption[]{\label{g2}
The real part of
the matrix elements of the self-energy operator of Si continued onto the real
axis for the first 8 bands at ${\bf k} = 0$. Inset: 
The matrix elements calculated along the imaginary axis for band 4 
(the valence band maximum) together with the fitted form (with two poles).
After Rojas, Godby, and Needs (1995).
}
\end{figure}

For comparison with experiment, the self-energy has to be calculated at real
frequencies. This is achieved by first calculating the self-energy and the
matrix element of the self-energy correction directly in real space $\langle
\psi _{{\bf k}n}|\Sigma \left( i\tau \right) -V^{\rm xc}|\psi _{{\bf k}n}\rangle 
$ and then Fourier transforming the result from imaginary time to imaginary
energy. The imaginary energy representation can then be analytically
continued to real energies by fitting the self-energy to the multipole form 
\begin{equation}
a_0+\sum_{i=1}^n\frac{a_i}{\omega -b_i} 
\end{equation} 
where the parameters $a_i$ and $b_i$ are in general complex. A good fit is
obtained with only $n=$ 2 with an RMS error of 0.2 \%. An example is shown
in Fig. \ref{g2}. The advantage of
calculating the self-energy along the imaginary axis is that one avoids the
sharp pole structures in both $G$ and $W.$ For quantities which require
frequency integration, it is then useful to perform the integration along
the imaginary axis whenever possible. However, when the detailed structure
of the self-energy is required along the real axis, knowledge of it at a few
points along the imaginary axis is not likely to be sufficient.

\noindent

One important aspect of the space-time method is its scaling with respect to
the system size
(Rojas, Godby, and Needs 1995). 
It is found numerically that the range of $W$ and $\Sigma $
is rather material independent so that the parameters $R_{\max }$ and $%
\Delta r$ do not change with system size. This means that the information
stored as well as the computational time for the non-interacting
response function should scale linearly with the
system size. However, the calculation of the dielectric function involves
matrix inversions which scale as N$^3$.
It is still favourable compared with
the conventional plane-wave basis approach which scales as N$^4$ but 
comparable to the localized basis approach which also scales as N$^3$
(Aryasetiawan and Gunnarsson 1994b).

An interesting mixed-space
approach for calculating the polarization function was recently
proposed by Blase {\em et al} (1995).
The polarization function is written as
\begin{equation}\label{Pmix} 
P({\bf r,r'},\omega)= \sum_{\bf q} \rm{exp}[i{\bf q}\cdot({\bf r-r'})]
P_{\bf q}({\bf r,r'},\omega)
\end{equation} 
where
\begin{eqnarray}\label{Pqmix} 
P_{\bf q}({\bf r,r}^{\prime },\omega ) 
&=&\sum_{{\rm spin}}\sum_{{\bf k}n}^{{\rm occ}%
}\sum_{n^{\prime }}^{{\rm unocc}}u _{{\bf k}n}^{*}({\bf r%
})u _{{\bf k+q}n^{\prime }}({\bf r})u _{{\bf k+q}
n^{\prime }}^{*}({\bf r}^{\prime })u _{{\bf k}n}({\bf r}^{\prime }) 
\nonumber \\
&&\times \left\{ \frac 1{\omega -\varepsilon _{{\bf k+q}n^{\prime
}}+\varepsilon _{{\bf k}n}+i\delta }-\frac 1{\omega +\varepsilon _{{\bf k+q}
n^{\prime }}-\varepsilon _{{\bf k}n}-i\delta }\right\}
\end{eqnarray} 
The function $P_{\bf q}({\bf r,r}^{\prime },\omega )$ is periodic in ${\bf r}$
and ${\bf r}'$ separately and it need be calculated within a unit cell only
which distinguishes this approach
from the direct real-space method where one of
the position variables is not restricted to the central cell.
The former approach scales as N$^3$, similar to localized-basis methods.
It was found that the crossover between the mixed-space and reciprocal-space
methods occurs for unit cells as small as that of Si.

The real-space or the mixed-space
approach is suitable for systems with large unit cells and a large
variation in the electron density, or open systems.

\section{Simplified {\em GW}}

{\em GW }calculations are computationally expensive and therefore it is
desirable to find simplifications which reduce the numerical effort but
still maintain the accuracy of the full calculations. So far there is no
simplified {\em GW} theory that is applicable to all systems. Most
simplified theories are for semiconductors and insulators. Although
some success has been achieved, none of these models is very reliable.
While band gaps can be reasonably well reproduced by these models, 
details of bandstructures are not satisfactorily described.
It is a major challenge to construct a good approximation for the
self-energy which includes both non-locality and energy dependence but which
is simple
enough to be applicable to complex systems without significant loss of
accuracy.

\subsection{The 
static Coulomb-hole and screened-exchange (COHSEX) approximation}

One of the earliest attempts to simplify {\em GW} self-energy is the static
COHSEX approximation (Hedin 1965a). 
It is obtained formally by setting $E-\varepsilon _{%
{\bf k}n}=0$ in equations (\ref{SEX}) and (\ref{COH}) yielding 
\begin{equation}
\Sigma _{{\rm SEX}}({\bf r,r}^{\prime })=-\sum_{{\bf k}n}^{{\rm occ}}\phi _{%
{\bf k}n}({\bf r})\phi _{{\bf k}n}^{*}({\bf r}^{\prime })W({\bf r,r}^{\prime
},\omega =0)  \label{sigmasex}
\end{equation}
\begin{equation}
\Sigma _{{\rm COH}}({\bf r,r}^{\prime })=\frac 12\delta ({\bf r-r}^{\prime
})\left[ W({\bf r,r}^{\prime },\omega =0)-v({\bf r-r}^{\prime })\right]
\label{sigmacoh}
\end{equation}
The first term is the exchange self-energy but with a statically screened
interaction. The second term is the Coulomb-hole term which is the
interaction energy between the quasiparticle and the potential due the
Coulomb hole around the quasiparticle, as a result of the rearrangement of
the electrons (screening). Both the screened-exchange and the Coulomb-hole
terms in equations (\ref{SEX}) and (\ref{COH}) are energy dependent and
non-local respectively
but in the static COHSEX approximation they are energy independent
and the Coulomb-hole term is local. The validity of the COHSEX approximation
relies on whether $E-\varepsilon _{{\bf k}n}$ is small compared to the
energy of the main excitation in the screened interaction which is
essentially the plasmon energy. Comparison to the results of the full
calculations shows that the COHSEX approximation consistently overestimates
the magnitude of the self-energy, by about 20 \% in Si, resulting in larger
band gaps in semiconductors (Hybertsen and Louie 1986). 
For $E$ corresponding to an occupied state,
most of the error resides in the Coulomb-hole term. 
The approximation $%
E-\varepsilon _{{\bf k}n}=0$ is more severe for the Coulomb-hole term than
for the screened-exchange term because the Coulomb-hole term involves a sum
over unoccupied states as well as over occupied states whereas the
screened-exchange term involves a sum over occupied states only.

Another source of error comes from the neglect of dynamical renormalization
or Z-factor in equation (\ref{zfactor}). When taking the matrix element $%
\langle \phi _{{\bf k}n}|\Sigma |\phi _{{\bf k}n}\rangle $ and setting $%
E-\varepsilon _{{\bf k}n}=0$, this approximately corresponds to calculating
the self-energy at $E=\varepsilon _{{\bf k}n}$ but the true self-energy
should be calculated at the quasiparticle energy, leading to the formula in
equation (\ref{qpenergy}). Indeed, a significant improvement can be obtained
if the Z-factor is taken into account when the self-energy itself is
calculated within the COHSEX approximation.

\subsection{Improving the COHSEX approximation}

An attempt to include dynamical renormalization into a simplified {\em GW}
scheme was made by Bechstedt {\em et al} (1992). 
In this scheme, the energy dependent
part of the self-energy is expanded to linear order around the LDA
eigenvalue. A proper calculation of the self-energy derivative, or
equivalently the Z-factor, requires unfortunately a full frequency dependent
response function. It is not sufficient to calculate the static response
function and its energy derivative at $\omega =0$. For the screened-exchange
term, on the other hand, knowledge of the first energy derivative of the
response function at $\omega =0$ is sufficient to expand $\Sigma _{\rm{SEX}}$ 
to first order around the LDA eigenvalue.

To take into account the energy dependence of the self-energy, the screened
interaction {\em W} is calculated within the plasmon-pole model. The static
dielectric function is modelled by (Bechstedt {\em et al} 1992)
\begin{equation}
\epsilon ({\bf q},\rho )=1+\left[ (\epsilon _0-1)^{-1}+\alpha \left( \frac q{%
q_{TF}}\right) ^2+\frac 34\left( \frac{q^2}{k_Fq_{TF}}\right) ^2\right] ^{-1}
\label{bechstedteps}
\end{equation}
where $k_F$ and $q_{TF}$ are the Fermi and Thomas-Fermi 
wave vectors respectively which
depend on the average electron density $\rho $. This model for $\epsilon $
interpolates between the free-electron gas result $1+(2\omega _p/q^2)^2$ at
high $q$, $1+\alpha (q_{TF}/q)^2$ at small $q$ (Thomas-Fermi theory) and the 
$q=0$ value $\epsilon _0$ for the semiconductor. It allows analytical
calculation of the screened Coulomb interaction. The coefficient $\alpha $
is obtained by fitting the model dielectric function to a full RPA
calculation for $\omega =0$. The values of $\alpha $ turn out to be material
independent for those semiconductors considered (Si, GaAs, AlAs, and ZnSe).

To take into account local field effects due to the inhomogeneity in the
charge density, an LDA ansatz is used (Hybertsen and Louie 1988a)
\begin{equation}
W({\bf r,r}^{\prime },\omega =0)=\frac 12\left[ W^h({\bf r-r}^{\prime },\rho
({\bf r}))+W^h({\bf r-r}^{\prime },\rho ({\bf r}^{\prime }))\right]
\label{hybertsenwmodel}
\end{equation}
where $W^h$ is the screened interaction for the electron gas. The sum rule,
that the total induced charge around a test charge is $-1+1/\epsilon _0,$ is
fulfilled but the induced charge density is allowed to vary according to
the local density. Using this model, the static Coulomb-hole term can be
calculated analytically (Bechstedt {\em et al} 1992)
\begin{equation}
V_{\rm{COH}}({\bf r})
=-\left( 1-\frac 1{\epsilon _0}\right) ^{\frac 12}\frac{%
q_{TF}\left( {\bf r}\right) }{\sqrt{\alpha }}\left[ 1+\frac{q_{TF}\left( 
{\bf r}\right) }{\alpha k_F\left( {\bf r}\right) }\sqrt{\frac{3\epsilon _0}{%
\epsilon _0-1}}\right] ^{-\frac 12}  \label{Vcoh}
\end{equation}
It is interesting to observe that the matrix element of $\Sigma
^{dyn}=\Sigma -\Sigma _{\rm{COHSEX}}$ 
and its energy derivative can be shown to
be independent of the state if local field effects are neglected. The state
dependence thus comes from the local field effects and it is found to be
rather weak. We note also the dependence on $q_{TF}\sim \rho^{1/3}$ as in the
LDA.
$\Sigma ^{dyn}$ gives an upward shift of $\sim$1.4 eV for
all the materials mentioned above. This means that good values for the band
gaps are obtained by using $\Sigma =\Sigma _{\rm{COHSEX}}$ 
but taking into
account the dynamical renormalization factor Z. The results for Si, GaAs,
AlAs, and ZnSe show agreement with the full {\em GW} calculations to within
0.2 eV for most of the states considered. Application was also made to GaN
recently (Palummo {\em et al} 1995). This scheme reduces the
computational effort by two orders of magnitude. 
Further tests on a wide range of semiconductors and insulators
are desirable to evaluate the validity of the number of approximations used in
the model.

\subsection{Extreme tight-binding models}

One of the major problems in {\em GW} calculations is the calculation of the
response function. In electron-gas-like materials such as the alkalis, it
is reasonable to model the dielectric function with a single plasmon and to
neglect off-diagonal elements. In semiconductors, however, local field
effects, which are described by the off-diagonal elements of the dielectric
matrix, are important. The local field is dominant at distances on the
interatomic scale so that high Fourier components are needed to describe it
in plane wave basis resulting in a large matrix. It is therefore
advantageous to calculate the dielectric matrix and its inverse directly in
real space. Real-space approach is also useful for systems with low
symmetry. Ortuno and Inkson (1978) proposed an extreme tight-binding model
where the valence and conduction bands were assumed to be flat so that the
band gap $E_g$ is the only parameter entering the model. Due to its
simplicity, the model allows analytic evaluation of the dielectric function
to a certain degree.

The wavefunctions are expanded in localized Wannier functions 
\begin{equation} 
\psi _{{\bf k}n}\left( {\bf r}\right) =\frac 1{\sqrt{N}}\sum_{\nu {\bf T}%
}e^{i{\bf k\cdot T}}\phi _{n\nu }\left( {\bf r-T}\right) c_{n\nu }\left( 
{\bf k}\right) 
\end{equation} 
where $\phi _{n\nu }$ is a Wannier function localized in bond $\nu $ and $%
{\bf T}$ is a lattice translational vector. In the two-flat-band model, the
coefficients $c_{n\nu }$ are equal to one. Since $\phi _{n\nu }$ is
localized in each cell and in each bond, we have 
\begin{equation}
\int d^3r\;
\phi _{n\nu }\left( {\bf r-T}\right) \phi _{n\mu }\left( {\bf r-T}^{\prime
}\right) \propto \delta _{\nu \mu }\delta _{{\bf TT}^{\prime }} 
\end{equation} 
Using this expansion in the expression for the polarization function and
using the flat-band approximation, $\varepsilon _{{\bf k}^{\prime }n^{\prime
}}-\varepsilon _{{\bf k}n}=E_g,$ one arrives at a simple expression 
for the dielectric function:
\begin{equation}
\epsilon \left( {\bf r,r}^{\prime },\omega \right) =\delta \left( {\bf r-r}%
^{\prime }\right) -N\left( \omega \right) \sum_{\nu {\bf T}}\int
d^3r^{\prime \prime }v\left( {\bf r-r}^{\prime \prime }\right) A_\nu \left( 
{\bf r}^{\prime \prime }-{\bf T}\right) A_\nu ^{*}\left( {\bf r}^{\prime }-%
{\bf T}\right) 
\end{equation} 
where 
\begin{equation}
N\left( \omega \right) =\frac{4E_g}{\omega ^2-(E_g-i\delta )^2} 
\end{equation} 
\begin{equation}
A_\nu \left( {\bf r-T}\right) =\phi _{{\rm cond},\nu }\left( {\bf r-T}%
\right) \phi _{{\rm val},\nu }^{*}\left( {\bf r-T}\right) 
\end{equation} 
This has the form of a separable matrix $\epsilon =1-{\bf BC}$ 
(Hayashi and Shimizu 1969, Sinha 1969) whose inverse
is given by $\epsilon ^{-1}=1+{\bf B}\left( 1-{\bf CB}\right) ^{-1}{\bf C}$.
Under certain approximations, the matrix $\epsilon$ can be inverted giving 
a screened interaction (Ortuno and Inkson 1979)
\begin{eqnarray}
W\left( {\bf r,r}^{\prime },\omega \right) &=&\int d^3r^{\prime \prime
}\epsilon ^{-1}\left( {\bf r,r}^{\prime \prime },\omega \right) v\left( {\bf %
r^{\prime \prime }-r}^{\prime }\right)  \nonumber \\
&=&v\left( {\bf r-r}^{\prime }\right) -\frac{4E_g}{E_g^2+\omega _p^2-\omega
^2}\sum_{\nu {\bf T}}D_\nu \left( {\bf r-T}\right) D_\nu ^{*}\left( {\bf r}%
^{\prime }-{\bf T}\right)  \label{Wsterne}
\end{eqnarray}
where 
\begin{equation}
D_\nu \left( {\bf r-T}\right) =\int d^3r^{\prime }v\left( {\bf r-r}^{\prime
}\right) A_\nu \left( {\bf r}^{\prime }-{\bf T}\right) 
\end{equation} 
and $\omega _p$ is the plasmon energy.
The approximation is good in the limit $E_g\gg \omega _p.$ The screened
potential is effective at $\omega =E_g$ and approaches a bare value at $%
\omega =\omega _p.$ The quantity $D_\nu \left( {\bf r-T}\right) $ represents
a dipole moment at ${\bf r}$ due to bond $\nu $ in cell ${\bf T.}$ The
physical interpretation of the second term in equation (\ref{Wsterne})
is that
an electron on a site interacts with other electrons through the Coulomb
interaction, inducing dipole moments on the other sites. These dipoles in
turn produce a potential which is screened by other induced dipoles by the
frequency-dependent factor arising from the inversion of the dielectric
function (Sterne and Inkson 1984).

Using the above screened interaction, the self-energy can be evaluated
analytically. Defining a state-dependent local potential 
\begin{equation}
\int d^3r^{\prime }\Sigma \left( {\bf r,r}^{\prime },\omega =E_n\right) \psi
_{{\bf k}n}\left( {\bf r}^{\prime }\right) =V^{\rm xc}_n\psi _{{\bf k}n}\left( 
{\bf r}\right) 
\end{equation} 
gives two potentials for the valence and conduction states respectively
(Sterne and Inkson 1984)
\begin{equation}
V_{{\rm COHSEX}}^{{\rm val}}\left( {\bf r}\right) =-\rho ^{1/3}\left( {\bf r}%
\right) \frac \gamma 2\left[ \frac{2\pi }3\right] ^{1/3}\left[ 1+\frac 1{%
\epsilon _0}\right] -\frac C2\left[ \frac{\epsilon _0-1}{\epsilon _0+\sqrt{%
\epsilon _0}}\right] 
\end{equation} 
\begin{equation}
V_{{\rm COHSEX}}^{{\rm cond}}\left( {\bf r}\right) =-\rho ^{1/3}\left( {\bf r%
}\right) \frac \gamma 2\left[ \frac{2\pi }3\right] ^{1/3}\left[ 1-\frac 1{%
\epsilon _0}\right] -\frac C2\left[ \frac{\epsilon _0-1}{\epsilon _0-\sqrt{%
\epsilon _0}}\right] 
\end{equation} 
where 
\begin{equation}
C=\int d^3rd^3r^{\prime }A_\nu ^{*}\left( {\bf r-T}\right) v\left( {\bf r-r}%
^{\prime }\right) A_{\nu ^{\prime }}\left( {\bf r}^{\prime }-{\bf T}\right) 
\end{equation} 
is an on-site exchange interaction and 
\begin{equation}
\epsilon _0\cong \frac{E_g^2+\omega _p^2}{E_g^2} 
\end{equation} 
Comparison with the LDA exchange-only potential, $V_x^{\rm LDA}=-\left[ 3/\pi
\right] ^{1/3}\rho ^{1/3},$ gives a value for $\gamma =2\left[ 9/\left( 2\pi
^2\right) \right] ^{1/3}.$ 
It is interesting to observe that although the
potentials have been derived from extreme tight-binding picture, one arrives
at a formula similar to that obtained from the electron gas (Sterne and Inkson
1984).

The method has been applied to C, Si, GaAs, Ge, and ZnSe (Jenkins, Srivastava,
and Inkson 1993). Good agreement with
the full $GW$ calculations are obtained for Si and GaAs but not for the
other two materials. Perhaps this is not surprising in view of the flat
extremal bands in Si and GaAs, making them ideal for the present approach.
Moreover, the pseudopotentials used in the LDA calculations are better for
the first two materials than for the last two so that the eigenfunctions
used to calculate the matrix elements of the self-energy are correspondingly
better. The LDA exchange only gap is 0.41 eV whereas the calculated value is
1.25 eV (experimentally 1.17 eV). The direct gap at the $\Gamma $ point is
also improved from 2.49 eV to 3.32 eV (experimentally 3.35 eV). In the case
of Ge the improvement is not as good as in the case of Si but while the LDA
predicts Ge to be a metal with a negative band gap of -0.19 eV, the method
gives a gap of 0.34 eV albeit too low compared to experiment (0.89 eV). In
diamond, the method improves the gap at the $\Gamma $ point from 5.46 eV in
the LDA (exchange only) to 7.87 eV (experimentally 7.4 eV). Similar
improvement is also obtained across the Brillouin zone. Finally, the results
for ZnSe are rather poor. The band gap is overestimated by $\sim $ 1 eV
(3.77 eV $vs$ 2.82 eV). The problem could be related to the difficulties of
treating $3d$ states within the pseudopotential approach resulting in
relatively poor wavefunctions.

A similar tight-binding approach was also proposed by Hanke and Sham (1988)
and Bechstedt and Del Sole (1988).
They derived an analytical model for $\Sigma$, $V^{\rm xc}$, and the gap
correction in insulators. They arrived at a similar $\rho^{1/3}$ formula
for the exchange-correlation potential but they also included a term
corresponding to the bare conduction-band exchange which is neglected in the
Sterne and Inkson model.
The method was applied to Si, diamond, and LiCl and the results are within a
few percent of the more elaborate calculations.
A more recent work uses orthogonalized linear combination of atomic orbitals
with applications to diamond, Si, Ge, GaAs, GaP, and ZnSe. The gaps are
generally within 10 \% of the experimental values (Gu and Ching 1994).

\subsection{Quasiparticle local density approximation (QPLDA)}

This approximation is based on the work of Sham and Kohn (Sham and Kohn
1966) who showed that for a system with a slowly varying density the
self-energy is short range in $|{\bf r-r}^{\prime }|$ 
although for larger energies the
self-energy is expected to be long range since screening is not very
effective at large energies. 
Numerical
calculations by Hedin (1965a) show that the self-energy of the
homogeneous electron gas quickly vanishes beyond $\left| {\bf r-r}^{\prime
}\right| =2r_s$ for $\omega \approx E_F$ 
and it depends therefore on the density in the
vicinity of $\left( {\bf r+r}^{\prime }\right) /2$ which suggests the
following approximation (Sham and Kohn 1965)
\begin{equation}
\Sigma \left( {\bf r,r}^{\prime },\omega \right) \approx \Sigma _h\left( 
{\bf r-r}^{\prime },\omega -\Delta \mu \left( \overline{n}\right) ;\overline{%
n}\right) 
\end{equation} 
where $\Sigma _h\left( {\bf r-r}^{\prime },E;n\right) $ is the self-energy
operator of the homogeneous electron gas with density $n,$ $\overline{n}%
=n\left[ \left( {\bf r+r}^{\prime }\right) /2\right] ,$ and $\Delta \mu =\mu
-\mu _h\left( \overline{n}\right) $ is the difference between the true
chemical potential and that of the homogeneous electron gas of density $%
\overline{n.}$ The local density approximation for the self-energy is
obtained by assuming the quasiparticle wavefunction in equation (\ref{qpeqn}%
) as a superposition of locally plane-wave-like functions 
(Sham and Kohn 1965)
\begin{equation}
\Psi \left( {\bf r,}\omega \right) \approx A({\bf r})
e^{i{\bf k}\left( {\bf r},\omega
\right) \cdot {\bf r}}
\end{equation} 
which when inserted into the quasiparticle equation gives the solution $%
\omega \left( E\right) =E$ if $k$ satisfies the local-density condition $%
k=k_{LD}$ with 
\begin{equation}
-E+\frac 12k_{LD}^2+V^H\left( {\bf r}\right) +\Sigma \left( k_{LD},E-\Delta
\mu \left( n\right) ;n\right) =0
\end{equation}
In obtaining the above equation, the ${\bf r}$ dependence of $A$ and
${\bf k}$ has
been neglected. For a slowly varying density, the local change in the
chemical potential from its homogeneous value is simply given by the
electrostatic potential (Thomas-Fermi) : 
\begin{equation}
V^H\left( {\bf r}\right) =\Delta \mu \left[ n\left( {\bf r}\right) \right] 
\end{equation}
and by definition 
\begin{eqnarray}
\mu _h\left( n\right)  &=&\frac 12k_F^2\left( n\right) +\mu ^{\rm xc}\left(
n\right) \nonumber \\
&=&\frac 12k_F^2\left( n\right) +\Sigma \left( k_F,\mu _h\left( n\right)
,n\right) 
\end{eqnarray}
Using the last two expressions, the condition for the local-density wave
vector becomes 
\begin{equation}
\frac 12\left( k_{LD}^2-k_F^2\right) =\left( E-\mu \right) -\left[ \Sigma
     _h\left( k_{LD},E-\Delta \mu \left( n\right) ;n\right) -\Sigma _h\left(
k_F,\mu _h\left( n\right) ;n\right) \right]   \label{ldcond}
\end{equation}
When operating on the locally plane-wave function, the effect of the
non-local operator $\Sigma _h\left( {\bf r-r}^{\prime },E\right) $ can be
reproduced exactly by a local operator $\Sigma _h\left( k_{LD},E;n\right)
\delta \left( {\bf r-r}^{\prime }\right) .$ The local density approximation
for the self-energy operator consists of using this local operator when
operating on the actual wave function at energy $E.$ It may happen that for
some values of $E$ equation (\ref{ldcond}) has no solution for positive $%
k_{LD}^2.$ In this case, one should analytically continue the self-energy
operator into complex momentum.
An early work of approximating the self-energy by a local potential
using information from the local density is given by Hedin and Lundqvist
(1971).

Instead of calculating the full self-energy $\Sigma _h\left( k,E;n\right) $
it is useful to write 
(Wang and Pickett 1983, Pickett and Wang 1984)
\begin{eqnarray}
\Sigma _h\left( k,E;n\right)  &=&\mu ^{\rm xc}\left( n\right) +\Sigma _h\left(
k,E;n\right) -\Sigma _h\left( k_F,\mu _h;n\right)  \nonumber\\
&=&\mu ^{\rm xc}\left( n\right) +\Delta \left( k,E;n\right) 
\end{eqnarray}
since $\mu ^{\rm xc}$ has been calculated rather accurately for the electron
gas. The GWA is then used to calculate the relatively small quantity $\Delta
.$ Hedin and Lundqvist (1971) calculated $\Delta(k_{\rm LDA})$
and for metals, the self-energy corrections in the local-density
approximation turn out to be very small. $\Delta $ is of the order of a few $%
\times 10^{-2}$ eV. For Ni, for example, the corrections are smaller than
0.1 eV (Watson {\em et al} 1976). The main reason for such small corrections is probably due to the
weak energy dependence of the electron-gas self-energy. Furthermore, most of
the effect of the self-energy is already accounted for by the local term $%
\mu ^{\rm xc}.$ The QPLDA self-energy is dominated by a constant contribution
and the assumption of slowly varying density is probably not good enough in
many cases.

Above, the homogeneous electron gas was used to obtain an approximate
self-energy for inhomogeneous systems. For semiconductors and insulators
one can also use a homogeneous insulator model for this purpose
which turns out to give
a significant improvement. The presence of an energy gap in these systems
brings in new physics not presence in the electron gas. The response
function needed to calculate the screened interaction is obtained from a
model semiconducting homogeneous electron gas. One simply introduces a gap
in the otherwise metallic spectrum of the electron gas, that is to say (Penn
1962, Levine and Louie 1982)
\begin{equation}
\epsilon _2\left( k,\omega \right) =\left\{ 
\begin{array}{c}
0,\quad \quad \quad \quad \quad |\omega |<\lambda E_F \\ 
\epsilon _2^{{\rm RPA}}\left( k,\omega _{-}\right) ,|\omega |>\lambda E_F
\end{array}
\right. 
\end{equation} 
where $\omega _{-}=\sqrt{\omega ^2-\lambda ^2E_F^2}$ sgn $\omega $. This
function satisfies the {\em f}-sum rule. The dielectric gap $\lambda E_F,$
which is in general larger than the minimum direct gap, is determined from
the experimental value of the static, long-wavelength dielectric constant $%
\epsilon _{0,}$%
\begin{equation}
\lambda E_F=\frac{\omega _p}{\sqrt{\epsilon _0-1}}
\end{equation} 
where $\omega _p=\sqrt{4\pi n}$ is the plasmon frequency. The
single-particle spectrum has a gap $E_g$ 
\begin{equation}
E\left( {\bf k}\right) =\frac 12k^2+\frac 12E_g{\rm sgn}\left( k-k_F\right) 
\end{equation}
The gap $E_g$ is taken to be the average value across the Brillouin zone of
the direct gap and it is numerically different from $\lambda E_F.$ It can be
shown that the valence bands are lowered and the conduction bands are raised
by the self-energy corrections, independently of the density sampled by the
wave functions. This means, the QPLDA will always lead to an increase in the
gap over the LDA value
(Wang and Pickett 1983, Pickett and Wang 1984).

The QPLDA was applied to Si and diamond with considerable success. For Si,
the zone-boundary transitions $X_4\rightarrow X_1,$ $L_3^{\prime
}\rightarrow L_3,$ and $L_3^{\prime }\rightarrow L_1$ are well reproduced by
the QPLDA. The zone-centre transition $\Gamma _{25^{\prime }}\rightarrow
\Gamma _{15}$ is underestimated by 0.25 eV, which is the worst case but the
transition $\Gamma _{25^{\prime }}\rightarrow \Gamma _{15}$ is reproduced
well. The calculated indirect gap of 0.93 eV still underestimates the
experimental value by 0.24 eV. The QPLDA gives a valence-band width of 12.5
eV, in good agreement with the experimental data 12.4 $\pm $ 0.6 eV (Grobman
and Eastman 1972) and 12.6 $\pm $ 0.6 eV (Ley {\em et al. }1972). For
diamond, the QPLDA gives an indirect gap of 5.74 eV, which is a slight
overestimate compared to the experimental value of 5.47 eV. The LDA value is
4.05 eV. The $\Gamma _{25^{\prime }}\rightarrow \Gamma _{15}$ gap is
experimentally about 7.3 eV and the QPLDA gives 7.36 eV (compared to the LDA
value of 5.51 eV). The valence-band width is 23.4 eV, in good agreement with
the experimental data of McFeely {\em et al. }(24.2 $\pm $ 1.0 eV) (McFeely 
{\em et al. }1974) but in disagreement with the data of Himpsel {\em et al}
(21 eV)
(Himpsel, van der Veen, and Eastman 1980). The LDA band width is 20.4 eV.

\subsection{LDA + $\delta \Sigma _{{\rm COHSEX}}$}

In metals, the screened interaction $W$ decreases rapidly for $\left| {\bf %
r-r}^{\prime }\right| >k_{TF}^{-1}$ but in insulators and semiconductors, it
decreases as $1/\epsilon _0\left| {\bf r-r}^{\prime }\right| $ for large $%
\left| {\bf r-r}^{\prime }\right| $ since the screening is not complete.
Accordingly, one writes (Gygi and Baldereschi 1989)
\begin{equation}
W\left( {\bf r,r}^{\prime },\omega \right) =W^{IEG}\left( {\bf r,r}^{\prime
},\omega \right) +\delta W\left( {\bf r,r}^{\prime },\omega \right) 
\end{equation} 
where $W^{IEG}$ is the short-range interaction potential of a metallic
inhomogeneous electron gas and $\delta W$ has the same long-range behaviour
as $W.$ The self-energy arising from the first term in $W$ has been shown by
Sham and Kohn (1966)
to be short ranged and depends only on the density in the
vicinity of ${\bf r}$ and ${\bf r}^{\prime },$ and is therefore 
approximated by a local potential: 
\begin{equation}
\Sigma _{GW}\left( {\bf r,r}^{\prime },\omega \right) =\mu ^{\rm xc}\left( {\bf r%
},\omega \right) \delta \left( {\bf r-r}^{\prime }\right) +\frac i{2\pi }%
\int d\omega ^{\prime }G\left( {\bf r,r}^{\prime },\omega +\omega ^{\prime
}\right) \delta W\left( {\bf r,r}^{\prime },\omega ^{\prime }\right) 
\end{equation} 
Pickett and Wang (1984)
found that the inclusion of energy dependence in the local
exchange-correlation potential had very little effects on the eigenvalues
obtained with an energy-independent LDA potential. It is then assumed that
for states close to the Fermi level, $\mu ^{\rm xc}$ takes its value at the
Fermi level $\mu ^{\rm xc}_{\rm LDA}\left( {\bf r}\right) .$ Furthermore, it is
assumed that $\delta W$ depends only on $\left| {\bf r-r}^{\prime }\right| ,$
which is strictly valid only when $\left| {\bf r-r}^{\prime }\right|
\rightarrow \infty $, but it has been found numerically that in non-metals
the local field effects become negligible as $\left| {\bf r-r}^{\prime
}\right| $ exceeds the interatomic distance (Hybertsen and Louie 1986,
Godby, Schl\"uter, and Sham 1988). 
One now makes the COHSEX
approximation on $\delta \Sigma ,$ rather than on the full self-energy: 
\begin{equation}
\delta \Sigma \left( {\bf r,r}^{\prime }\right) =-\rho \left( {\bf r,r}%
^{\prime }\right) \delta W\left( {\bf r-r}^{\prime }\right) +\delta \Sigma
_{\rm{COH}} 
\end{equation} 
where the first term is the screened exchange contribution and 
\begin{equation}
\delta \Sigma _{\rm{COH}}
=\frac \Omega {2\left( 2\pi \right) ^3}\int d^3q\delta
W\left( {\bf q}\right) 
\end{equation} 
is the Coulomb hole contribution which is a constant, shifting all
eigenvalues by the same amount. The Fourier transform of $\delta W$ is given
by 
\begin{equation}
\delta W\left( q\right) =\frac{4\pi }{\Omega q^2}\left[ \epsilon
_{SC}^{-1}\left( {\bf q,q,}\omega =0\right) -\epsilon _M^{-1}\left( {\bf q}%
,\omega =0\right) \right] 
\end{equation} 
where $\epsilon _{SC}^{-1}\left( {\bf q,q,}0\right) $ is the diagonal part
of the inverse dielectric matrix for the semiconductor calculated in the RPA
and $\epsilon _M^{-1}(q,0)$ is the inverse of the static Lindhard dielectric
function (Lindhard 1954) of a homogeneous electron gas.

Applications to diamond, Si, Ge, GaAs, AlAs, and GaP give in general
agreement to within 0.2 eV with the results of the full calculations. Larger
deviations (0.3-0.4 eV) occur for the $X_{4v}\rightarrow X_{1c}$ gap in
diamond and the $\Gamma _{25^{\prime }v}\rightarrow \Gamma _{2^{\prime }c}$
gaps in Ge. The Coulomb-hole contribution $\delta E_{\rm{COH}}$ 
is positive and
cancels approximately the effect of the screened exchange term in the
valence bands. As a result, the net effect of the self-energy correction is
to shift the conduction band upwards and leave the valence bands essentially
unchanged. This is also in agreement with the full calculations.

\section{Applications}

\subsection{Alkali metals}

\begin{figure}[bt]
\unitlength1cm
\begin{minipage}[t]{15.0cm}
\rotatebox{181}
{\centerline{\epsfxsize=3.375in \epsffile{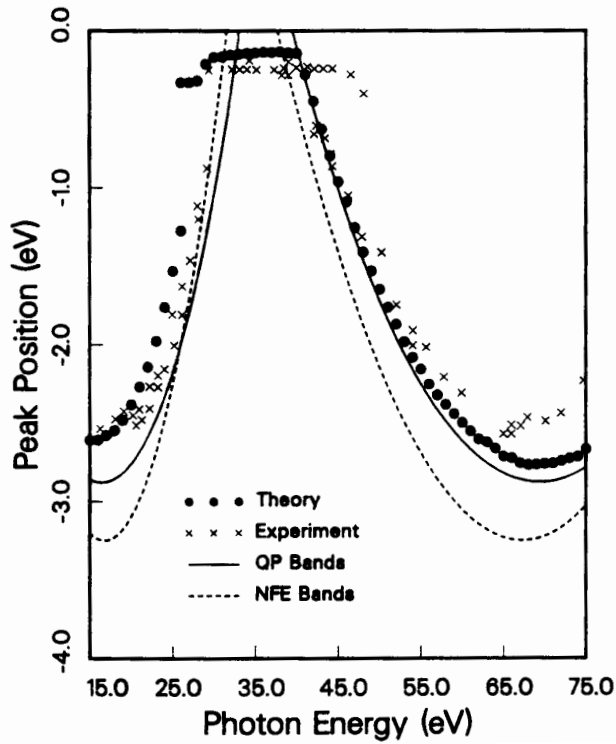}}}
\vskip0.3cm
\caption[]{\label{figalkali}
Peak position as a function of photon energy for photoemission
at normal angle from the Na (110) surface. The crosses show experimental
results (Jansen and Plummer 1985), the filled circles show the 
photoemission calculations
of Shung and Mahan (1987), the full line the quasi-particle energies
and the dashed lines the NFE theory. (After Shung and Mahan 1987). 
}
\end{minipage}
\hfill
\end{figure}

The early calculations in the GWA were performed for 
the electron gas, because of its simplicity.
(For a review see Hedin and Lundqvist 1969).
The alkali metals
are therefore of particular interest, being the systems which 
are the closest approximation to the electron gas. For these systems
the correlation effects are only moderately strong,
and the GWA could therefore be expected to be relatively
accurate. 
It therefore created a lot of interest when the first high-resolution
angular resolved photoemission experiments for the alkali metals
appeared (Jensen and Plummer 1985, Lyo and Plummer 1988, 
Itchkawitz {\it et al} 1990). Two unexpected observations were
made. First the band width was smaller than expected. For Na the
occupied part of the band was found to be 2.65 eV broad
(Lyo and Plummer 1988). This is consistent with an old angular-integrated 
measurement by Kowalczyk {\it et al} (1973). This band width is
 substantially
smaller than the width 3.2 eV predicted by nearly free electron (NFE)
theory. Figure \ref{figalkali} shows the experimental peak position (crosses)
as a function of the photon energy together with the prediction
of NFE theory (dashed line).
 The  band narrowing is about a factor of two larger than 
the narrowing (0.27 eV)
predicted by $GW$ calculations for the electron gas of 
the appropriate 
density (Hedin 1965a). Although this error is not very large in absolute
terms,  it raised questions about the accuracy 
of the GWA for these systems. In addition      a set
of essentially dispersionless states were observed close
to the Fermi energy, which was completely unexpected according to the 
NFE band structure. This raised the interesting issue that the alkali
metals may have charge density waves (Overhauser 1985).

A $GW$ calculation was performed by Northrup {\it et al} (1987, 1989)
for Na metal and by Surh {\it et al} (1988) for K, using a 
generalized plasmon pole approximation.  
These calculations gave essentially the same band narrowing (0.31 eV) for Na
as a $GW$ calculation for the electron gas.
A much better agreement with experiment was, however, obtained by using 
an improved dielectric function (Northrup {\it et al} 1987, 1989)
\begin{equation}\label{alkali1}
\epsilon^{-1}=1+v\lbrack 1-P(v+K^{\rm xc})\rbrack^{-1}P,
\end{equation}
where $P$ is the independent particle polarizability and 
$K^{\rm xc}=\delta V^{\rm xc}/\delta \rho$ with $V^{\rm xc}$ the (LDA) 
exchange-correlation potential and $\rho$ the density. 
Equation (\ref{alkali1}) gives the 
appropriate dielectric function within the density functional 
formalism. Using this $\epsilon$ instead of the RPA $\epsilon$, the band      
width narrowing increased from 0.31 eV to 0.57 eV. Finally, 
a calculation was performed using a Green function with a certain 
self-consistency, namely with the LDA eigenvalues replaced by 
the quasi-particle energies. This led to a further increase of 
band width narrowing to 0.71 eV.  
Lyo and Plummer (1988) also observed large effects of including the
corrections to the dielectric function in equation (\ref{alkali1}).

The theoretical justification for including such corrections 
without simultaneously adding vertex corrections is, however, weak. 
We notice that the total energy of the system can be expressed in
terms of the dielectric function. The quasi-particle energies can
then be obtained by differentiating with respect to the occupation
numbers (Rice 1965). If the dielectric function has the form of 
equation (\ref{alkali1}), it was shown that there is also a vertex 
correction of the type (Rice 1965, Ting 1975, Mahan 1994) 
\begin{equation}\label{alkali2}
\Gamma={1\over 1-PK^{\rm xc}}.
\end{equation}
Mahan and Sernelius (1989) have extensively tested various  
corrections to the dielectric function, including the corresponding 
vertex corrections when calculating the self-energy. They found that
for the electron gas
the vertex corrections cancel most of the effects of the corrections
of the dielectric function, and the final results are rather close
to the original GWA. Del Sole, Reining, and Godby (1994)
obtained similar conclusions for Si.
The results for the electron gas suggest that there are   
discrepancies between appropriate self-energy calculations and the 
peak positions in the photoemission experiments.

\noindent

The relation between the peak position in photoemission and the 
quasiparticle energies was studied by Mahan and coworkers in a series of 
papers, focussing on surface and quasiparticle life-time effects
(Shung and Mahan 1986, 1988, Shung {\it et al} 1987).
They studied a model which includes   the rapidly varying potential 
in the surface region as well as lattice potential in the bulk. 
The effects of these potentials on the photoemission process were included,
while the lattice part of the potential was neglected when calculating 
the states.
The self-energy was calculated using the Rayleigh-Schr\"odinger perturbation
theory, which gives results that differ slightly from the traditional
$GW$ results. For instance, the Na band width is reduced by 0.37 eV due
to this self-energy.
Mahan and coworkers furthermore included the effects of
the imaginary part of the self-energy of the emitted electron, by using  
wave function $\phi^>$ for these electrons which decayed exponentially 
inside the surface.
Finally, the self-energy of the initial state was included as well.
Due to the exponential decay of  $\phi^>$,
the electron momentum perpendicular to the surface is not conserved,
and there are nonvertical (non $k_{\bot}$-conserving) transitions.

Mahan and coworkers found that the broadening of the initial
state and the instrumental resolution as well as the interference 
between bulk and surface photoemission shift the apparent peak positions
towards lower binding energies relative to the quasiparticle energies
by about 0.2-0.4 eV
(Shung and Mahan 1988). Including these effects as well as the
quasiparticle self-energy shifts leads to the filled circles in figure 
\ref{figalkali}. There is a substantial band narrowing relative 
to the quasiparticle energies (full line) and the agreement
with experiment is generally quite good. This suggests that the GWA 
gives good quasiparticle energies for Na, and that the main
reason for the discrepancy between these energies and the photoemission
peak positions can be explained by considering the details of the 
photoemission process.

Mahan and coworkers
also gave an explanation for the unexpected features
close to the Fermi energy (see the essentially dispersionless 
structure in figure \ref{figalkali}). These features   occur
for photon energies for which there are now vertical, energy-conserving
transitions available. Due to the exponential decay of  $\phi^>$,
nonvertical but
energy-conserving transitions are available. It was shown that 
interference between surface and bulk emission puts most of the weight
of  these transitions close to the Fermi energy (Shung and Mahan 1986).
figure \ref{figalkali} illustrates that the theory can almost completely
explain the experimental features, with just a few experimental points at 
larger photon energy unexplained.
Thus there seems to be no need to assume a charge-density wave to explain
this structure.

\subsection{Semiconductors and Insulators: sp systems}

The LDA systematically underestimates the band gaps in semiconductors and
insulators. In table \ref{bandgaps}
the calculated LDA band gaps of some materials are
compared with the experimental gaps. The discrepancies range from 30-100 \%
and for Ge the LDA conduction and valence bands in fact overlap when
relativistic corrections are included. 
Also, individual bands away from the Fermi level can
be in error up to 50 \%. It is natural to ask if the band-gap problem
originates from the error in the LDA. Exchange-correlation potentials $V^{\rm
xc}
$ calculated from {\em GW }self-energies turn out to be similar to the LDA $%
V^{\rm xc}$ (Godby, Schl\"uter, and Sham 1988)
which indicates that even the exact $V^{\rm xc}$ probably does not give
the correct gap but this is still an open question. 
The $V^{\rm xc}$ may be a non-analytic function of the particle
number (Almbladh and von Barth 1985b, Perdew and Levy 1983, Sham and
Schl\"uter 1983). 
That is to say, $V^{\rm xc}$ with an extra electron, $V^{\rm xc}_{\rm N+1}$, may
have an additional constant compared to $V^{\rm xc}_N$ and this constant may be
large. 
Moreover, experience with empirical potentials shows that a
local potential cannot in general give both the correct band
structure and the ground-state electron density (Kane 1971).

In table \ref{bandgaps}
the band gaps of some materials calculated within the GWA are
shown. The agreement with experiment is very good, in most cases to within
0.1 eV. Although it is not strictly true, the self-energy correction is
approximately an upward rigid shift of the conduction band relative to the
valence band, the so called scissor operator, i.e. cutting the band
structure along the band gap and shifting the conduction band rigidly
upwards.
The scissor operator is accurate to 0.1, 0.2, 0.2, and 0.4 eV in
Si, GaAs, AlAs, and diamond respectively
(Godby, Schl\"uter, and Sham 1988).
The validity of the scissor
operator in Si is somewhat fortuitous, due to an almost complete
cancellation between the strong energy dependence and non-locality.

The experimental values are obtained from optical measurements or
photoemission and inverse photoemission experiments. The latter corresponds
closer to the theoretical values whereas
the former may contain excitonic binding energy which should be
subtracted off but unfortunately it is unknown. In optical experiments, the
excited electron does not leave the system and may therefore form an exciton
with the corresponding hole.

To discuss in more details the features in the self-energy which are
important for the quasiparticle energies, we consider Si as a prototype
since it has been studied extensively. The main features are energy
dependence and non-locality. We first consider non-locality within the
COHSEX approximation. A measure of non-locality in the self-energy is its
range, the distance $|{\bf r-r}^{\prime }|$ beyond which the self-energy is
approximately zero. This range $r_h$ is approximately given by the
corresponding value for a jellium with the average density of Si ($r_s=2$)
and $r_h\sim 2r_s$. In Si more than 99 \% of the matrix
element of the self-energy in a state $\psi $, 
$\langle \psi |\Sigma |\psi \rangle $, 
originates from $|{\bf r-r}^{\prime }|<r_h$.
One expects
non-locality to be important when $r_h$ is comparable to or greater than the
extent or wavelength of the wavefunction which is the case in Si. The matrix
element in a non-local potential can be very sensitive to the nodal
structure of $\psi $. This is contrary to the case when $\Sigma $ 
is local
such as $V^{\rm xc}$ since it is then $|\psi |^2$ that enters into the integral.
In Si, non-locality has the effect of widening the gap. This can be
understood as follows. The top of the valence band is bonding p whereas
the bottom of the conduction band is antibonding p. Therefore
non-locality has a larger effect on the conduction band than on the valence
band since the antibonding state has an extra node which means a smaller
wavelength than that of the bonding state. The presence of an extra node in
the antibonding state reduces the matrix element 
$\langle \psi |\Sigma |\psi \rangle $ relative to 
$\langle \psi |V^{\rm xc}|\psi \rangle $ and therefore the
conduction band is pushed upwards. Non-locality has a smaller effect on the
valence state so the net effect is a widening of the gap
(Godby, Schl\"uter, and Sham 1988).

\noindent
\begin{figure}[bt]
\unitlength1cm
\begin{minipage}[t]{15.0cm}
\rotatebox{1}
{\centerline{\epsfxsize=4.0in \epsffile{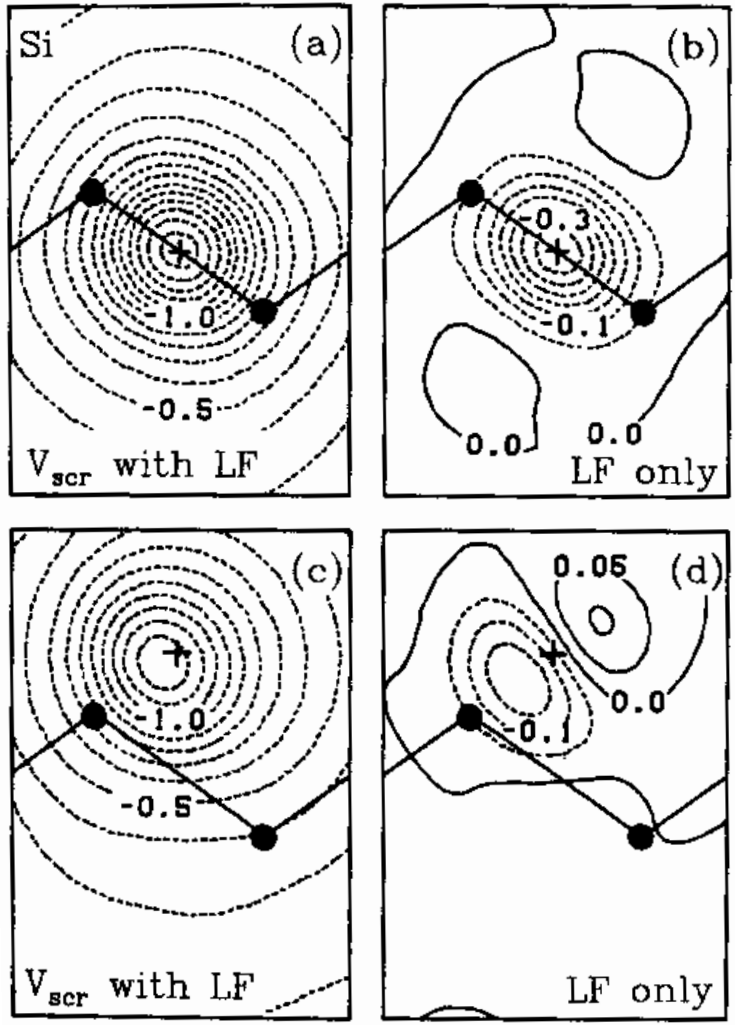}}}
\vskip0.3cm
\caption[]{\label{hl5}
(a) and (c) show 
the screening potential in response to a single electron at ${\bf r}'$
(indicated by +) in the ($1\bar 10$) plane of Si in units of Ry.
(b) and (d) show the contribution from local fields only. The bond chain is
indicated by a straight line.
After Hybertsen and Louie (1986).
}
\end{minipage}
\hfill
\end{figure}

The non-locality arises from the density matrix (exchange charge) 
\begin{equation}
\rho _x({\bf r,r}^{\prime })=\sum_i^{\rm occ}
\psi _i({\bf r})\psi _i^{*}({\bf r}%
^{\prime })  \label{exchangecharge}
\end{equation}
and the screened potential $W({\bf r,r}^{\prime },\omega )$. In
semiconductors, the screening is not complete because due to the gap, a
finite energy is required to excite a particle-hole pair. The range of the
screened interaction and its nodal structure are then determined by $\rho _x$%
. Apart from non-locality, anisotropy also plays an important role. In a
homogeneous system, $\rho _x$ and $W$ only depend on $|{\bf r-r}^{\prime }|$%
. In an inhomogeneous system, the screening potential $W-v$ as a function of 
${\bf r}^{\prime }$ may strongly depend on the location ${\bf r}$ of the
test charge. In Si, for example, there is a large accumulation of charge in
the bonding region as opposed to the antibonding region. It is to be
expected that the screening potential of a test charge located in the
bonding and antibonding regions will be very different, as shown in 
figure \ref{hl5}.
This local field effect , which is entirely missing in the
homogeneous case, is very important in covalent materials. Local fields are
crucial in determining the strength of the screening hole but not its shape
and they contribute directly to the differing strengths of $\Sigma $ at
different points in the unit cell and therefore to the band-gap correction
figure \ref{hl5} (Hybertsen and Louie 1986).
In Si local fields account for more than one-third of the screening
potential in the region around the bond as can be seen in
figure \ref{hl5}.

\noindent
\begin{figure}[bt]
\unitlength1cm
\begin{minipage}[t]{15.0cm}
\rotatebox{180.25}
{\centerline{\epsfxsize=4.5in \epsffile{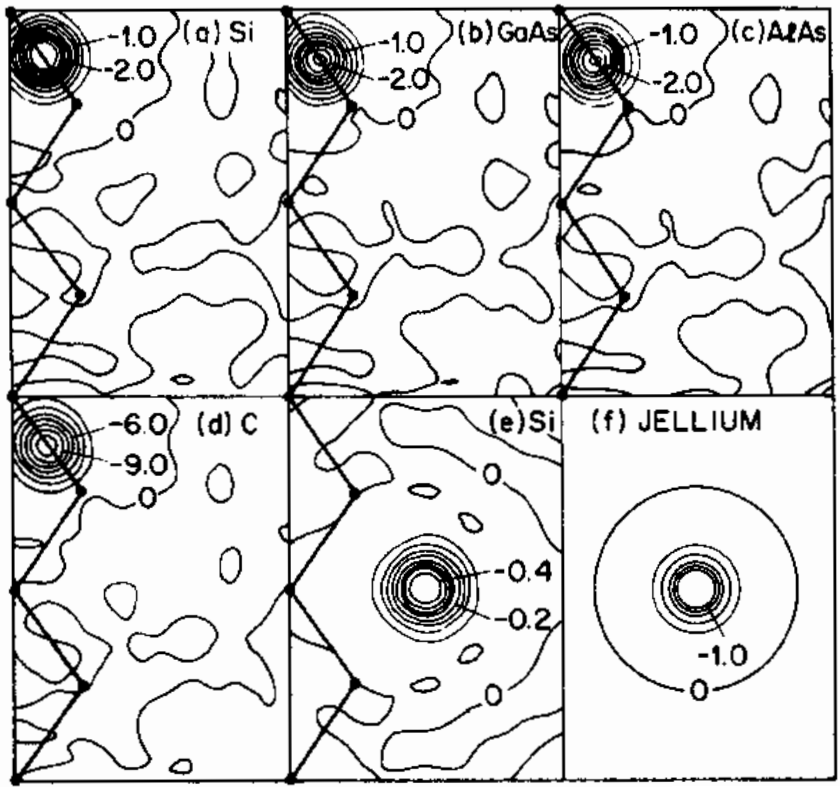}}}
\vskip0.3cm
\caption[]{\label{g7}
Contour plots of the self-energy $\Sigma({\bf r,r}',\omega$=midgap) in 
eV a.u.$^{-3}$ for ${\bf r}$ fixed at the bond centre and ${\bf r}'$ shown in
the (110) plane for (a) Si, (b)GaAs, (c) AlAs, and (d) diamond. For Si the
corresponding plots with ${\bf r}$ fixed at the tetrahedral interstitial site
is also shown (e). For comparison, the self-energy operator of jellium with
$r_s=2.0$ (the average electron density of Si) is shown in (f) (from Hedin
(1965a)). After Godby, Schl\"uter, and Sham (1988).
}
\end{minipage}
\hfill
\end{figure}

The local fields shift the centre of the screening potential and
increase it in the bonding region and reduce it in the interstitial. The
local field effect is one important feature in Si which distinguishes
covalent-bond semiconductors from the alkalis. The screening potential in
these materials is therefore considerably more complex than in the alkalis.
Calculations for Si show that the local field effect in $W$ is confined to a
region near ${\bf r}$ as shown in figure \ref{hl5}. This is not surprising
since the long-range part of the potential is contained in the diagonal
element $W({\bf q})$ corresponding to small ${\bf q}$. In the plane-wave
basis, the local fields are described by the off-diagonal elements of the
dielectric matrix. Nevertheless, $\Sigma ({\bf r,r}^{\prime },\omega =0)$ is
almost spherical as a function of ${\bf r}^{\prime }$ for a fixed ${\bf r}$
and it is reasonably well reproduced by jellium $\Sigma $ with the same
average density as that of Si. This is illustrated in figure
\ref{g7}. However, the interaction of the wavefunction
with the non-locality in $\Sigma $ is not contained within the jellium model.

\begin{figure}[bt]
\unitlength1cm
\begin{minipage}[t]{15.0cm}
\rotatebox{180}
{\centerline{\epsfxsize=2.0in \epsffile{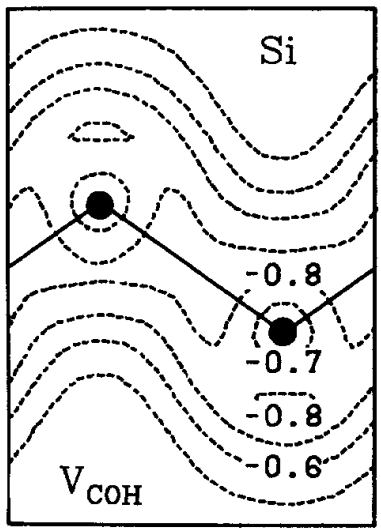}}}
\vskip0.3cm
\caption[]{\label{hl6}
The local potential corresponding
to the Coulomb hole part of the electron self-energy in the COHSEX approximation
in the $(1\bar 10)$ plane of Si in units of Ry. 
After Hybertsen and Louie (1986).
}
\end{minipage}
\hfill
\end{figure}

The local field effect has a large influence on $\Sigma _{\rm{COH}}$. 
In fact, in
a homogeneous system $\Sigma _{\rm{COH}}$ is a constant within the COHSEX
approximation (equation (\ref{sigmacoh})). 
Thus, if local field effect is neglected, $\Sigma _{\rm{COH}}$ has
no effects on the band dispersion. $\Sigma _{\rm{COH}}$ 
is deeper in the bonding
region and shallower in the antibonding region as shown in figure
\ref{hl6}. Since the valence state is concentrated in the bonding region and
the conduction state in the antibonding region, $\Sigma _{\rm{COH}}$ 
makes a
large contribution to the gap. $\Sigma _{\rm{COH}}$ is, however, a local
potential within the COHSEX approximation, a feature which is common to $%
V^{\rm xc}$ as opposed to $\Sigma _{{\rm SEX}}$ 
which is non-local. This mechanism of
gap opening by $\Sigma _{\rm{COH}}$ 
could in principle be accounted for by $%
V^{\rm xc} $. The local field effect on $\Sigma _{{\rm SEX}}$, 
on the other hand, is
small, since the local field effect is rather localized and 
$\Sigma _{{\rm SEX}}$
is dominated by the large bare Coulomb interaction for small $|{\bf r-r}%
^{\prime }|$. The local field contribution to $\Sigma _{{\rm SEX}}$ 
is about 25 \%
of that to $\Sigma _{\rm {COH}}$ 
but of opposite sign (Hybertsen and Louie 1986). Non-locality is essential
in determining the correct quasiparticle energies and in particular the band
gap.

\noindent

The COHSEX approximation is a static approximation to the self-energy.
Strictly speaking, the quasiparticle energy should be obtained from the
position of the peak in the spectral function. This procedure requires
knowledge of the energy dependence of the self-energy at least around the
quasiparticle energy. The degree of the energy dependence is related to the
renormalization factor Z 
(weight of the quasiparticle, equation (\ref{zfactor})): the smaller Z, the
stronger the energy dependence. The value of Z for semiconductors is
0.7-0.8 (Hybertsen and Louie 1986). 
The slope of the real part of the self-energy around the chemical
potential is negative and the renormalization factor Z reduces the
self-energy correction, as shown in equation (\ref{qpenergy}).
Thus, if the energy dependence of
the self-energy is neglected, that is if the self-energy were calculated at
the LDA instead of quasiparticle energy, the band-gap correction would be
overestimated. This is in agreement with the results of the COHSEX
approximation which approximately corresponds to neglecting the
renormalization factor Z. A similar conclusion is reached if the self-energy
is approximated by its value at $\omega =0$. In this case the self-energy
correction for the valence state would be underestimated whereas for the
conduction state overestimated, leading again to an overestimated band gap.
A much better agreement is obtained if the self-energy correction $(\Sigma
_{\rm{COHSEX}}-V^{\rm xc})$ 
is simply multiplied by Z. We note that the COHSEX
approximation without local field effect generally gives a gap in better
agreement with experiment, although not for Si. This means that local field
effect, or in a certain sense non-locality, tends to cancel energy
dependence. The importance of energy dependence is illustrated by plotting $%
\langle \psi _{{\bf k}n}|\Sigma (E_{{\bf k}n})-V^{\rm xc}|\psi _{{\bf k}%
n}\rangle $ and $\langle \psi _{{\bf k}n}|\Sigma (0)-V^{\rm xc}|\psi _{{\bf k}%
n}\rangle $ as a function of $E_{{\bf k}n}$. 
The first quantity is very much like a
step function (scissor operator) while the second quantity shows a strong
energy dependence (Figure 5b and 6b of Godby, Schl\"uter, and Sham 1988).
The effect of energy dependence is therefore to
alter greatly the dispersion of the individual bands.

In general,
the energy dependence of $\Sigma $ leads to a strongly state-dependent
self-energy correction
$\Delta \Sigma =\Sigma -V^{\rm xc}$ 
within each band as well as across the gap. 
The weak state dependence of $\Delta \Sigma $ for Si within a band,
resulting in a scissor operator, 
is therefore a coincidence. In diamond, for instance, the scissor operator
approximation is not as good as in Si.

\noindent
\begin{figure}[bt]
\unitlength1cm
\begin{minipage}[t]{15.0cm}
\rotatebox{180}
{\centerline{\epsfxsize=2.5in \epsffile{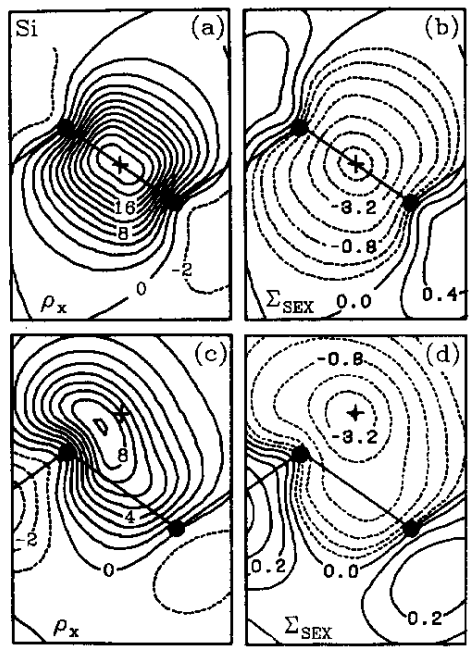}}}
\vskip0.3cm
\caption[]{\label{hl7}
(a) and (c): The exchange charge in the $(1\bar 10)$ plane of Si with
${\bf r}'$ fixed (indicated by +) in units of electrons/cell.
(b) and (d): The screened-exchange part of the
self-energy $\times |{\bf r-r}'|$ in the COHSEX approximation in units of
a.u. Ry./cell and the contours increase in powers of 2.
After Hybertsen and Louie (1986).
}
\end{minipage}
\hfill
\end{figure}

The highest occupied state in the exact
DFT gives the exact ionization energy (Almbladh and von Barth 1985a).
Assuming that the LDA $V^{\rm xc}$ is close
to the exact one, we expect the self-energy correction to shift the conduction
band but not the top of the valence band.
If the RPA is used to obtain the electron gas data (von Barth-Hedin potential
1972), it is indeed found that the the top of the valence band is almost the
same within the LDA and the GWA (Godby, Schl\"uter, and Sham 1988).
More accurate electron gas data obtained from {\em GW} calculations
(Lundqvist and Samathiyakanit 1969) or Quantum Monte Carlo simulations 
(Ceperley and Alder 1980), however, shift the top of the valence band upward
by about 0,5 eV. 
Thus, vertex corrections (corrections beyond the GWA)
may shift the {\em GW} bandstructure by 0.5 eV upwards 
(Godby, Schl\"uter, and Sham 1988) or the LDA may not be sufficiently
accurate.

Other authors have also repeated $GW$ calculations for Si band structure
with similar results (Hamada, Hwang, and
Freeman 1990, Rohlfing, Kr\"uger, and Pollman 1993).
Calculations of the quasiparticle bandstructure and the band gap of many
semiconductors and insulators have been performed by a number of authors.
Quasiparticle band structures of six II-VI compounds
ZnS, ZnSe, ZnTe, CdS, CdSe, and CdTe have been calculated by 
Zakharov {\em et al} (1994).
We list in table \ref{bandgaps} the results of these calculations.
Quasiparticle energies for the F-center defect in LiCl have also been
calculated by Surh, Chacham, and Louie (1995)
and self-energy calculations of carrier-induced band-gap narrowing in Si
may be found in the work of Oschlies, Godby, and Needs (1992,
1995). Berggren and Sernelius (1981) also studied the band-gap narrowing 
in doped Si and Ge as a function of impurity concentration.
Application of the GWA
to metal-insulator transition of Si in the diamond structure
(Godby and Needs 1989) suggests that the metalization occurs at a much smaller
volume than in the LDA which in turn indicates that the the Fermi surface
obtained from the Kohn-Sham DFT is not necessarily the same as that of the
real system as shown by Sch\"onhammer and Gunnarsson (1988) for model systems.
Recently, application to a two-dimensional crystal found good agreement
between the quasiparticle energies in the GWA and Quantum Monte Carlo results
(Engel, Kwon, and Martin 1995).

\begin{table}[t]
\begin{center}
\begin{tabular}{l l l l}
\hline \\
          & LDA       & GWA           &  Expt. \\
\hline \\
AlAs      & 1.37      & 2.18$^b$      & 2.32$^\alpha$ \\

Al$_{0.5}$Ga$_{0.5}$As
          & 1.12      & 2.06$^e$      & 2.09$^\alpha$\\

AlN (wurtzite)
          & 3.9       & 5.8$^d$       & 6.2$^\alpha$ \\
AlN (zinc-blende)
          & 3.2       & 4.9$^d$       &  \\

AlP       & 1.52      & 2.59$^e$      & 2.50$^\alpha$\\

AlSb      & 0.99      & 1.64$^e$      & 1.68$^\alpha$\\

CdS (zinc-blende)
          & 1.37$^i$, 0.83$^g$      & 2.83$^i$, 2.45$^g$      
                                    & 2.55$^{\eta ,\alpha}$\\
CdS (wurtzite)
          & 1.36      & 2.79$^i$      & 2.59$^\alpha$\\

CdSe (zinc-blende)
          & 0.76      & 2.01$^i$      & 1.90$^\alpha$\\
CdSe (wurtzite)
          & 0.75      & 1.91$^i$      & 1.97$^\alpha$\\

CdTe (zinc-blende)
          & 0.80      & 1.76$^i$      & 1.92$^\alpha$\\
CdTe (wurtzite)
          & 0.85      & 1.80$^i$      & 1.60$^\alpha$\\

Diamond   & 3.90      & 5.6$^a$,
                        5.33$^b$,
                        5.67$^c$      & 5.48$^\alpha$  \\

GaAs      & 0.67      & 1.58$^b$,
                        1.32$^c$,
                        1.22$^e$      & 1.52$^\alpha$, 1.63$^\gamma$\\

GaN (wurtzite)
          & 2.3       & 3.5$^d$       & 3.5$^\alpha$\\
GaN (zinc-blende)
          & 2.1       & 3.1$^d$       & 3.2$^\delta$, 3.3$^\epsilon$\\

GaP       & 1.82      & 2.55$^e$      & 2.39$^\alpha$\\

GaSb      &-0.10      & 0.62$^e$      & 0.80$^\alpha$\\

Ge        & $<$0      & 0.75$^a$,
                        0.65$^c$      & 0.744$^\alpha$ \\

InAs      &-0.39      & 0.40$^e$      & 0.41$^\alpha$\\

In$_{0.53}$Ga$_{0.47}$As
          & 0.02      & 0.80$^e$      & 0.81$^\alpha$\\

InP       & 0.57      & 1.44$^e$      & 1.42$^\alpha$\\

InSb      &-0.51      & 0.18$^e$      & 0.23$^\alpha$\\

LiCl      & 6.0       & 9.1$^a$       & 9.4$^\alpha$   \\

Li$_2$O   & 5.3       & 7.4$^h$       & 6.6$^\theta$\\
MgO       & 5.0       & 7.7$^f$       & 7.83$^\zeta$\\

Si        & 0.52      & 1.29$^a$, 
                        1.24$^b$,
                        1.25$^c$      & 1.17$^\beta$ \\

SiC($\beta$)
          & 1.31      & 2.34$^c$      & 2.39$^\alpha$\\
ZnS (zinc-blende)
          & 2.37      & 3.98$^i$      & 3.80$^\alpha$\\
ZnS (wurtzite)
          & 2.45      & 4.03$^i$      & 3.92$^\alpha$\\
ZnSe (zinc-blende)
          & 1.45      & 2.84$^i$      & 2.96$^\alpha$\\
ZnSe (wurtzite)
          & 1.43      & 2.75$^i$      & 2.87$^\alpha$\\
ZnTe (zinc-blende)
          & 1.33      & 2.57$^i$      & 2.71$^\alpha$\\
ZnTe (wurtzite)
          & 1.48      & 2.67$^i$      & \\
\hline\\
\end{tabular}
\end{center}
\caption{ \label{bandgaps}
Minimum band gaps of semiconductors and insulators
which have been calculated within the GWA.
The energy is in eV.
}
\vspace{12pt}
$^\alpha${\em Numerical Data and Functional Relationships in Science and
Technology} 1982\\
$^\beta$Baldini and Bosacchi 1970\\
$^\gamma$Aspnes 1976\\
$^\delta$Lei {\em et al} 1992a,b, Eddy {\em et al} 1993\\
$^\epsilon$Paisley {\em et al} 1989, Sitar {\em et al} 1992\\
$^\zeta$Whited, Flaten and Walker 1973\\
$^\eta$Cardona, Weinstein, and Wolff 1965\\
$^\theta$Rauch 1940\\
\\
$^a$Hybertsen and Louie 1986\\
$^b$Godby, Schl\"uter, and Sham 1988\\
$^c$Rohlfing, Kr\"uger, and Pollman 1993\\
$^d$Rubio {\em et al} 1993\\
$^e$Zhu and Louie 1991 (The number in the bracket corresponds to a calculation
    using a model dielectric matrix)\\
$^f$Sch\"onberger and Aryasetiawan 1995\\
$^g$Rohlfing, Kr\"uger, and Pollmann 1995a\\
$^h$Albrecht, Onida, and Reining 1997\\
$^i$Zakharov {\em et al} 1994\\
\end{table}

\subsubsection{Core polarization}

A disadvantage of using pseudopotentials is a difficulty of including the
core states in the calculation of the exchange potential as well as the
polarization. Within a pseudopotential scheme, it is inevitable that core
electrons are treated in an approximate manner. The core electron charge
densities are frozen at their atomic values in the reference configuration
used for constructing the pseudopotential. The core electrons and their
potential are then eliminated so that there is no possibility for them to
relax. Core relaxation gives rise to crystal field distortion and strong
mixing between 3d{\em \ }and 3p states. These single-particle effects are
small and can be included {\em a posteriori }by comparison with all-electron
calculations. It is, however, important to take into account many-body
effects arising from core relaxation since they can be large. In atoms with
easily polarizable core the inclusion of core relaxation leads to an
increase in the ionization energies, a contraction of the valence shell, a
reduction of polarizabilities and oscillator strengths of the valence
electrons.

The self-energy may be broken up into three terms 
(Hedin 1965b, Hedin and Lundqvist 1969):
\begin{equation}
\Sigma =iG_cW+iG_vW_v+iG_vvR_cv
\end{equation}
where $W=W_v+vR_cv.$ The first term is the core-valence and core-core
(screened) exchange which is essentially the same as the bare exchange since
screening is ineffective for small distance. The second term is the
self-energy of the valence electron and the last term is the screened
polarization potential due to the core electrons acting on the valence
electrons. In pseudopotential calculations, the first and last term together
are approximated by the LDA. The absolute contribution from these terms is
estimated to be $\sim $1 eV for atomic Na (Hedin and Lundqvist 1969)
and solid Al (Arbman and von Barth 1975) but the difference $\Sigma
-V^{\rm xc}$, which is a more relevant quantity, is much smaller. For s-p
semiconductors, this difference can be significantly larger. As a result,
the calculated direct gaps and also the orderings and splittings of the
conduction bands in Ge and GaAs, materials of technological interest, are in
disagreement with experiment. It is important to get the orderings of the
conduction band right since they affect the transport properties and
life-time of thermally excited carriers. Atomic calculations in Ge estimate
the error for the 4s and 4p states to be 0.3 and 0.04 eV respectively
(Hybertsen and Louie 1986). If
this error is taken into account, the result for the band gap becomes even
better. In transition metals it is crucial to include the core electrons in
the calculations of the bare exchange since the error can be as large as $%
\sim $1 eV.

While the core-valence exchange can be treated straightforwardly in the
Hartree-Fock theory, core-valence correlation is more complicated. A good
approximation for taking into account core polarization is provided by the
core polarization potential (CPP) method used in quantum chemistry
(M\"uller, Flesch, and Meyer 1984). The
physical idea behind this method is that core polarization functions are
characterized by sharp high-frequency excitations and rather insensitive to
the valence environment. This allows evaluation of the core polarization
function for isolated atoms and neglect of frequency dependence. Consider a
valence electron in the presence of a core. The electron polarizes the core
resulting in polarization ${\bf p=}\alpha _c\nabla \left( 1/r\right) $ where 
${\bf r}$ is the position of the electron with respect to the core. From
classical theory of electrostatic, the electric field arising from this
polarization at ${\bf r}^{\prime }$ is given by 
\begin{equation}
{\bf E}\left( {\bf r}^{\prime }\right) =\frac{3\left( {\bf p}\cdot {\bf r}%
^{\prime }\right) {\bf r}^{\prime }-r^{\prime 2}{\bf p}}{r^{\prime 5}}
\end{equation} 
The potential experienced by another electron at ${\bf r}^{\prime }$ due to
core polarization is then 
\begin{equation}
V_{e-e}\left( {\bf r,r}^{\prime }\right) =-\alpha _c\frac{{\bf r\cdot r}%
^{\prime }}{\left( rr^{\prime }\right) ^3}  \label{vee}
\end{equation}
assuming that both $r$ and $r^{\prime }$ are large. For a set of cores and
many valence electrons, the core polarization potential is 
\begin{equation}
V_{CPP}=-\frac 12\sum_c\alpha _c{\bf f}_c\cdot {\bf f}_c  \label{vcpp}
\end{equation}
where ${\bf f}_c$ is the electric field acting on core $c$ due to the
valence charges at $i$ and all other cores. 
\begin{equation}
{\bf f}_c=\sum_i\frac{{\bf r}_{ci}}{r_{ci}^3}C\left( r_{ci}\right)
-\sum_{c^{\prime }\neq c}\frac{{\bf R}_{cc^{\prime }}}{R_{cc^{\prime }}^3}%
Z_{c^{\prime }}
\end{equation} 
where $C$ is a cut-off function, $Z_{c^{\prime }}$ is the net charge of core 
$c^{\prime }$ and ${\bf R}_{cc^{\prime }}={\bf R}_{c^{\prime }}-{\bf R}_c.$
Inserting ${\bf f}_c$ into equation (\ref{vcpp}) yields 
(M\"uller, Flesch, and Meyer 1984)
\begin{eqnarray}
V_{CPP} &=&-\frac 12\sum_c\alpha _c\left\{ \sum_i\frac 1{r_{ci}^4}C^2\left(
r_{ci}\right) +\sum_{i\neq j}\frac{{\bf r}_{ci}\cdot {\bf r}_{cj}}{\
r_{ci}^3r_{cj}^3}C\left( r_{ci}\right) C(r_{cj})\right.   \nonumber \\
&&\quad \;\;\;\;\;\;\;\;\left. -2\sum_i\sum_{c^{\prime }\neq c}\frac{{\bf r}%
_{ci}\cdot {\bf R}_{cc^{\prime }}}{r_{ci}^3R_{cc^{\prime }}^3}Z_{c^{\prime
}}C(r_{ci})+\sum_{c^{\prime },c^{\prime \prime }\neq c}\frac{{\bf R}%
_{cc^{\prime }}\cdot {\bf R}_{cc^{\prime \prime }}}{\ R_{cc^{\prime
}}^3R_{cc^{\prime \prime }}^3}Z_{c^{\prime }}Z_{c^{\prime \prime }}\right\} 
\label{vcppfull}
\end{eqnarray}
The first term is the CPP in atoms with a single valence electron. Without
the cut-off function, it would diverge at small $r$.
The cut-off function $C$ is semi-empirical with a
parameter related to the size of the core (Biermann 1943, Biermann and
L\"ubeck 1948, Bates 1947). The first term is a one-particle
operator which is added to the pseudopotential. The second term is the
two-electron interaction which was qualitatively discussed above. The third
term is an indirect interaction of a valence electron with another core.
This term is repulsive and cancels the attractive potentials from the first
and last term. The latter is a core-core interaction and, together with the
third term, essential to make sure vanishing long-range interaction with a
neutral atom.

The effective interaction among the {\em valence} electrons is given by 
(Hedin and Lundqvist 1969)
\begin{equation}
W_C=v+v\sum_cR_cv+v\sum_cR_cv\sum_{c^{\prime }\neq c}R_{c^{\prime }}v+\ldots
\label{Wc}
\end{equation}
$R_c$ is the full or self-consistent response function of core $c$ given by 
\begin{equation}
vR_cv=V_{e-e} 
\end{equation} 
Thus, the screened interaction becomes
\begin{eqnarray}
W &=&\epsilon ^{-1}v \nonumber\\
&=&\left[ 1-W_CP^0\right] ^{-1}W_C
\end{eqnarray}
where $P^0$ is the valence RPA polarization function. Using this CPP
formalism, various transition energies for Si, Ge, AlAs, and GaAs were
calculated by Shirley, Zhu, and Louie (1992). 
The results for the fundamental bandgaps of Si, Ge, AlAs, and GaAs
shown in table \ref{corepol} are in
systematically better agreement with experiment compared with previous
calculations where core relaxation effects were taken into account within
the LDA only. The correction can be as large as 0.4 eV for the fundamental
gap in GaAs. Notable also (not shown in the table)
is the correct $L-\Gamma -X$ ordering of
conduction-band states in Ge obtained in the CPP approach. $X-L$ and $%
X_{6c}-X_{7c}$ splittings in GaAs are also improved 
(Shirley, Zhu, and Louie 1992).

The CPP formalism is relatively easy to implement in many-body valence
calculations without increasing computational cost significantly. Apart from
its use within the pseudopotential approach, it can also be used in
all-electron calculations with frozen core.

\vfill\eject
\begin{table}[h]
\begin{center}
\begin{tabular}{l l l l}
\hline \\
          & LDA       & CPP           &  Expt. \\
\hline \\
Si                                &           &            &\\ 
$\Gamma_{8v}\rightarrow 0.85X_{5c}$   & 1.29      & 1.16      & 1.17 \\

Ge                                &           &            &\\ 
$\Gamma_{8v}\rightarrow \Gamma_{7c}$  & 0.53      & 0.85       & 0.89\\
$\Gamma_{8v}\rightarrow L_{6c}$       & 0.58      & 0.73       & 0.744\\

AlAs                              &           &            &\\ 
$\Gamma_{8v}\rightarrow X_{6c}$       & 2.09      & 2.01       & 2.24\\

GaAs                              &           &            &\\ 
$\Gamma_{8v}\rightarrow X_{6c}$       & 1.02      & 1.42       & 1.52\\
\hline\\
\end{tabular}
\end{center}
\caption{ \label{corepol}
Fundamental
bandgaps of Si, Ge, AlAs, and GaAs calculated within the LDA and the GWA
including core polarization within the CPP formalism 
(Shirley, Zhu, and Louie 1992) compared with
experiment (Madelung O 1984).
Energies are in eV.
}
\end{table}

\subsection{Transition metals}

The success of the GWA in semiconductors has encouraged applications to more
complicated systems of transition metals and their compounds. From numerical
point of view, transition metals require a new approach
for calculating the response function and the self-energy. The
conventional plane-wave basis is not suitable in this case because the
localized nature of the 3d states results in a prohibitively large number of
plane waves. From theoretical point of view, it has been argued that the
localized nature of the 3d states makes atomic approach as more
suitable for explaining the characteristic properties of transition metals.
The presence of a Fermi surface as shown by de Haas-van Alphen measurements,
on the other hand, strongly suggests that the itinerant character of the 3d
electrons should be taken into account. Moreover,
the band widths in transition metals
are not too small and the ratio between the Hubbard $U$ and the
band width is of order one.This is crucial when we consider screening of a
photoemission hole where 3d electrons from neighbouring cells can take part
in the screening whereas such possibility is absent in the atomic case. It
seems then that RPA-type of approach such as the GWA is meaningful for these
systems. {\em GW }calculations for transition metals have not been
extensively performed. We concentrate therefore on two materials Ni 
(Aryasetiawan 1992a, Aryasetiawan and von Barth 1992b) 
and NiO (Aryasetiawan and Gunnarsson 1995)
for which full {\em GW }calculations have been done in some details and on
MnO for which a model {\em GW} calculation has been performed 
(Massidda {\em et al} 1995a).

\subsubsection{Nickel}

Among the transition elements, Ni is the most anomalous case in many
respects. The LDA bandstructure deviates significantly from angle-resolved
photoemission data.\ The occupied 3d-band width is 30 \% smaller than that
of the LDA (3.3 eV {\em vs }4.5 eV) (H\"ufner {\em et al} 1972, Himpsel,
Knapp, and Eastman 1979) and the experimental exchange splitting
is half the LDA\ value (0.25-0.30 eV {\em vs }0.6 eV)
(Eberhardt and Plummer 1980). Furthermore there is
the famous 6 eV satellite below the Fermi level (H\"ufner {\em et al} 1972,
H\"ufner and Wertheim 1973,
Kemeny and Shevchik 1975) which is entirely missing in
the LDA or in any single-particle theory. These discrepancies are related to
excited-state properties. An indication that single-particle theories would
have difficulties in describing quasiparticle properties in Ni is the fact
that the two lowest atomic configurations $3d^94s$ and $3d^84s^2$ are almost
degenerate, differing by only 0.025 eV (Moore 1958). 
Another indication of many-body
effects in Ni is the unusually large quasiparticle widths 
(Eberhardt and Plummer 1980) - up to 2 eV at
the bottom of the 3d band - which implies strong interaction between the
quasiparticles and the rest of the system resulting in short life-times. The
photoemission process introduces an additional 3d hole to an already
existing one, causing on-site many-body correlations not amenable to a
single-particle treatment. This is in contrast to Cu where there is only one
3d hole after photoemission and where the LDA bandstructure is good, apart
from a somewhat too high position (0.4 eV) of the 3d band relative to the 4s
band (see e.g. Jones and Gunnarsson 1989). 
Ground-state properties such as equilibrium lattice constant, bulk
modulus, and magnetic moment are well reproduced by the LDA with the
exception of the cohesive energy where the LDA value is about 1 eV too small
(Moruzzi, Janak, and Williams 1978).

\begin{figure}[bt]
\unitlength1cm
\begin{minipage}[t]{15.0cm}
\rotatebox{180.00}
{\centerline{\epsfxsize=3.5in \epsffile{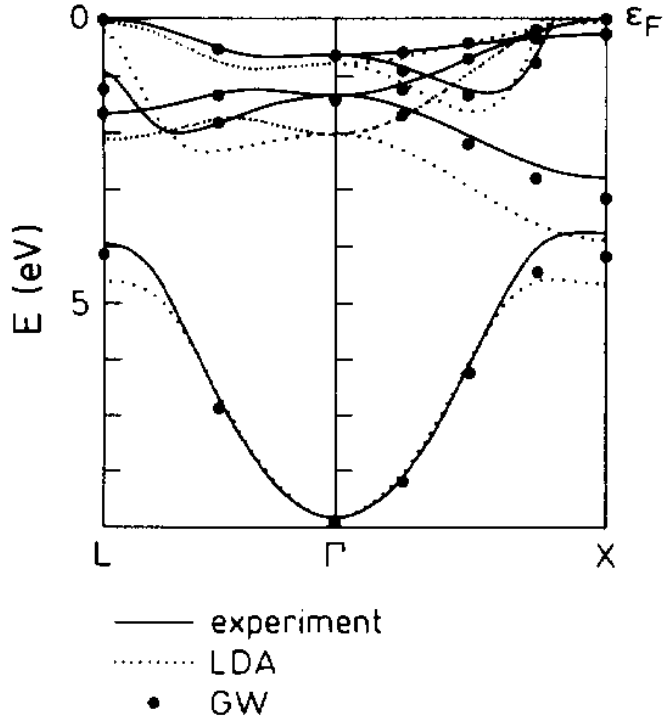}}}
\vskip0.3cm
\caption[]{\label{a4}
The band structure of Ni along $\Gamma=(0,\;0,\;0)$ and
$X=(1,\;0,\;0)$ and along $\Gamma$ and $L=(0.5,\;0.5,\;,0.5)$.
The solid curves are the experiment and the dotted curves are the LDA 
(from M{\aa}rtensson and Nilsson 1984). The filled circles are the
quasiparticle energies in the GWA. After Aryasetiawan (1992a) and Aryasetiawan
and von Barth (1992b).
}
\end{minipage}
\hfill
\end{figure}

The {\em GW} results for Ni may be summarized as follows
(Aryasetiawan 1992a, Aryasetiawan and von Barth 1992b): The LDA
bandstructure is much improved, in particular the 3d-band width is narrowed
by almost 1 eV, as shown in figure \ref{a4}. 
The quasiparticle life-times are also given rather well by
the GWA but the exchange splittings remain essentially unchanged from their
LDA values and the 6 eV satellite is not reproduced.

\begin{figure}[bt]
\unitlength1cm
\begin{minipage}[t]{15.0cm}
\rotatebox{-0.50}
{\centerline{\epsfxsize=3.5in \epsffile{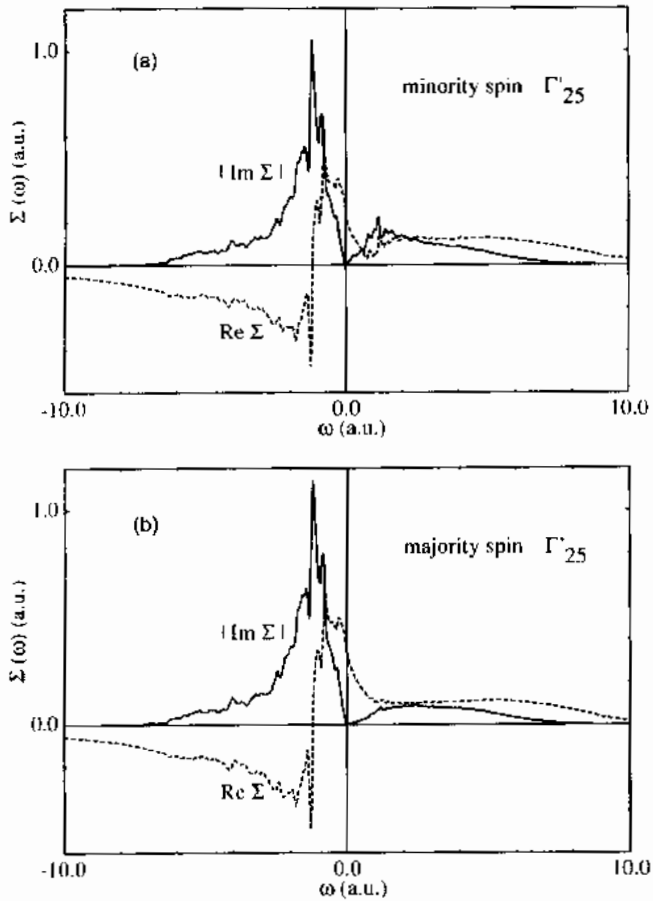}}}
\vskip0.3cm
\caption[]{\label{a1}
The real and imaginary parts of the correlation part of the self-energy
for the minority and majority spin state $\Gamma_{25}'$ (The lowest 3d
state at the $\Gamma$ point). The units are in a.u. (Hartree=27.2 eV).
After Aryasetiawan (1992a).
}
\end{minipage}
\hfill
\end{figure}

The self-energies show a number of interesting features. 
As an illustration,
the self-energy as a function of frequency for the $\Gamma'_{25}$
state is shown in figure \ref{a1}.
The imaginary part of the self-energy
is significantly more complicated compared with those of the alkalis or
semiconductors. The latter are typically characterized by a large peak
associated with a plasmon excitation but they otherwise show no other
distinct structures. The frequency structure of the imaginary part of the
self-energy is determined essentially by the imaginary part of the screened
interaction $W$, as may be seen in equation (\ref{imsignegw}).
In alkalis or $%
s-p$ semiconductors {\rm Im }$W$ is dominated by a plasmon peak which 
is then mirrored in 
${\rm  \mathop{\rm Im} }$ $\Sigma^c .$ 
Similarly in Ni, there is a strong similarity between 
{\rm Im} $W$ (figure \ref{ag3}) and {\rm Im }$\Sigma^c $. In transition metals,
there is no well-defined plasmon excitation, rather it merges with the
single-particle excitations forming a broad spectrum. There is a two-peak
structure at about 20 and 30 eV which is probably due to plasmon excitation.
An estimate based on the electron gas formula gives a plasmon energy of 30.8 eV
when the 3d electrons are included in the density. This coincides rather
well with the second large peak in the two-peak structure. A smaller
structure at around 5-6 eV originates from transitions from the occupied
valence band to states just above the Fermi level which constitute a large
density of states. Interesting to observe is the behaviour of {\rm Im }$%
\Sigma^c $ at large frequencies. The hole and particle parts show similar
behaviour and they therefore tend to cancel each other when one performs a
Hilbert transform to obtain the real part of the self-energy. This justifies
the use of energy cut-off in the calculation of the response function. It
also agrees with our physical intuition that the main contribution to the
self-energy should come from energies up to the plasmon energy. For states
lying a few eV below the Fermi level, the hole part (negative energy) of 
{\rm Im }$\Sigma^c $ has larger weight than the particle part (positive
energy). This simply reflects the fact that the hole (occupied) states have
larger overlap and correlation with other occupied states resulting in
larger correlated part of the self-energy for the hole part. As we go
towards the Fermi level, the hole and particle parts become of almost equal
weight. This is the reason why the self-energy correction for states at the
bottom of the band is larger than for those at the top of the band,
resulting in band narrowing. {\rm Im }$\Sigma^c $ around the Fermi level shows
a quadratic Fermi liquid behaviour but it becomes linear rather quickly.

The real part of the self-energy is obtained by Hilbert transforming the
imaginary part. A notable feature is a large derivative at around the Fermi
level which implies large renormalization of the quasiparticle weights.
Typical values for the renormalization factor is 0.5 for the 3d states
(Aryasetiawan 1992a). This
is significantly
smaller than in the electron gas (0.7) (Hedin and Lundqvist 1969)
or semiconductors (0.8) (Hybertsen and Louie 1986) which
reflects a larger loss of single-particle character of the quasiparticles.
The s states on the other hand have renormalization factor $\sim $ 0.7,
comparable to those in the alkalis and semiconductors. It is in agreement
with our physical picture that the 3d states are more correlated than the
4s-4p states.

In comparison with semiconductors, the self-energy corrections in Ni is
considerably more complicated. While in semiconductors the self-energy
correction is approximately a scissor operator which increases the band gap
by shifting the conduction band upwards and leaving the valence band
unchanged, the self-energy correction in Ni shows a rather strong state and
energy dependence. The self-energy correction can be positive or negative
depending on the state and its magnitude varies throughout the Brillouin
zone. For example, at the $X$-point the bottom of the 3d band experiences a
self-energy correction of 0.8 eV while the top of the band is almost
unchanged. The correction to the state $\Gamma _{25}^{\prime }$ is positive
whereas at $L_2^{\prime }$ state it is negative. The state dependence of the
self-energy correction is demonstrated clearly in figure \ref{a4}
by the lowest
valence band which is of 4s character at the bottom and a mixture between s
and d at the top. The self-energy correction is approximately zero at the
bottom of the band and positive at the band edges. The free-electron-like s
states are well described by the LDA but the description of the more
localized d states is less satisfactory. 

As can be seen in figure \ref{a4}, the 3d band width
is reduced by almost 1 eV.
The exchange splittings on the other hand remain essentially
unchanged from their LDA values. The discrepancy between the LDA
exchange splitting and the experiment is rather small, 0.3 eV, which is 
slightly larger than
the numerical accuracy (0.1-0.2 eV) but
the results seem to indicate inadequacy in the GWA itself.
{\em GW} calculations on transition metal atoms 
also show that strong correlations among 3d electrons of opposite spin
are not well accounted for by the GWA (Shirley and Martin 1993).
The quasiparticle widths or the inverse life-times are
in reasonable agreement with available experimental data (Aryasetiawan 1992a).
The large width at the bottom of
the 3d band indicates a strong interaction between the quasiparticles and the
rest of the system.


\noindent

As mentioned before, the 6 eV
satellite is not reproduced in the GWA. 
That the 6 eV satellite is missing in the GWA can be seen directly in the
imaginary part of the self-energy. For the existence of a satellite, there
should be a strong peak at around 5-6 eV reflecting the presence of a stable
excitation. But as can be seen in figure \ref{a1}, 
such peak is missing.

3p-resonance photoemission measurements at 67 eV photon
energy corresponding to the binding energy of the 3p core
exhibit an asymmetric (Fano) resonant enhancement of the satellite
and the main 3d
line shows a strong antiresonance (Guillot {\em et al} 1977). 
This is explained as an Auger process
where a 3p core electron is excited to fill the empty 3d states followed by
a super Coster-Kronig decay 
\begin{equation}
3p^63d^94s+\hbar \omega \rightarrow 3p^53d^{10}4s\rightarrow 3p^63d^84s+e 
\end{equation} 
Although there is some indication that the 6 eV structure might arise from
single-particle states (Kanski, Nilsson, and Larsson 1980)
the 3p resonance photoemission provides a strong evidence of the many-body
character of the 6 eV satellite.
The standard explanation for the presence of the satellite (Penn 1979, Liebsch
1979, 1981) is that a
3d hole created in a photoemission experiment
introduces a strong perturbation due to its localized
nature, causing another 3d electron to be excited to an empty state just
above the Fermi level. In atomic picture, the state with two 3d holes
correspond to the configuration $3d^74s^2$ which is separated from the main
line configuration $3d^84s$ by more than 6 eV but which is reduced
considerably by metallic screening. The two holes scatter each other
repeatedly forming a ''bound state'' at 6 eV. In a simple picture, the
photon energy is used to emit a d electron and to excite another into an
empty d state so that the emitted electron appears to have a lower binding
energy (satellite). The excited electrons mainly come from the bottom of the
d band since they have the largest mixing with the s-p states and therefore,
according to the dipole selection rule, a large transition to the empty d
states. This then has been argued as the source of band narrowing. The {\em %
GW} calculations, however, show that the largest contribution to band
narrowing comes mainly from screening rather than two-hole interactions.

\noindent

The reduction in the exchange splittings is related to the satellite. For
simplicity we consider a two-level model with fully occupied majority
channel and one occupied minority channel. A photoemission hole in the
majority channel can induce another hole in the minority channel but not
vice versa since the majority channel has no empty states. Thus, the effects
of two-hole interactions are larger for the majority than for the minority
channel resulting in a reduction in the exchange splitting. Calculations
based on the Hubbard model within the t-matrix approach (Kanamori 1963)
confirm this
picture (Liebsch 1979, 1981, Penn 1979). 
These model calculations also assign the reduction in the band width
as originating from the two-hole interactions which is not in complete
agreement with
the {\em GW} calculations. These results may be reconciled as follows. In
the t-matrix calculations, it was found that there was no value of U that
gave the correct satellite position and the band width. To get the correct
band width, the value of U was such that the satellite energy became too
large (Liebsch 1981). 
If we assume that the t-matrix theory gives the correct physical
description for the satellite, the appropriate value of U would not give a
large reduction in the band width, but this is taken care of by the GWA.
Thus one concludes that self-energy calculations which include the diagrams
of the GWA and t-matrix theory may give both the correct band width and the
satellite structure. Diagrammatic comparison between the GWA and the
t-matrix theory reveals that hole-hole interactions described by the
t-matrix are not included in the GWA except to second order only. Direct
comparison between {\em GW} calculations on real systems and Hubbard model
calculations is, however, difficult if not impossible. This is because the
Hubbard U cannot be easily related to the screened interaction in the%
{\em GW} calculations.

An extension of the t-matrix
theory including Fadeev's three-body interaction (Fadeev
1963) was made by Igarashi (1983, 1985) and by Calandra and Manghi (1992,
1994). The theory has been applied to Ni (Igarashi {\em et al} 1994,
Manghi, Bellini, and Arcangeli 1997)
and NiO (Manghi and Calandra 1994).

\subsubsection{Nickel Oxide}

NiO is a prototype of the Mott-Hubbard insulators. It was pointed
out by Mott in the late forties 
that a system with an on-site Coulomb energy
larger than the single-particle bandwidth tends to become an insulator and
that single-particle theory is bound to give a wrong prediction for the
state of the system (Mott 1949). 
Indeed, the LDA predicts NiO to be a metal when the
calculation is performed in a paramagnetic state (Mattheiss 1972a,b). 
Slater (1974) suggested that a
gap could be opened by an interplay between antiferromagnetism and
crystal-field splittings. A detailed work along this direction can be found
in a paper by Terakura {\em et al} (Terakura {\em et al} 1984). 
The LDA
does produce a gap in an antiferromagnetic state but of only 0.2 eV in
contrast to the experimental gap of 4.0 eV (Powell and Spicer 1970,
H\"ufner {\em et al} 1984, Sawatzky and Allen 1984). 
As expected, the free-electron
like O p band is well described by the LDA but the magnetic moment is too
small (1.0 $\mu _{{\rm B}}$) compared to experiment (1.7-1.9 $\mu _{{\rm B}}$%
) (Alperin 1962, Fender {\em et al} 1968, Cheetham and Hope 1983).
Clearly there is something seriously wrong with the LDA. A more
convincing evidence is provided by CoO, where the number of electrons in the
paramagnetic structure is odd, making it impossible for any single-particle
theory to predict CoO as an insulator without doubling the unit cell. 
Experimentally, CoO is an
antiferromagnetic insulator so that one might argue like Slater that
single-particle theory could still give the correct prediction if the
calculation is performed in an antiferromagnetic structure. One realizes,
however, that the difference in magnetic energy distinguishing the
paramagnetic and antiferromagnetic states is only a fraction of an eV, which
is much smaller than the band gap. This means that the result of a
theoretical calculation for the band gap should not depend on whether the
calculation is performed in a paramagnetic or an antiferromagnetic structure.

The basic physics of the Mott-Hubbard insulators was explained by Mott
several decades ago (Mott 1949). 
From the tight-binding limit, switching on hopping
matrix elements causes the formation of a band of width $w$ centred around
the atomic eigenvalue. The possibility of occupying states with lower energy
favours electron hopping but it costs a Coulomb energy $U$ for an electron
to hop from one site to the neighbouring site. If $U$ is larger than $w,$
the gain in kinetic energy is overwhelmed by the loss in Coulomb energy and
the system prefers to be an insulator with a gap approximately given by $U,$
splitting the lower and upper Hubbard bands. While the Mott
picture is essentially correct, there are a number of experimental data
which cannot be explained. The value of $U,$ for instance, is estimated to
be 8-10 eV which is much larger than the experimental gap. More recent
studies initiated by Fujimori, Minami, and Sugano (Fujimori {\em et al} 1984)
and Sawatzky and Allen (Sawatzky and Allen 1984) based on the cluster
approach and Anderson impurity model show that the gap in NiO is a
charge-transfer gap. If an electron is removed from a Ni site the number of
holes increases leading to a state with high energy due to an increase in
the Coulomb interaction among the holes. The hole created on the Ni site may
be filled by the transfer of an electron from an O site. Although it costs
some energy transfer this leads to a state with a small binding energy . The
states at the top of the valence band therefore have a large O p character.
The lowest conduction state is of d character as in the Mott picture and the
gap is therefore formed between the valence O p and conduction Ni d states.
Much of the Ni d weight goes into a satellite located below the O p,
opposite to the Mott and the Slater pictures. This model is able to explain
experimental data which would otherwise be difficult to explain by the Mott
and Slater pictures. The most convincing evidence supporting the
charge-transfer picture is the 2p resonant photoemission experiment in CuO
(Tjeng {\em et al} 1991) which has a similar electronic structure as that of
NiO. In this experiment, a 2p core electron is excited and the remaining
hole is subsequently filled by a valence electron. Since dipole transition
matrix element is largest between p and d states, resonance in the valence
energy region can be identified as the position of d states which turns out
to be below the O p band, rather than above as in the Mott and Slater
pictures.

\noindent
\begin{figure}[bt]
\unitlength1cm
\begin{minipage}[t]{15.0cm}
\rotatebox{1.00}
{\centerline{\epsfxsize=4.0in \epsffile{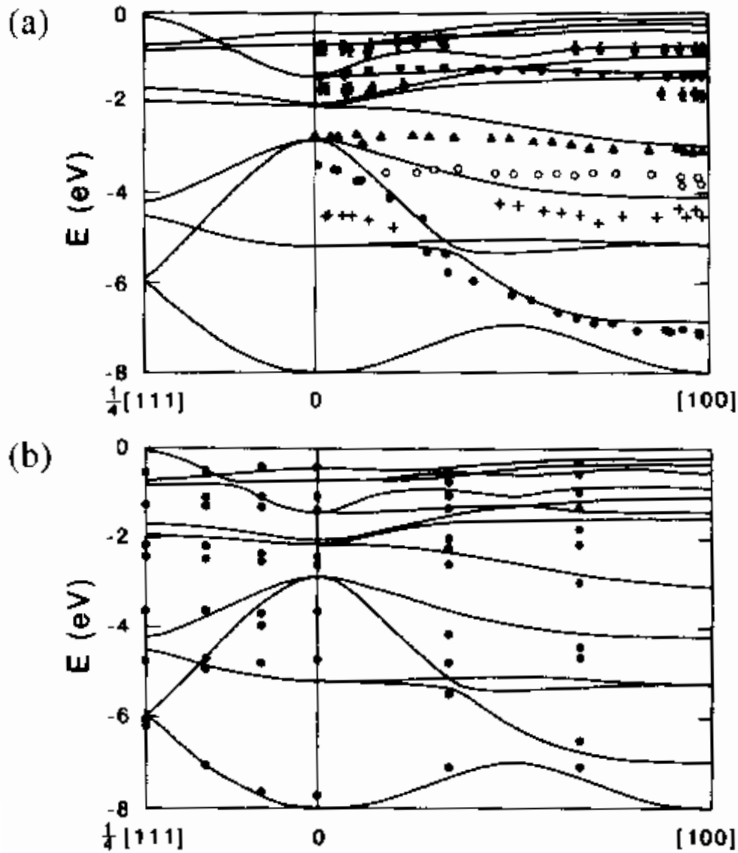}}}
\vskip0.3cm
\caption[]{\label{ag1}
(a) Comparison between the LDA (solid line) and the experimental band
structure of Antiferromagnetic II NiO. 
From Shen {\em et al} (1990, 1991a, 1991b).
(b) Comparison between the LDA band structure 
(solid line) and the quasiparticle band
structure in the GWA (filled circles) for NiO. After Aryasetiawan and Gunnarsson
(1995).
}
\end{minipage}
\hfill
\end{figure}

The LDA antiferromagnetic bandstructure is shown in 
figure \ref{ag1}. The highest
valence state is formed by the majority $e_g$ and minority $t_{2g}$ Ni
states and the lowest conduction band is formed by the minority $e_g$ state. 
{\em GW} calculation for NiO gives a gap of $\sim $5.5 eV 
(Aryasetiawan and Gunnarsson 1995). Starting from the
LDA antiferromagnetic bandstructure with a gap of 0.2 eV, a
{\em GW} calculation shifts the Ni $e_g$ conduction upwards, increasing the
gap to $\sim $1 eV. 
This upward shift of the $e_g$ conduction band leads to a substantial change
of the character of the wavefunctions, reducing the amount of minority
$e_g$ character in the occupied states.
To include a limited self-consistency, we introduce a new non-interacting
Hamiltonian $H^0$ with a non-local potential 
$\Delta |e_g\rangle \langle e_g|$ 
where $\Delta $ is chosen so that the band gap obtained from $H^0$
agrees with the gap obtained from the previous {\em GW} calculation.
This $H^0$ is then used to generate a new $G^0$ and a new self-energy in the
GWA. This procedure is iterated to self-consistency.
The non-local potential modifies
the eigenvalues as well as the wave functions used to construct the zeroth
order Green function $G^0$. The raising of the unoccupied majority $e_g$
band by the self-energy correction reduces the hybridization with the O p
band and has the effects of raising the bottom of the O p band and pushing
down the top of the O p band at the $\Gamma $ -point resulting in better
agreement with photoemission data. In addition, the width of the unoccupied $%
e_g$ band is reduced. The reduction in hybridization also reduces the
magnitude of the exchange interaction of the $e_g$ band with the occupied
states which has the consequence of widening the gap. Thus, it is important
that the wave functions are also modified in the self-consistent procedure.
The final position of the unoccupied $e_g$ band is just below the Ni 4s. As
a check, the calculation has also been performed in the ferromagnetic state.
A gap of $\sim $ 5.2 eV was obtained, close to the antiferromagnetic value.
In contrast to the Slater model, the gap does not depend on the
antiferromagnetic ordering and the results correctly predict that NiO
remains an insulator above the N\'{e}el temperature. The {\em GW }%
calculation clearly improves the LDA gap markedly and it is in reasonable
agreement with the experimental value of 4.0 eV.
An estimate of the magnetic moment yields a value of 1.6 $\mu_B$ (Aryasetiawan
and Gunnarsson 1995) in good agreement
with the experimental value of 1.7-1.9 $\mu_B$.

\noindent
\begin{figure}[bt]
\unitlength1cm
\begin{minipage}[t]{15.0cm}
\rotatebox{0}
{\centerline{\epsfxsize=3.5in \epsffile{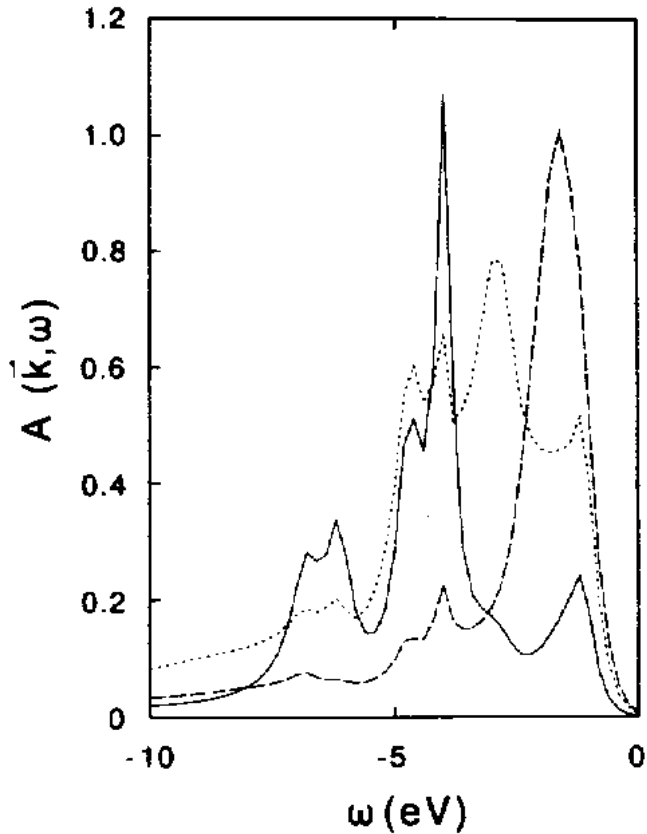}}}
\vskip0.3cm
\caption[]{\label{ag2}
The spectral function of antiferromagnetic NiO at ${\bf k}=
\frac{2}{3}(1\;0\;0)2\pi/a$ within the GWA projected on the O p orbitals
(solid line) and the Ni d orbitals (The dotted line corresponds to the
majority spin and the dashed line the minority spin).
After Aryasetiawan and Gunnarsson (1995).
}
\end{minipage}
\hfill
\end{figure}

To study the character of the states at the top of the valence band, the
spectral function has been calculated. Projection of the spectrum into the
Ni 3d and O p orbitals shows that there is an increase of the O p
character at the top of the valence band but the main character is
primarily Ni 3d. The satellite at -8 eV (Shen {\em et al} 1990, Shen 
{\em et al} 1991a, Shen {\em et al} 1991b) is not reproduced at the final
self-consistent spectrum but a more detailed study of the satellite
structure reveals that it is rather sensitive to the starting Hamiltonian
(Aryasetiawan and Karlsson 1996). Starting from the LDA Hamiltonian in fact
gives a satellite at about -10 eV but this satellite diminishes in intensity
as the gap opens up. The origin of this behaviour can be traced back to the
presence of a plasmon-like peak at low energy which is related to the
incorrect LDA bandstructure. As the gap opens up, this peak structure
becomes broadened and consequently the satellite structure diminishes. As in
the case of Ni, it appears that the satellite structure is due to
short-ranged correlations which are not properly taken into account by the
RPA. T-matrix approach could be appropriate and might remove some Ni d
weight from the top of the valence band to the satellite region but it is
not clear how this could increase the charge transfer from the O p.

\subsubsection{Manganese Oxide}

A calculation on MnO based on a simplified {\em GW }scheme described in
section 4 has been performed by Massidda {\em et al} (1995a). The electronic
structure of MnO is the simplest among the transition metal oxides and in
some respects similar to NiO. As in the case of NiO, the LDA gives a too
small band gap of 1.0 eV compared with the experimental value of 3.8-4.2 eV.
The magnetic moment is also somewhat too small (LDA 4.3 $\mu _{{\rm B}}$,
experiment 4.6-4.8 $\mu _{{\rm B}}$) although the relative discrepancy is
not as large as in NiO. The larger LDA band gap is due to the fact that in
MnO the majority spin is fully occupied and the minority spin is empty
resulting in a large magnetic moment so that the exchange splitting is also
large and dominates the ligand-field splitting and band broadening due to
intersublattice coupling. In NiO the magnetic moment is smaller and the
exchange splitting is comparable to the ligand-field splitting and band
broadening.

The semiempirical model {\em GW} calculation gives a gap of 4.2 eV which
compares well with the experimental value of 3.8-4.2 eV. The LDA magnetic
moment is also improved to 4.52 $\mu _{{\rm B}}$. There is an increase of O
p character and a decrease of Ni 3d character at the top of the valence
band, which are percentagewise large but small in absolute term, so that the
main character is still primarily Ni 3d. The results are qualitatively
similar to the full {\em GW} calculations on NiO described above.
Calculation on CaCuO$_2$ using the model $GW$ scheme was also done recently
(Massidda {\em et al} 1997).

\subsubsection{3d semicore states}

It is well-known that the LDA eigenvalues for localized states are usually too
high compared to experiment. The discrepancies can be several eV. For example,
the Zn semicore 
3d states in ZnSe are too high by 2.5 eV, the Ga semicore 3d states in GaAs
by 4 eV and the Ge 3d semicore states by as much  as 5 eV.
In free atoms these deviations are even larger and in, for instance,
a free Zn atom the 3d eigenvalue is about 6.5 eV too high.
This raises interesting questions about whether or not GWA can describe
these shallow core levels and why the LDA eigenvalues are substantially 
worse for free atoms than for solids, although the semicore states 
are almost completely localized.

For {\it deep} core levels, an important contribution to the 
GWA self-energy comes from the polarization (relaxation) of the
more weakly bound electrons (Hedin and Lundqvist 1969, Lundqvist 1969).
This relaxation is a classical
electrostatic effect, which can be described in $\Delta$SCF calculations,
performing ground-state calculations with and without the core hole. 
Explicit $\Delta$SCF calculations 
for free atoms were performed by Hedin and Johansson (1969), who 
obtained quite accurate results.  This suggests that GWA may 
be rather accurate for such deep core levels, since it includes 
similar physics as the $\Delta$SCF calculations.

\noindent
\begin{figure}[bt]
\unitlength1cm
\begin{minipage}[t]{15.0cm}
\rotatebox{181}
{\centerline{\epsfxsize=3.5in \epsffile{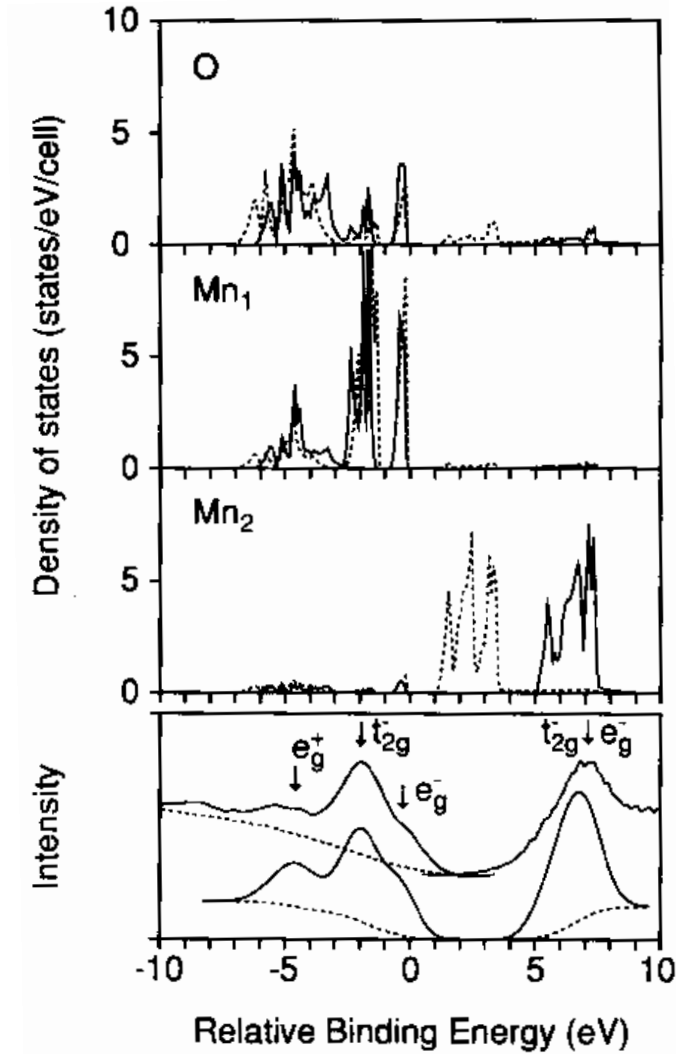}}}
\vskip0.3cm
\caption[]{\label{m2}
Partial densities of spin-up states of antiferromagnetic MnO. Top panel: the
total O 2p projection into the O spheres. Second and third panels: d
projection into the two inequivalent Mn spheres. Solid (dashed) lines
correspond to the model {\em GW} (spin polarized LDA) calculations. 
Bottom panel, upper curves: Inverse photoemission spectrum (from van Elp
{\em et al}
1991) and the difference between the on- and off-resonance photoemission
spectra (from Lad and Henrich 1988), representing the Mn contribution. Lower
curves: {\em GW} d-projected density of states into both Mn spheres.
Dotted lines illustrate the integrated-intensity background.
After Massidda {\em et al}.
}
\end{minipage}
\hfill
\end{figure}

It is not a priori clear that GWA can describe semicore states. Unlike the
deep core states, the {\it shallow} core states are screened by 
states which are almost degenerate with the hole, which may 
introduce new effects. 
Calculations for semicore states in ZnSe, GaAs and Ge (Aryasetiawan and 
Gunnarsson 1996) showed, however, that GWA improves the LDA eigenvalues  
significantly also for these core states, leaving a 
discrepancy of only 0.5-1.0 eV to experiment (too small binding energy).     

The errors in the LDA eigenvalues are to a large extent due the
unphysical interaction of an electron with itself. This interaction
is only incompletely cancelled by the exchange correlation potential
(Gunnarsson, Lundqvist, and Wilkins 1974). This leads to an ionization 
energy which is too small. In addition relaxation effects are neglected,
which tends to overestimate the ionization energy and therefore partly 
cancels the error from the self-interaction.
As discussed above, essential relaxation effects are included in
 $\Delta$SCF calculations. In addition, the LDA total energies are 
rather accurate and the self-interaction errors are rather small.
This is due to an exact sum-rule satisfied by the LDA
(Gunnarsson and Lundqvist, 1976). While this sum-rule
is very important for total energy calculations, 
it does not necessarily imply a good exchange-correlation potential
or  accurate eigenvalues. It  is therefore crucial that $\Delta$SCF 
calculations involve total energy differences rather than eigenvalues. 
 
To understand the different accuracy of the LDA eigenvalues for a 
solid and a free atom, Aryasetiawan and Gunnarsson (1996)
performed transition state calculations in the Slater transition 
state approach (Slater 1974). These calculations showed that
in the solid the creation of a core hole leads to a substantial
charge transfer to the site where the hole was created 
(Lang and Williams 1977, Zunger and Lindefelt 1983).
Due to this charge transfer the relaxation energy is larger than for a free 
atom. The cancellation of the error  in the eigenvalues due to the
self-interaction and the error due to the neglect of relaxation effects 
therefore becomes more complete
in the solid than in a free atom. This explains why the errors in
the eigenvalues are much smaller in the solid  (Aryasetiawan and 
Gunnarsson 1996). 

The  $\Delta$SCF gave ionization energies for the 3d semicore states which were
about 1 eV too large. The discrepancy between  $\Delta$SCF and GWA
may be due to an error in the LDA, which tends to overestimate the 
exchange-correlation energy between a 3d electron 
and the 3s-3p core (Gunnarsson and Jones 1980) which is about 1 eV in
the series K-Cu (Harris and Jones 1978). It could also be due to the
difference in the RPA screening and the static LDA screening.
Transition state calculations for a number of Zn compounds 
have also been performed by Zhang, Wei, and Zunger (1995).
We observe, however, that the Slater transition state approach requires that
the semicore states are sufficiently localized to form a bound state in the
transition state calculations. The bound state has no dispersion, in contrast
to the {\em GW} results which show the full band structure. Thus, if a
bound state is not formed in the transition state calculation, the result
would be identical to that of the LDA. The problem can be illustrated for Cu
metal where the 3d band is 0.5 eV too high. The transition state would give a
single number rather than a band assuming that a bound state is formed in the
first place. A {\em GW} calculation on the other hand lowers the position of
the band by 0.2-0.3 eV while maintaining the LDA band structure (Aryasetiawan
and Gunnarsson 1997).

A model {\em GW} calculation for ZnO by Massidda {\em et al} (1995b)
also improved
the LDA result from -5.4 eV to -6.4 eV but a significant error remains when
compared to experimental results -8.6 eV and -7.5 eV.

\subsection{Surfaces}

\subsubsection{Jellium surface}

{\em GW} calculations for a model jellium-vacuum interface
have been done in some details. The problem of interest here is how the
surface barrier looks like. According to classical electrostatic, the image
potential seen by an electron in the vacuum far from the surface behaves
like 
\begin{equation}\label{vimage} 
V_{image}=-\frac 1{4\left( z-z_0\right) } 
\end{equation} 
where $z$ is the coordinate normal to the surface and $z_0$ is the position
of the effective image plane. In the LDA, the image potential is known to
decay exponentially (Lang and Kohn 1973). 
This unphysical behaviour is due to the fact that in
the LDA the exchange-correlation hole is determined by the local density and it
does not feel the surface directly. 
The exchange-correlation hole obeys a sum-rule that it must integrate to one
but since the density is small outside the surface it means that the hole
becomes very extended. In fact about half the hole resides far inside the
surface. Due to the very extended structure of the hole, the resulting 
$V^{\rm xc}$ decays exponentially outside the surface.
The situation is similar to the atomic case where the LDA $%
V^{\rm xc}$ decays exponentially instead of decaying as $1/r$ as in the exact $%
V^{\rm xc}.$ (Almbladh and von Barth 1985a).
Similarly, the exact DFT exchange-correlation potential should be
capable of reproducing the correct behaviour of the image potential which is
important in many applications. For example, the LDA potential cannot
produce the Rydberg series of image states. Binding energies and life-time
of surface states bound by image potential depend crucially on the $1/z$
dependence as do tunneling currents in the scanning-tunneling microscope and
positions of alkali ions adsorbed on metal substrates. Interpretation of
inverse photoemission data relies on the existence of the image tail of the
surface barrier.

\begin{figure}[bt]
\unitlength1cm
\begin{minipage}[t]{15.0cm}
\rotatebox{180.75}
{\centerline{\epsfxsize=4.5in \epsffile{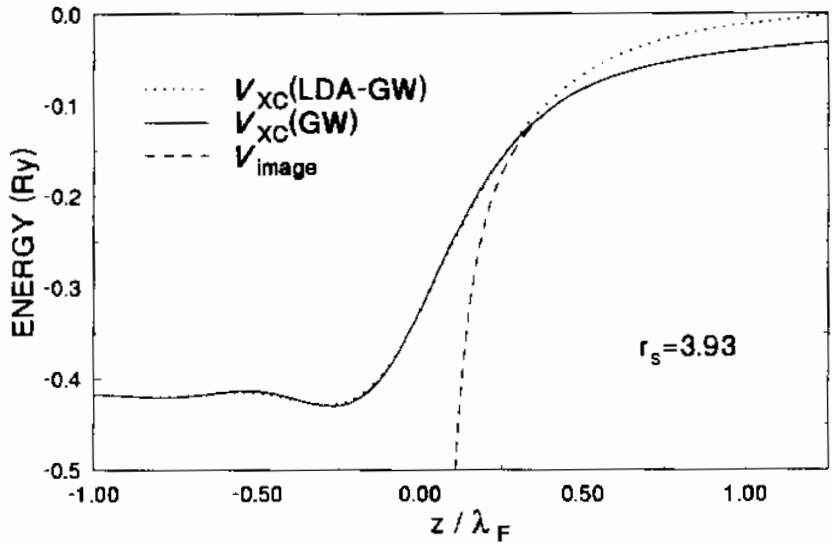}}}
\vskip0.3cm
\caption[]{\label{e1}
$V^{\rm xc}(z)$ at a jellium surface for $r_s=3.93$ ($\lambda_F=12.9$ a.u.).
The solid curve is the solution to equation (\ref{shamschluter}) using the GWA
for $\Sigma$, and the dotted curve is the corresponding LDA potential.
The dashed curve is the image potential $V^{im}(z)=-1/4(z-z_0)$.
After Equiluz {\em et al} (1992).
}
\end{minipage}
\hfill
\end{figure}

By calculating the self-energy for the surface, a $V^{\rm xc}$ can be derived
simply by taking the trace of the following Dyson equation 
(Sham and Schl\"uter 1983,1985, van Leeuwen 1996)
\begin{equation}
G=G^{DF}+G^{DF}\left( \Sigma      -V^{\rm xc}\right) G 
\end{equation} 
and noting that the exact density is equal to the exact DF density (i.e. the
trace of $G$ is equal to the trace of $G^{DF})$. This yields 
\begin{eqnarray}
&&\int d^3r_1V^{\rm xc}\left( {\bf r}_1\right) \int d\omega G^{DF}\left( {\bf r,r%
}_1,\omega \right) G\left( {\bf r}_1,{\bf r,}\omega \right) 
\nonumber \\
&=&\int d^3r_1\int d^3r_2\int d\omega G^{DF}\left( {\bf r,r}_1,\omega
\right) \Sigma      \left( {\bf r}_1,{\bf r}_2,\omega \right) G\left( {\bf r}%
_2,{\bf r},\omega \right)
\label{shamschluter}
\end{eqnarray}
The exchange-correlation potential obtained from the above equation using
the {\em GW} self-energy exhibits the correct asymptotic behaviour as can
be seen in figure \ref{e1}. 
It can also be seen from the figure that it is the
correlation potential originating from the Coulomb correlation that
determines the asymptotic behaviour of the image potential. The exchange
potential, on the other hand, is numerically found to decay as $v^x\sim
-a/z^2$ and to contribute significantly to the determination of $z_0$. The
position of the effective image plane $z_0$ deduced from $V^{\rm xc}$ is
therefore different from the one for a classical test charge. Thus, for $%
r_s=2.07$ the value of $z_0$ deduced from the image tail of $V^{\rm xc}$ is $%
z_0=0.72\pm 0.1$ a.u. measured from the jellium edge while the value
obtained from a linear response calculation is $z_0=1.49$ a.u. 
(Eguiluz {\em et al} 1992).

The Kohn-Sham eigenvalues calculated from the $V^{\rm xc}$ deduced from $\Sigma
     _{GW}$ turn out to very close to the quasiparticle energies obtained
from the Dyson equation (Deisz, Eguiluz, and Hanke 1993). 
The difference $E_{QP}-E_{KS}=0.02$ eV for $%
q_{\parallel }$ $=0$ which is much smaller than the binding energy of the
state - about $0.5$ eV. Furthermore, the Kohn-Sham and quasiparticle
eigenfunctions are practically identical (the overlap is $>0.999$ ). This
does not, however, imply that the physics of the self-energy at the surface
can be completely described by a local potential. The imaginary part of the
self-energy associated with damping is intrinsically non-local and energy
dependent. It cannot be mimicked by a local complex potential.
As the electron moves away from the surface, the
maximum of Im $\Sigma      $ remains at the surface, reflecting a high
degree of non-locality. The range of this non-locality is comparable with
the decay length of the electron density in the vacuum (Deisz, Eguiluz, and
Hanke 1993).
It was also found that Im $\Sigma      $ deviates from the quadratic behaviour
$\simeq (E-E_F)^2$ when the electron is in the vacuum outside the surface and
the departures grow as the electron-surface separation increases. The effect
is attributed to the suppression of one-electron decay channels near the Fermi
level (Deisz and Eguiluz 1997).

\noindent

\subsubsection{Si(1 0 0) surface}

There have also been studies for more detailed microscopic models
of surfaces. In particular, 
the Si(100) surface has been extensively studied experimentally
and theoretically, because of the technological importance of Si.
At room temperature there is a $2\times 1$ reconstruction which
at low temperatures goes over in a $c(4\times 2)$ reconstruction
(see, e.g., Johansson {\it et al.} for references to experiment).
The surface atoms are believed to form buckled dimers, where
one atom moves out of the surface and the other into the surface.
The room temperature photoemission spectrum has been measured
by Johansson {\it et al.} (1990).

\begin{figure}[bt]
\unitlength1cm
\begin{minipage}[t]{15.0cm}
\rotatebox{181}
{\centerline{\epsfxsize=4.5in \epsffile{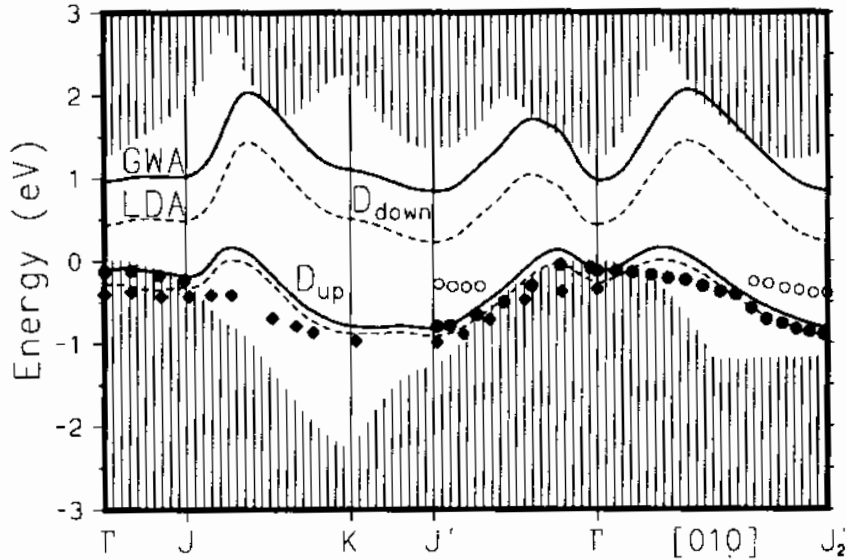}}}
\vskip0.3cm
\caption[]{\label{si}
Calculated dangling-bond bands. Solid lines, GWA energies,
dashed lines LDA energies. The experimental results are   
shown by diamonds (Uhrberg {\it et al} 1981) and circles
(filled and open) (Johansson {\it et al.} 1990).
(After Rohlfing {\it et al.} 1995b).
}
\end{minipage}
\hfill
\end{figure}

The electronic structure has been calculated in the GWA by 
Northrup (1993) ($c(4\times2$)), by Kress, Fiedler, and Bechstedt 1994
($2\times 1$) and by Rohlfing {\it et al.} (1995b)($2\times 1$). 
In figure \ref{si} we show the results of Rohlfing {\it et al.}
(1995b). These calculations were performed using a Gaussian basis 
set and a super cell containing eight Si layers. The figure shows 
the LDA bulk bands (dashed) and  $GW$ surface states. The $GW$ surface band
agrees rather well with an experimental band, while another 
experimental band has no correspondence in the theoretical calculation.
It is interesting that the calculation of Northrup, using the low temperature
structure, produced two bands in close agreement with experiment.
Actually, the experimental samples may have some domains with 
this $c(4\times 2)$ structure (Johansson {\it et. al} 1990).
The $GW$ calculation shifts the conduction band by about +0.50-+0.65 eV
relative to the valence band (Rohlfing {\it et al.} 1995b).
According to optical experiments, the bottom of the valence band
at $\Gamma$ is 1.1 eV above the top of the valence band in good
agreement with the theoretical result 0.95 eV. For the indirect
band gap there are experimental estimates in the range 0.44 eV 
to 0.9 eV compared with the theoretical result 0.7 eV (Rohlfing
{\it et al.} 1995b).

\noindent

In this context we also mention that there has been work
within the GWA for interfaces, e.g., for calculating Schottky
barriers. This work will not be discussed further here, 
but we refer the reader to Charlesworth {\it et al.} (1993)
and Godby and Sham (1994)
for an example of such work and further references.
For other work on semiconductor surfaces we refer to
Hybertsen and Louie (1988b) and Bechstedt and Del Sole (1990).
Rohlfing, Kr\"uger, and Pollmann (1996) performed calculations on
clean, H and S terminated Ge(001) surface and
Zhang {\em et al} (1988)
have calculated the valence band off-set of AlAs-GaAs(001).

\subsection{Clusters}

Electronic excitations and optical spectra in clusters have been studied
mainly within the configuration interaction approach 
(Bonacic-Koutecky {\em et al} 1992, Bonacic-Koutecky, Fantucci, 
and Koutecky 1990a,b). While it gives
accurate results, its applications are limited to systems containing a small
number of electrons, typically less than 10. GWA provides an alternative
for calculating excitation properties in clusters with relatively large number
of electrons which cannot be handled by the CI method.

{\em GW }calculations have been
performed for a jellium-sphere model for alkali metals 
(Saito {\em et al} 1989) and more recently for
the real cluster ${\rm Na}_{{\rm 4}}$ (Onida {\em et al} 1995).
In alkali-metal clusters, it is known
that the ionization energies calculated within the LDA are too low compared
to experiment (Ishii, Ohnishi, and Sugano 1986, Saito and Cohen 1988)
and the discrepancy becomes worse the smaller the cluster. The
size dependence of the ionization energy in the LDA is too weak. This
discrepancy is attributed to self-interaction which is not taken into
account properly in LDA eigenvalues. That the discrepancy gets worse for
smaller clusters is intuitively clear since the larger the clusters the more
they resemble electron gas on which the LDA is based. On the other hand, the
LDA gives the correct sequence of eigenvalues for the valences states in the
jellium-sphere model which is $1s$, $1p,$ $1d$, $2s,$ $1f,$ $2p,$ etc.
(Saito {\em et al} 1989).

In the jellium-sphere model, the ionic charges are smeared to form a sphere
of a uniform positive background. {\em GW} calculation for this model
corresponding to ${\rm Na}_{{\rm 20}}$ lowers the occupied LDA eigenvalues
and increases the unoccupied ones, thus increasing the energy gap similar to
the results for bulk semiconductors and insulators. The ionization energy as
a function of cluster size is also in better agreement with experiment
although the absolute values are somewhat too large
(Saito {\em et al} 1989). This could be due to
the absence of core polarization in the jellium-sphere model which gives a
positive contribution to the self-energy. The $GW$ results are actually
rather close to the HF results, the reason being that screening due to
long-range correlations is much less important in a small finite system than
in a solid (Saito {\em et al} 1989).

{\em GW} calculation for real ${\rm Na}_{{\rm 4}}$ yields similar results
(Onida {\em et al} 1995).
The unoccupied states are raised by between 0.75 and 0.90 eV while the
highest occupied molecular orbital (HOMO) and the lowest occupied molecular
orbital (LUMO) state are lowered by 1.55 and 1.40 eV respectively giving a
HOMO-LUMO gap of 3.0 eV compared with the LDA value of 0.55 eV. Again the $%
GW $ result is close to the HF value of 3.4 eV for the reason discussed
above. Indeed calculation of the static dielectric function for this cluster
shows a metallic behaviour for small distances but it drops to less than
unity starting at about 6 a.u.. For comparison, the value of the dielectric
function for metallic sodium at 5 a.u. is about 50 whereas in the cluster it
is 1.2. This phenomenon of antiscreening is typical of a small system. If a
positive test charge is introduced at the centre of the cluster, say,
electrons will surround the test charge and the surface of the cluster will
therefore be positively charged. Thus as the distance from the centre
increases, the screening quickly vanishes and becomes negative
(Onida {\em et al} 1995). A similar
behaviour is also observed in the ${\rm C}_{{\rm 60}}$ molecules 
(Gunnarsson, Rainer, and Zwicknagl 1992).

\noindent
\begin{figure}[bt]
\unitlength1cm
\begin{minipage}[t]{15.0cm}
\rotatebox{0}
{\centerline{\epsfxsize=4.5in \epsffile{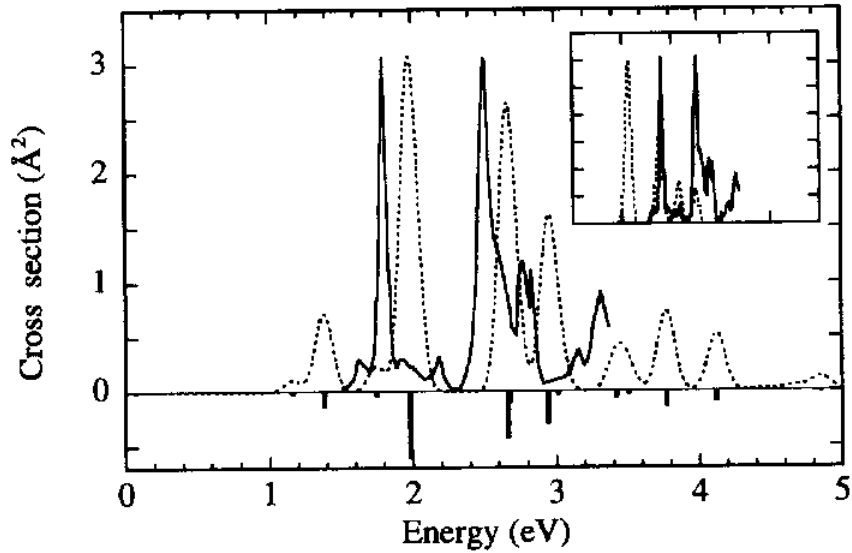}}}
\vskip0.3cm
\caption[]{\label{o2}
Calculated absorption spectra of Na$_4$ including self-energy and excitonic
effects (dotted line) in arbitrary units and using a Gaussian broadening of
0.06 eV. The solid line is the experimental photodepletion cross sections from
Wang {\em et al} (1990).
The vertical bars show the unbroadened spectrum and the inset shows the LDA
results.
After Onida {\em et al} (1995).
}
\end{minipage}
\hfill
\end{figure}

\subsection{Fullerenes}    

\begin{figure}[bt]
\unitlength1cm
\begin{minipage}[t]{15.0cm}
\rotatebox{181}
{\centerline{\epsfxsize=4.5in \epsffile{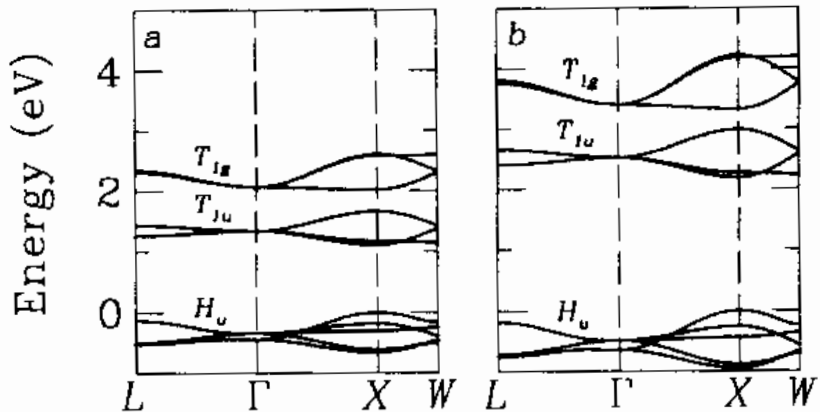}}}
\vskip0.3cm
\caption[]{\label{c60}
LDA (a) and $GW$ (b) band structures for solid C$_{60}$ in the $Fm{\bar 3}$
structure \\
(After Shirley and Louie 1993).
}
\end{minipage}
\hfill
\end{figure}

Solid C$_{60}$ (fullerite) is a molecular solid, where the 
orbitals of a free C$_{60}$ molecule essentially keep their
character and the hopping between the molecules only leads to a 
small broadening of the discrete molecular states.
The band structure of C$_{60}$ is shown in figure \ref{c60}.
In undoped C$_{60}$ the $H_u$ band is occupied and the $T_{1u}$
band is empty. The LDA (left part of the figure) band gap is
about 1 eV (Erwin and Pickett 1991, Troullier and Martins 
1992, Satpathy {\it et al.} 1992), which is substantially 
smaller than the experimental value, 2.3 eV (Lof {\it et al.} 1992), 
obtained from photoemission and inverse photoemission.

\noindent

Shirley and Louie (1993) have performed a $GW$ calculation for solid
C$_{60}$ on a fcc lattice with all molecules having the same
orientation ($Fm{\bar 3}$ structure). This differs somewhat 
from the experimental
$T=0$ structure Pa3, where the molecules take four different 
orientations. This difference should not be important for the 
present discussion. Shirley and Louie (1993) used the Levine and 
Louie (1982) model for the static dielectric function together with 
a plasmon pole approximation. 
Their results are shown in figure \ref{c60}. The band gap is increased 
to 2.15 eV in good agreement with the experimental results.
The band widths were also increased by about 30 $\%$. There are
no reliable experimental results for the dispersional band widths.
It was concluded that undoped solid C$_{60}$ is a standard
band insulator with dispersive bands.

The calculation by Shirley and Louie was performed for the undoped
C$_{60}$ solid. In A$_3$C$_{60}$ (A= K, Rb) the $T_{1u}$ band is
half-filled. The Hubbard $U\sim 1.5$ eV (Lof {\it et al.} 1992), 
describing the interaction between 
two electrons on the same molecule, is large 
compared with the width ($W\sim 0.6$ eV) of the partly filled $T_{1u}$
band. One may therefore expect strong correlation effects for 
these systems. Although it is not clear if the GWA can
describe such strong correlation effects, it is interesting to 
perform such a calculation.

To screen the Hubbard $U$,
Gunnarsson (1997) considered a model dielectric function
\begin{equation}\label{eq:c60.1}
\epsilon({\bf q},\omega)=\epsilon_0-{\omega_{0p}^2\over m^{\ast}\omega^2},
\end{equation}
which describes the coupling to a plasmon at $\omega=\omega_{0p}/\sqrt
{m^{\ast}\epsilon_0}$. This plasmon is due to the oscillations
of the three electrons donated by the alkali atoms to the $T_{1u}$
band. The  model (\ref{eq:c60.1}) is therefore complimentary to 
the calculation
of Shirley and Louie (1993), since this $T_{1u}$ plasmon does not
exist for the undoped system, while the physics considered by 
Shirley and Louie is neglected in Eq. (\ref{eq:c60.1}). 
This model of the dielectric function was combined with a 
tight-binding (TB) model (Satpathy {\it et al.} 1992) for the band structure,
which reproduces the LDA $T_{1u}$ band well.  

First a Hartree-Fock (HF) calculation was performed for this 
model. It was found that in HF the width of the $T_{1u}$
band is increased by about 75 $\%$. Next a $GW$ calculation was performed. 
Including the coupling to the $T_{1u}$ plasmons was found to reduce 
the $T_{1u}$ width to a value 35 $\%$ 
smaller than the original TB width. Actually for a model of this type
one can show under rather general assumptions that the band width is
always reduced if there are no other bands above or below the band 
considered (Gunnarsson 1997). The density of the $T_{1u}$ electrons
is very small and corresponds to the electron gas parameter $r_s\sim 7$.
Thus the density is substantially smaller than for the free-electron
like metals. It is then not too
surprising that the quasi-particle weight $Z\sim 0.4-0.5$ is also
smaller than for these metals.  
This means that much of the spectral weight appears in satellites.
If the narrowing of the $T_{1u}$ band due to the coupling to the $T_{1u}$
plasmon is combined with the broadening found by Shirley and Louie
due to other couplings, the net result is a small change 
of the band width.

\section{Self-consistency}

The set of Hedin's equations 
(\ref{hedin1},\ref{hedin2},\ref{hedin3},\ref{hedin4}),
in the original formulation of the self-energy
expansion in powers of the screened interaction $W$ constitutes a
self-consistent cycle (Hedin 1965a, Hedin and Lundqvist 1969). 
Within the GWA, starting from a (usually)
non-interacting Green function $G_0$ one calculates the polarization
function $P_0=-iG_0G_0$ and the corresponding screened interaction $W_0.$
The self-energy is then obtained from $\Sigma      _0=iG_0W_0.$ In most $GW$
calculations that have been performed so far, 
$\Sigma      _0$ is taken to be the
final self-energy. The interacting Green function $G$ obtained from the
Dyson equation $G=G_0+G_0\Sigma      _0G$ is, however, not necessarily the same
as $G_0.$ To achieve self-consistency, the Green function obtained from the
Dyson equation should be used to form a new polarization function $P=-iGG,$
a new screened interaction $W,$ and a new self-energy $\Sigma      $ which in
turns yields a new Green function through the Dyson equation. This process
is continued until $G$ obtained from the Dyson equation is the same as $G$
used to calculate the self-energy. Self-consistency is evidently an
important issue since it guarantees that the final results are independent
of the starting Green function. Moreover, according to the Baym-Kadanoff
theory (1961), a self-consistent $GW$ scheme 
ensures conservation of particle
number and energy when the system is subjected to an external perturbation. 
Conservation of particle number means that 
the continuity equation 
\begin{equation}
-\partial _tn({\bf r},t)=\nabla {\bf \cdot j(r},t)
\end{equation} 
is satisfied.
Conservation of energy means that the energy change when an external
potential is applied to the system is equal to the work done by the system
against the external potential when calculated using the self-consistent $G$
(Baym and Kadanoff 1961, Baym 1962).
Moreover, self-consistency ensures that
\begin{equation}
N=\frac 1\pi tr\int_{-\infty }^\mu d\omega \,{\rm
\mathop{\rm Im}
\,}G(\omega )
\end{equation}
gives the correct total number of particles.
when $n$ and ${\bf j}$ is obtained from the self-consistent Green function.
The first self-consistent $GW$ calculation was probably by
de Groot, Bobbert, and van Haeringen (1995) for a model quasi-one dimensional
semiconducting wire. The relevance of this model to real solids is, however,
unclear.

Five aspects may be distinguished in relation to self-consistency
(von Barth and Holm 1996):

\begin{enumerate}
\item  Modification of quasiparticle wavefunctions.

\item  Shift of quasiparticle energies.

\item  Modification of quasiparticle weights (Z factors).

\item  Modification of quasiparticle life-times.

\item  Modification of the screening properties of the system.
\end{enumerate}

These aspects were recently studied in detail by von Barth and Holm (1996)
and by Shirley (1996)
for the electron gas. Naturally, the first aspect cannot be
addressed for the electron gas since the quasiparticle wavefunctions remain
plane waves. The results of these studies are

\begin{itemize}
\item  The band width is increased from its non-self-consistent value,
worsening the agreement with experiment.

\item  The weight of the quasiparticles is increased, reducing the weight in
the plasmon satellite.

\item  The quasiparticles are narrowed, increasing their life-time.

\item  The plasmon satellite is broadened and shifted towards the Fermi
level. In fact, it almost disappears at full self-consistency.
\end{itemize}

\noindent
\begin{figure}[bt]
\unitlength1cm
\begin{minipage}[t]{15.0cm}
\rotatebox{179}
{\centerline{\epsfxsize=3.5in \epsffile{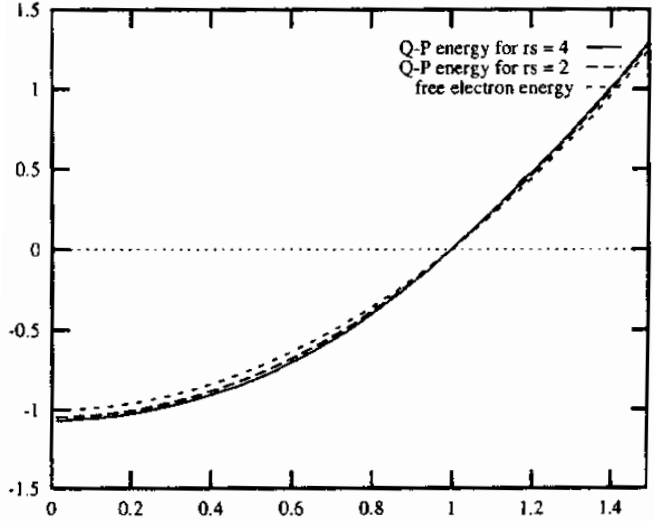}}}
\vskip0.3cm
\caption[]{\label{b1}
The quasiparticle dispersion for the electron gas for $r_s=2$ and $4$
from partial self-consistent $GW_0$ calculations
($W_0$ is held fixed at the RPA whereas $G$ is allowed to vary to
self-consistency)
compared with the free-electron dispersion. The
largest change in the band width occurs for $r_s=4$.
After von Barth and Holm (1996).
}
\end{minipage}
\hfill
\end{figure}

As discussed below, the main effects of self-consistency  are caused by
allowing the quasiparticle weight to vary
(von Barth and Holm 1996). These results are true
for the case when the screened interaction $W$ is fixed at the RPA level
(calculated using the non-interacting $G_0$) and only the Green function is
allowed to vary to self-consistency and also for 
the case when both $G$ and $W$ are
allowed to vary (full self-consistency case). The increase in the
band width is
disturbing and can be understood as follows. Let us consider the first case
with fixed $W=W_0$ for simplicity. First we note that the $GW$ result
for the band width
after one iteration is close to the free electron one. This means that
there is almost a complete cancellation between exchange and correlation.
After one iteration the quasiparticle
weight is reduced to typically 0.7 and the rest of the weight goes to the
plasmon satellite. Correspondingly, when the new $G$ is used to calculate a
new {\rm Im }$\Sigma^c ,$ its weight is transferred away from low energy to
high energy, due to the sum rule (von Barth and Holm 1996)
\begin{equation}
\int_{-\infty }^\infty d\omega |\mbox{Im} 
\Sigma^c \left( {\bf k},\omega \right)| =\sum_{%
{\bf q}}\int_0^\infty d\omega |\mbox{Im} W_0\left( {\bf q},\omega \right)| 
\end{equation} 
which shows that the left-hand-side is a constant depending only on the prechosen $W_0$
but independent of ${\bf k}$ and self-consistency. For a state at the Fermi
level, this has little effect since ${\rm 
\mathop{\rm Im}
\,}\Sigma^c $ has almost equal weights for the hole $(\omega \leq \mu )$ and
the particle part $(\omega >\mu )$ which cancel each other when calculating $%
{\rm 
\mathop{\rm Re}
\,}\Sigma ^c,$ as can be seen in equation (\ref{realsig}).
But for the state at the bottom
of the valence band, ${\rm 
\mathop{\rm Im}
\,}\Sigma^c $ has most of its weight in the hole part so that the shifting of
the weight in ${\rm 
\mathop{\rm Im}
\,}\Sigma^c $ to higher energy causes ${\rm 
\mathop{\rm Re}
\,}\Sigma ^c$ to be less positive than its non-self-consistent value. 
A similar effect is found for the exchange part which becomes less negative
but because the bare Coulomb 
interaction has no frequency dependence, 
the renormalization factor
has a smaller effect on $\Sigma^x$ so that the reduction in $\Sigma^x$ 
is less than the reduction in $\Sigma^c$.
The net effect is then an increase in the band width.
The shifting of the weight in ${\rm 
\mathop{\rm Im}
}\Sigma^c $ to higher energy has immediate consequences of increasing the
life-time and the renormalization weight Z (through a decrease in 
\mbox{$\vert$}
$\partial {\rm 
\mathop{\rm Re}
}\Sigma ^c/\partial \omega |)$ of the quasiparticles and of broadening the
plasmon satellite, compared to the results of one iteration
(von Barth and Holm 1996).

When $W$ is allowed to vary (full self-consistency) the results become even
worse: the band width becomes even wider and the plasmon satellite
becomes broad and featureless, in contradiction to experiment. The
quasiparticle weight is increased further. These results can be explained by
the disappearance of a well-defined plasmon excitation in $W.$ The quantity $%
P=-iGG$ no longer has a physical meaning of a response function, rather it
is an auxiliary quantity needed to construct $W.$ Indeed, it does not
satisfy the usual $f-$ sum rule. The equations 
${\rm Re}\epsilon ({\bf q},\omega _p)={\rm Im}\epsilon ({\bf q},\omega _p)=0$ 
determining the plasmon energy are not satisfied any more since the
Green function now always has weight around $\omega =\omega _p.$ This has
the effect of transferring even more weight in ${\rm 
\mathop{\rm Im}
\,}\Sigma^c $ to higher energy with the consequences discussed in
the previous paragraph.
Shirley (1996) included the second order self-energy diagram (vertex
correction)
and found that it cancelled the effects of self-consistency to some degree.

\noindent
\begin{figure}[bt]
\unitlength1cm
\begin{minipage}[t]{15.0cm}
\rotatebox{181}
{\centerline{\epsfxsize=3.5in \epsffile{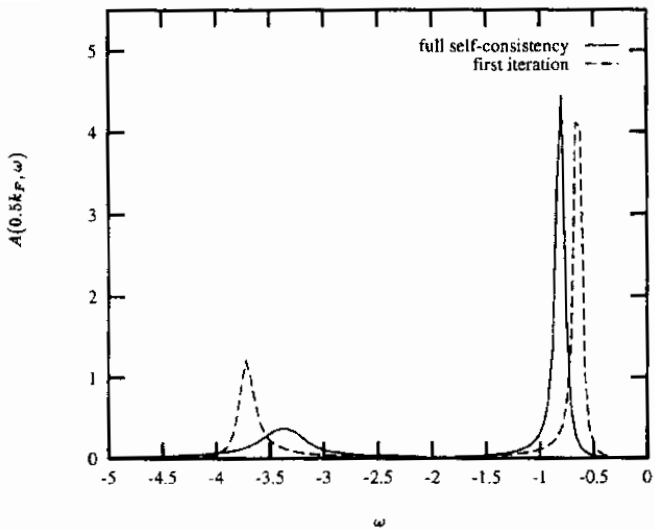}}}
\vskip0.3cm
\caption[]{\label{b10}
The spectral function from a partial self-consistent $GW_0$ calculation
(see previous figure) compared to that of the first iteration for $k=0.5 k_F$ 
and $r_s=4$. The self-consistent quasiparticle energy is lowered (band
broadening) compared to the non-self-consistent one whereas the satellite
position is somewhat improved. After von Barth and Holm (1996).
}
\end{minipage}
\hfill
\end{figure}

A very interesting outcome of the full self-consistent calculation 
(Holm and von Barth 1997) is that
the total energy calculated with the Galitskii-Migdal formula (1958)
turns out to
be strikingly close to the total energy calculated from a much more
elaborate quantum Monte Carlo technique (Ceperley and Alder 1980). 
For $r_s=$ 2 and 4 quantum Monte-Carlo gives 0.004
and -0.155 Rydberg respectively while the self-consistent $GW$ gives 0.005 and
-0.156 Rydberg. This unexpected result may be related to the fact that
the self-consistent $GW$ scheme is energy conserving and it is partly explained
by consideration of the Luttinger-Ward energy functional (1960)
which is
variational with respect to $G$ and its minimum is equal to the value
obtained from the Galitskii-Migdal formula. What is not clear is why the
first order energy diagram (giving the $GW$ self-energy upon taking
functional derivative with respect to $G)$ appears to represent a very good
energy functional. Furthermore, the chemical potential calculated from $\mu
=\partial E/\partial N$ is in agreement with the value obtained from $\mu
=k_F^2/2+\Sigma      (k_F,\mu)$ and the particle density $n=2\sum_{{\bf k}%
}\int_{-\infty }^\mu d\omega A({\bf k},\omega )$ yields $n=k_F^3/(3\pi ^2) $
, i.e. particle number is conserved, as proven by Baym (1962). 
It has also been
proven that with a fixed $W=W_0$ particle number is also conserved
(Holm and von Barth 1997).

The conclusion is that it is not a good idea to perform fully
self-consistent $GW$
calculations for quasiparticle energies. 
If full self-consistency is not introduced, an important question is how to
choose $H_0$ determining $G_0$. Farid (1997) argued that $H_0$ should be
chosen so that its ground-state density is the same as the one resulting from
the $GW$ calculation.
It could be more favourable
to perform partially self-consistent calculations by, for instance, fixing
$W$ at the RPA level and modifying
the quasiparticle energies in $G_0$, or
by choosing a single-particle
Hamiltonian such that the resulting quasiparticle energies
obtained from a $GW$ calculation are the same as the those of the 
single-particle Hamiltonian. 
In any case, efforts should be directed towards
finding vertex corrections (beyond $GW$ ).

\section{Vertex corrections}

By vertex corrections we mean corrections to the self-energy beyond the
$GW$-RPA approximation and corrections to the response function beyond the
RPA. Vertex corrections to the RPA response function naturally involve
interaction between screening electrons not taken into account in the RPA.
This interaction must include the effect of exchange and correlation, which
makes the interaction depend upon the relative spin states of the electrons.
Vertex corrections to the self-energy on the other hand involve interaction
of the hole with its surroundings not taken into account in the GWA. A
consistent treatment should include vertex corrections in both the
response function and the self-energy (Ward 1950). 
This preserves conservation laws and
is a natural outcome of the perturbation expansion of the self-energy in the
functional derivative technique of Baym and Kadanoff (1961) and Baym (1962).
For example, calculations of the optical spectra of Si indicate a cancellation
between vertex corrections in the response function and the renormalization of
the quasiparticle energies (Bechstedt {\em et al} 1997, Del Sole and Girlanda
1996). 
Mahan and Sernelius (1989) found a similar effect in the calculations of
band widths of the electron gas.
Conserving approximations including vertex corrections
with exchange effects only have been considered by
Hong and Mahan (1994) for the electron gas.

In the GWA, the screened interaction is calculated within the RPA which
takes into account primarily long-range correlation which gives rise to
collective excitations (plasmons). The photoemission hole is coupled to one
plasmon only in the GWA. Vertex corrections may be loosely divided into the
short-range and long-range parts. Short-range vertex corrections improve the
description of the quasiparticles and low-energy satellites whereas
long-range vertex corrections improve description of high-energy satellites
(plasmons).

The RPA pair distribution function, which is the probability of finding another
electron at a certain distance from a given electron, is negative for small
distances, which is unphysical (Pines 1961). 
Exchange and correlations should therefore
increase the probability of finding another electron at small distances. 
The same applies to holes.
For a test charge, the screening becomes more effective when the effects of
exchange and correlations are taken into account. But for an electron, the
RPA screening is {\em reduced} when the effects of exchange and correlations
between the screened electron and the surrounding electrons are taken into
account.
The physical interpretation of vertex corrections is as follows: an
electron pushes away other electrons in its vicinity
creating a screening hole around the electron.
Taking into account exchange and correlations of the screening holes
increases the screening since the probability of the holes getting closer
together is increased leading to more screening holes.
If we now take into account the effects of exchange and correlations
between the screened electron and the other electrons, then
screening will be reduced because the probability of finding electrons 
at small distances to the screened electron is increased, leading to
stronger effective interactions between the electrons. The net effect is
that the RPA screening is reduced.
Thus, vertex corrections for electrons will in
general reduce screening. Vertex corrections in the response function can
also take the form of interaction between electrons and holes in
electron-hole pairs created by the perturbation in the system due to the
presence of a photoemission hole. This interaction may actually create an
additional bound state (exciton) much lower in energy, of the order of a few
eV, than the plasmon energy.

Vertex corrections to the self-energy are particularly important for systems
with localized states, such as those containing 3d and 4f 
orbitals. This is because the electron-hole pairs created in the
screening process can be rather localized and therefore
interact strongly with
the localized photoemission hole. This interaction can significantly modify
the quasihole energy and its weight as well as creating new collective
excitations appearing as low energy satellites. This type of vertex
corrections is short range in nature. Long-range vertex corrections modify
the structure of the plasmon satellites and may create multiple plasmons as
observed in the alkalis. In the GWA, the photoemission hole only couples to
one plasmon, resulting in one-plasmon satellite. The hole, however, can in
general interact with several plasmons during its propagation, producing
multiple plasmon satellites.

\subsection{Direct evaluation of the second-order self-energy}

Formally, the GWA is the first-order term in the expansion of the
self-energy in the screened interaction {\em W}. It is then thought that the
simplest vertex correction is the second-order self-energy. This procedure,
however, warrants some precautions. First, the physical meaning of the
second-order term is not clear. Second, this second-order term when
evaluated with a frequency-dependent interaction can give a self-energy with
wrong analytic properties which result in a negative density of states, as
shown in the electron-gas case (Minnhagen 1974).

This type of vertex correction was calculated for the band gap of Si both
with a bare and a screened interaction (Daling and van Haeringen 1989,
Daling {\em et al} 1991, Bobbert and van Haeringen 1994). 
The second-order self-energy with
the bare Coulomb interaction was found to be factors 18 and 37 smaller than
the first-order self-energy for $\Gamma _{15}$ and $\Gamma _{25}^{^{\prime
}} $ states, which correspond to a correction of $\sim$1 \% with
respect to the Hartree-Fock direct gap. A calculation using a screened
interaction yields a correction to the gap of the order of $\sim$4 \%.
Since this calculation was performed for the gap states only, it is not
clear if the density of states became negative for some energies. Both
calculation indicate in any case that the second-order term is small. This
is encouraging since it suggests that higher-order terms are probably small
too since the GWA already gives results in agreement with experiment. An
interesting result is that the second-order vertex correction does not shift
the absolute position of the LDA valence-band maximum which in the
approximation used is too low by 0.5 eV. It is speculated that higher order
vertex corrections could account for the required shift
(Bobbert and van Haeringen 1994).

\subsection{Vertex corrections based on the LDA exchange-correlation
potential}

The set of equations (\ref{hedin1},\ref{hedin2},\ref{hedin3},\ref{hedin4})
derived by Hedin provides a systematic way of
developing a perturbation series for the self-energy in powers of the
screened interaction {\em W}. In the original derivation, the zeroth-order
Green function is taken to be the Hartree one. However,
the band structure and wavefunctions in the Hartree theory 
are less accurate compared with those
of the LDA, making Hartree Green function a poor starting point. In
practical calculations, one uses a Green function constructed from the LDA
band structure. The response function is also calculated using the LDA Green
function. 

Consistent vertex corrections 
were derived by Rice (1965)
in relation to the dielectric function as discussed in the alkali
section. A similar approach was also made Ting, Lee, and Quinn (1975) and Mahan
(1994).
Alternatively, consistent vertex corrections can also be derived from the
the set of equations (\ref{hedin1}, \ref{hedin2}, \ref{hedin3},
\ref{hedin4}) where one regards $V^{\rm xc}_{\rm LDA}$ as a
self-energy correction to the Hartree approximation, albeit an {\em ad hoc}
one (not based on a systematic diagrammatic
expansion). 
Based on this starting point, the vertex function $\Lambda $ can
be easily evaluated yielding (Hybertsen and Louie 1986,
Del Sole, Reining, and Godby 1994)
\begin{equation}
\Lambda (1,2,3)=\delta (1,2)\delta \left( 1,3\right) -i\int d\left(
5,6,7\right) K^{\rm xc}\left( 1,5\right) G\left( 5,6\right) G\left( 7,5\right)
\Lambda \left( 6,7,3\right)   \label{vertexlda}
\end{equation}
where 
\begin{equation}
K^{\rm xc}\left( 1,5\right) =\frac{\delta V^{\rm xc}\left( 1\right) }{\delta \rho
\left( 5\right) }  \label{kxc}
\end{equation}
remembering that $\rho \left( 1\right) =-iG\left( 1,1^{+}\right) $. The new
self-energy with vertex corrections has the form of the GWA but with a new
screened potential 
\begin{equation}
\widetilde{W}=v\left[ 1-P^0\left( v+K^{\rm xc}\right) \right] ^{-1}
\label{effectivew}
\end{equation}
corresponding to a dielectric function 
\begin{equation}
\widetilde{\epsilon }=1-P^0\left( v+K^{\rm xc}\right)   \label{effeps}
\end{equation}
The same result was already discussed in section V.A where the self-energy
was expressed as a corrected dielectric function $\epsilon$ 
(equation (\ref{alkali1})) and a vertex correction $\Gamma$ 
(equation (\ref{alkali2})). The product $\epsilon^{-1}\Gamma$ is equal to
$\tilde\epsilon ^{-1}$ with $\tilde\epsilon$ given above. 
This dielectric function may also be derived straightforwardly from
time-dependent LDA and may therefore be interpreted as the dielectric
function that screens the external potential felt by an electron, as opposed
to a test charge
(Hybertsen and Louie 1986). 
This distinguishes itself from the RPA (time-dependent
Hartree) dielectric function in that the induced charge does not only
generate the Hartree potential but also an exchange-correlation potential.
It is clear that one could start with a different local zeroth-order
self-energy other than $V^{\rm xc}_{\rm LDA}$ and arrive at a similar formula. Note
that in the formula for the self-energy in equation (\ref{hedin1}), 
the vertex function $\Lambda 
$ enters both {\em W} (through {\em P}) and $\Sigma$. A problem with
starting with a local zeroth-order self-energy is that the self-energy with
the vertex corrections becomes asymmetric in 
${\bf r}$ and ${\bf r}^{\prime } $.

Application of such scheme to the electron gas gives small changes compared
with the original GWA (Mahan and Sernelius 1989).
Similarly, for Si it yields practically the same gap (0.70 eV)
and valence band width (11.4 eV) as those of the standard {\em GW}
calculation
(Del Sole, Reining, and Godby 1994). 
However, the absolute position of the top of the valence band
is shifted 10 meV upwards by the {\em GW}+vertex corrections and 400 meV
downwards by the standard {\em GW} calculation. This could improve the
calculations of band off-sets at interfaces. A calculation with the vertex
corrections in the response function {\em R} only but not in $\Sigma      $, that
is with 
\begin{equation}
W=v\left[ 1+v\left[ 1-P^0\left( v+K^{\rm xc}\right) \right] ^{-1}P^0\right] 
\end{equation} 
\begin{equation}
R=\left[ 1-P^0\left( v+K^{\rm xc}\right) \right] ^{-1}P^0,
\end{equation} 
gives a smaller gap (0.57 eV) and a smaller band width (10.9 eV)
(Del Sole, Reining, and Godby 1994). The latter
result is similar to the case of the
alkalis (Northrup, Hybertsen, and Louie 1987, Surh, Northrup, and 
Louie 1988). The top of
the valence band is shifted downward by 0.4 eV like in the standard GWA. The
result for the band width should be taken with caution because the
plasmon-pole approximation used in the calculations. The worse results
obtained by including vertex corrections in {\em R} only, is consistent with
the violation of conservation laws.

\subsection{The cumulant expansion}

A diagrammatic approach for including vertex corrections is provided by
the cumulant expansion method.
One of the first applications of the cumulant expansion method was in
studying the X-ray spectra of core-electrons in metals (Nozi\`{e}res and de
Dominicis 1969, Langreth 1970). 
Later on the method was extended to valence states by Bergersen, Kus,
and Blomberg (1973)
and Hedin (1980).
The core-electron problem is modelled by a Hamiltonian 
consisting of a core electron interacting with 
a set of plasmons: 
\begin{eqnarray}
H &=&\varepsilon c^{\dagger }c+
\sum_{{\bf q}}\omega
_{{\bf q}}b_{{\bf q}}^{\dagger }b_{{\bf q}}
+
\sum_{{\bf q}
}cc^{\dagger }g_{{\bf q}}\left( b_{{\bf q}}+b_{{\bf q}}^{\dagger }\right)
\end{eqnarray}
where $c$ is the annihilation operator for the core electron with energy
$\varepsilon$,
$b_{{\bf q}}^{\dagger }$ is the creation operator for a
plasmon of wave vector ${\bf q}$ and energy $\omega _{{\bf q}}.$
and the last term is the coupling of the core electron to the plasmon field.
The Hamiltonian can be solved exactly and it can be shown
that the cumulant expansion also gives the exact solution
(Langreth 1970):
\begin{equation}
A_{\pm }\left( \omega \right) =\sum_{n=0}^\infty \frac{e^{-a}a^n}{n!}\delta
\left( \omega -\varepsilon -\Delta \varepsilon \mp n\omega _p\right) 
\end{equation}
where $+$ refers to absorption spectrum and $-$ to emission spectrum. $%
a=\sum_{{\bf q}}g_{{\bf q}}^2/\omega _p^2$ and $\Delta \varepsilon =a\omega
_p$ is the shift in core energy due to the interaction with the plasmon
field. 
It is
assumed that the plasmon excitations have no dispersion although this
assumption is not necessary.
The spectra consist therefore of the main quasiparticle peak at $%
\omega =\varepsilon +\Delta\varepsilon $ 
and a series of plasmon excitations at
multiples of the plasmon energy below the quasiparticle peak which is in
accordance with experiment. This is in contrast to the {\em GW} spectra which
only has one plasmon excitation located at too high energy, typically 1.5
$\omega_p$ below the quasiparticle peak 
(Hedin, Lundqvist, and Lundqvist 1970).
More recently the cumulant expansion method was applied to a model
Hamiltonian with electron-boson interaction and the cumulant was calculated
to higher order (Gunnarsson, Meden, and Sch\"onhammer 1994).

In the cumulant expansion approach, the Green function for the hole $\left(
t<0\right) $ is written as (Langreth 1970, Bergersen, Kus, and Blomberg
1973, Hedin 1980)
\begin{eqnarray}
G\left( k,t<0\right) 
&=&
i\theta \left( -t\right)\langle N|\hat c^\dagger_k(0)\hat c_k(t)|N\rangle
\nonumber\\
&=&i\theta \left( -t\right) {\rm e}^{-i\varepsilon
_kt+C^h\left( k,t\right) } 
\end{eqnarray} 
and the hole spectral function is 
\begin{eqnarray}
A\left( k,\omega \leq \mu \right) 
&=&\frac 1\pi \mbox{Im}G\left( k,\omega \leq \mu \right)  
\nonumber \\
&=&\frac{1}{2\pi}\int_{-\infty}^\infty dt e^{i\omega t} 
\langle N|\hat c^\dagger_k(0)\hat c_k(t)|N\rangle
\nonumber \\
&=&\frac 1\pi \mbox{Im}\,i\int_{-\infty }^0dt\,{\rm e}^{i\omega t}{\rm e}%
^{-i\varepsilon _kt+C^h\left( k,t\right) }  \label{Akw}
\end{eqnarray}
where $k$ denotes all possible quantum labels and $C^h\left( k,t\right) $ is
defined to be the cumulant. 
Expanding the exponential in powers of the cumulant
we get 
\begin{equation}
G\left( k,t\right) =G_0\left( k,t\right) \left[ 1+C^h\left( k,t\right)
+\frac 12[C^h\left( k,t\right) ]^2+\ldots \right] 
\end{equation} 
where $G_0\left( k,t\right) =i\exp \left( -i\varepsilon _kt\right) .$ In
terms of the self-energy, the Green function for the hole can be expanded as 
\begin{equation}
G=G_0+G_0\Sigma G_0+G_0\Sigma G_0\Sigma G_0+\ldots 
\end{equation} 
To lowest order in the screened interaction $W,$ the cumulant is obtained by
equating 
\begin{equation}
G_0C^h=G_0\Sigma G_0 
\end{equation} 
where $\Sigma =\Sigma _{GW}=iG_0W.$ If $G_0$ corresponds to, 
e.g. $G_{\rm LDA} $,
then $\Sigma =\Sigma      _{GW}-V^{\rm xc}.$ 
The first-order cumulant is therefore (Hedin 1980,
Almbladh and Hedin 1983, Aryasetiawan, Hedin, and Karlsson 1996)
\begin{eqnarray}
\label{cumulant}
C^h\left( k,t\right) &=&i\int_t^\infty dt^{\prime }\int_{t^{\prime }}^\infty
d\tau \;{\rm e}^{i\varepsilon _k\tau }\Sigma \left( k,\tau \right) 
\end{eqnarray}
The cumulant may be conveniently divided into a quasiparticle part and a
satellite part: $C^h=C^h_{QP}+C^h_S$
where 
\begin{eqnarray}
C^h_{QP}(k,t)&=&(i\alpha_k+\gamma_k)+(-i\Delta \varepsilon _kt+\eta _k)t
\\
C^h_S(k,t)&=&
\int_{-\infty }^\mu d\omega \frac{{\rm e}^{i\left( \varepsilon _k-\omega
-i\delta \right) t}}{\left( \varepsilon _k-\omega -i\delta \right) ^2}\Gamma
\left( k,\omega \right)
\end{eqnarray}
with 
\begin{equation}
i\alpha_k+\gamma_k=
\left. \frac{\partial \Sigma \left(
k,\omega \right) }{\partial \omega }\right| _{\omega =\varepsilon
_k}
\end{equation} 
\begin{equation}
\Delta \varepsilon _k={\rm P}\int_{-\infty }^\infty d\omega \frac{\Gamma
\left( k,\omega \right) }{\varepsilon _k-\omega }\,,\;\;\;\eta _k=\pi
\Gamma \left( k,\varepsilon _k\right) 
=|\rm{Im}\Sigma(k,\varepsilon _k)|
\end{equation} 
$\eta _k$ is the inverse life-time of the quasiparticle
and $\Gamma(k,\omega)$ is the spectral function of the self-energy which is
proportional to $\rm{Im}\Sigma(k,\omega)$.
A similar derivation can be carried out for the particle Green function 
\begin{equation}
G\left( k,t>0\right) =-i\theta \left( t\right) {\rm e}^{-i\varepsilon
_kt+C^p\left( k,t\right) } 
\end{equation} 
The result is 
\begin{equation}
C^p\left( k,t\right) =-i\Delta \varepsilon _kt-\eta _kt+\left. \frac{%
\partial \Sigma \left( k,\omega \right) }{\partial \omega }\right|
_{\omega =\varepsilon _k}+\int_\mu ^\infty d\omega \frac{{\rm e}^{i\left(
\varepsilon _k-\omega +i\delta \right) t}}{\left( \varepsilon _k-\omega
+i\delta \right) ^2}\Gamma \left( k,\omega \right) 
\end{equation} 

It is physically appealing to extract the quasiparticle part from the Green
function: 
\begin{equation}
G_{QP}^h\left( k,t\right) =i\theta \left( -t\right) 
{\rm e}^{i\alpha_k+\gamma_k}
{\rm e}^{(-iE_k+\eta _k)t}\,,\;\;\;E_k=\varepsilon _k+\Delta
\varepsilon _k 
\end{equation}
The spectral function for this quasiparticle can be calculated analytically
(Almbladh and Hedin 1983):
\begin{equation}
A_{QP}\left( k,\omega <\mu \right) =\frac{{\rm e}^{-\gamma _k}}\pi \frac{%
\eta _k\cos \alpha _k-\left( \omega -E_k\right) \sin \alpha _k}{\left(
\omega -E_k\right) ^2+\eta _k^2}  \label{AQP}
\end{equation}
From equation (\ref{Akw}) we have
$\langle N|\hat c^\dagger_k(0)\hat c_k(t)|N\rangle
={\rm e}
^{-i\varepsilon _kt+C^h\left( k,t\right) } 
$ for $t<0$.
By analytically continuing to $t>0$ 
the spectral
function in equation (\ref{Akw}) can be rewritten as 
\begin{equation}
A\left( k,w\right) =\frac 1{2\pi }\int_{-\infty }^\infty dt\,{\rm e}%
^{i\omega t}{\rm e}^{-i\varepsilon _kt+C^h\left( k,t\right) } 
\end{equation}
where for $t>0$ we have $C^h(k,t)=C^{h*}(k,-t)$.
The total spectra can be written as a sum of $A_{QP}$ and a convolution
between the quasiparticle and the satellite part: 
\begin{eqnarray}
A\left( k,\omega \right) &=&A_{QP}\left( k,\omega \right) +\frac 1{2\pi
}\int_{-\infty }^\infty dt\,{\rm e}^{i\omega t}{\rm e}^{(-iE_k+\eta
_k+C^h\left( k,0\right) )t}\left[ {\rm e}^{C_S^h\left( k,t\right) }-1\right]
\nonumber \\
&=&A_{QP}\left( k,\omega \right) +A_{QP}\left( k,\omega \right) *A_S\left(
k,\omega \right)  \label{convolution}
\end{eqnarray}
where 
\begin{eqnarray}
A_S\left( k,\omega \right) &=&\frac 1{2\pi }\int dt\,{\rm e}^{i\omega
t}\left\{ {\rm e}^{C_S^h\left( k,t\right) }-1\right\}  \nonumber \\
&=&\frac 1{2\pi }\int dt\,{\rm e}^{i\omega t}\left\{ C_S^h\left( k,t\right)
+\frac 1{2!}\left[ C_S^h\left( k,t\right) \right] ^2+\ldots \right\}
\label{cumulants}
\end{eqnarray}
The second term $A_{QP}*A_S$ is responsible for the satellite structure. The
Fourier transform of $C_S^h$ can be done analytically 
(Aryasetiawan, Hedin, and Karlsson 1996)
\begin{eqnarray}
C^h_S\left( k,\omega <0\right) 
&=&\frac{\Gamma \left( k,\varepsilon _k+\omega \right) -\Gamma \left(
k,\varepsilon _k\right) -\omega \Gamma ^{\prime }\left( k,\varepsilon
_k\right) }{\omega ^2}
\end{eqnarray}

As follows from equations (\ref{AQP}) and (\ref{convolution}), 
the quasiparticle
energy in the cumulant expansion is essentially determined by $E_k,$ which
is the quasiparticle energy in the GWA.

 \noindent
 \begin{figure}[h]
 \unitlength1cm
 \begin{minipage}[t]{8.5cm}
 \centerline{\epsfxsize=3.in \epsffile{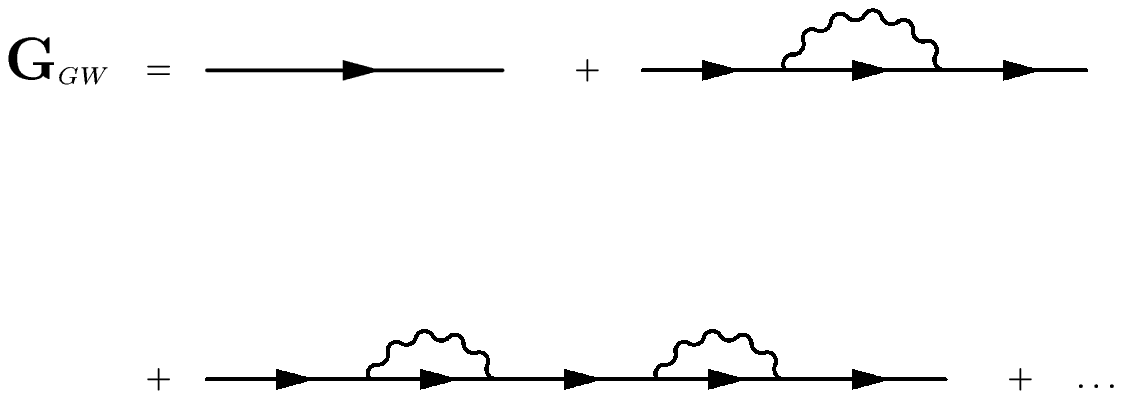}}
 \centerline{\epsfxsize=3.in \epsffile{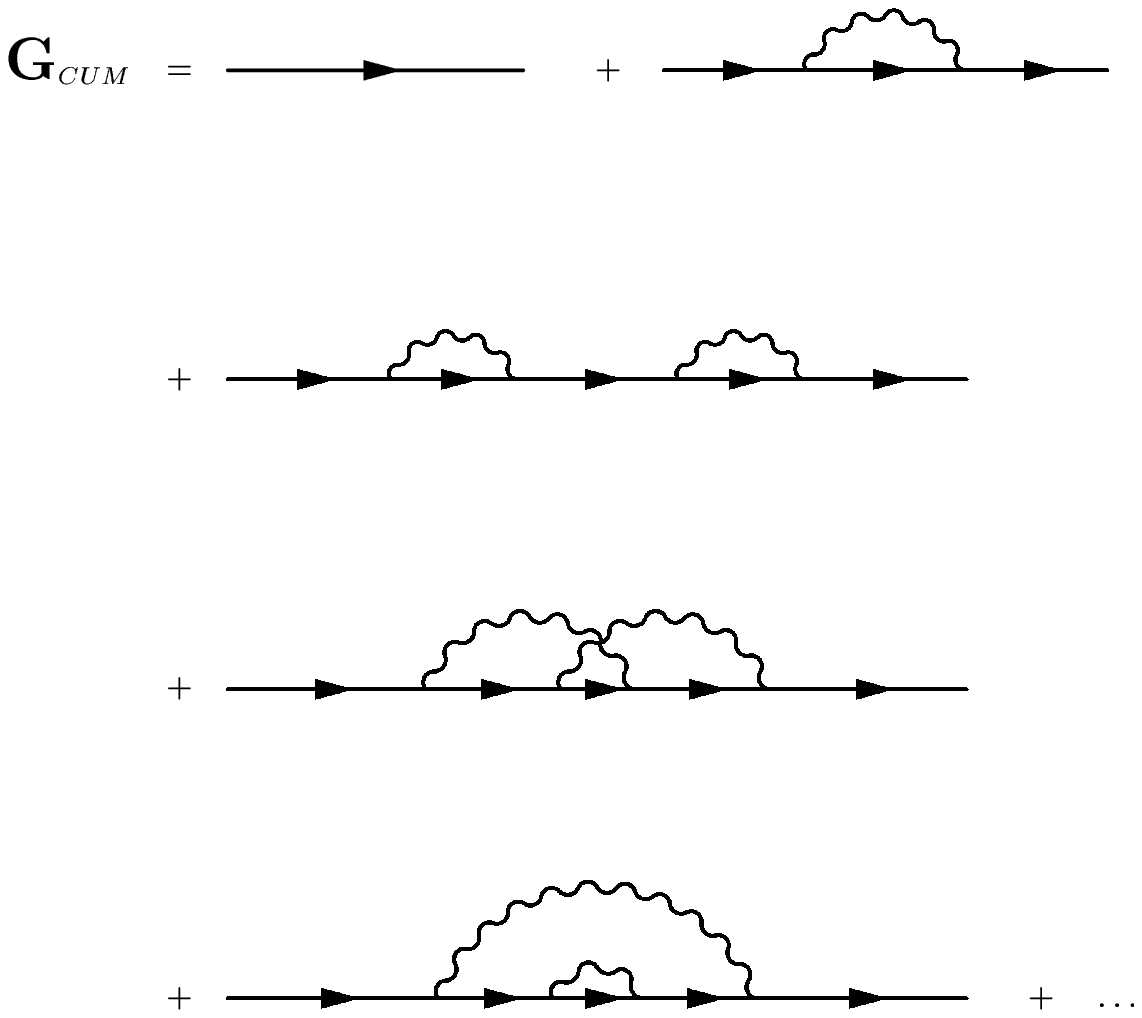}}
 \vspace{4cm}
 \caption{Diagrammatic expansion for the Green function to second order
 in the GWA and the cumulant expansion respectively. The solid lines
 represent non-interacting Green functions $G_0$, and the wiggly lines
 represent the screened interaction $W$.
 }
 \label{fig:diag}
 \end{minipage}
 \hfill
 \end{figure}

\noindent
\begin{figure}[bt]
\unitlength1cm
\begin{minipage}[t]{15.0cm}
\rotatebox{180}
{\centerline{\epsfxsize=3.in \epsffile{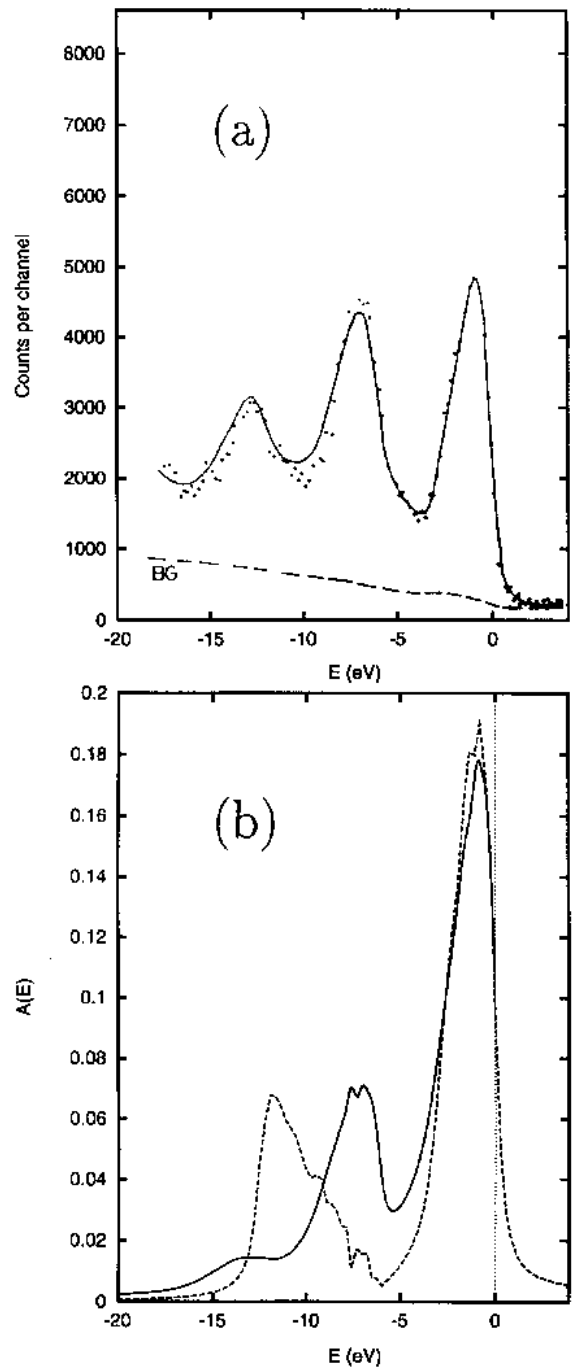}}}
\vskip0.3cm
\caption[]{\label{a1ab}
(a) The experimental spectral function of Na (dots).
The solid line is a synthetic spectrum obtained by convoluting the density of
states from a band structure calculation and the experimental core level
spectrum. BG is the estimated background contribution.
From Steiner, H\"ochst, and H\"ufner (1979).
(b) The calculated total spectral function of Na for the occupied states.
The solid and dashed lines correspond to the cumulant expansion
and the GWA, respectively. After Aryasetiawan, Hedin, and Karlsson (1996).
}
\end{minipage}
\hfill
\end{figure}

By comparing the diagrammatic expansions in the GWA and the cumulant expansion
we can get some idea about the vertex corrections.
In figure \ref{fig:diag} the Green function diagrams are shown to
second order in the screened interaction, which should be sufficient for our
purpose. The cumulant expansion diagrams are obtained by considering the
three possible time-orderings of the integration time variables $t^{\prime }$
in ${C^h}^2\left( k,t\right) $ with $C^h\left( k,t\right) $ 
given by equation (\ref{cumulant}). The cumulant expansion contains
second-order diagrams which are not included in the GWA. It is these
additional diagrams that give rise to the second plasmon satellite and they
are quite distinct from the second-order diagram common to both
approximations. The interpretation of the latter diagram is that a hole
emits a plasmon which is reabsorbed at a later time and the hole returns to
its original state before plasmon emission. This process is repeated once at
a later time. Thus there is only one plasmon coupled to the hole at one
time. In contrast, the other two diagrams, not contained in the GWA,
describe an additional plasmon emission before the first one is reabsorbed,
giving two plasmons coupled to the hole simultaneously. Similar
consideration can be extended to the higher-order diagrams.
If self-consistency is taken into account then the second second-order diagram
is also included in the GWA. 

The cumulant expansion contains only boson-type diagrams describing emission
and reabsorbtion of plasmons but it does not contain diagrams corresponding
to interaction between a hole and particle-hole pairs. 
For this reason, the
cumulant expansion primarily corrects the satellite description whereas the
quasiparticle energies are to a large extent determined by the GWA as
mentioned before.
Interaction between hole-hole and particle-hole
is described by the ladder diagrams which in a Hubbard model study
were found to improve the low-energy satellite
(Verdozzi, Godby, and Holloway 1995).

The cumulant expansion was applied recently to calculate the photoemission
spectra in Na and Al (Aryasetiawan, Hedin, and Karlsson 1996). The
experimental spectra consist of a quasiparticle peak with a set of plasmon
satellites separated from the quasiparticle by multiples of the plasmon
energy. The spectra in the GWA shows 
only one plasmon satellite located at a too high
energy, approximately 1.5 $\omega_p$ below the quasiparticle which is
similar to the core electron case.
The cumulant expansion method remedies this problem and yields spectra in
good agreement with experiment regarding the position of the satellites. The
relative intensities of the satellites with respect to that of 
the quasiparticle are still in discrepancy. This is likely due to
extrinsic effects
corresponding to the interaction of the photoemitted
electron with the bulk and the surface on its way out of the solid resulting
in energy loss. These are not taken into account in the sudden
approximation.

When applied to valence electrons with band dispersions the cumulant
expansion does not yield the exact result anymore as in the core electron
case. Surprisingly, the numerical results show that the cumulant expansion
works well even in Al with a band width of $\sim$11 eV.
Considering its simplicity, it is a promising approach for describing
plasmon satellites.

\section{Summary and conclusions}

The GWA has been applied by now to a large number of systems, ranging
from atoms, simple metals, semiconductors, transition metals, clusters,
and surfaces and interfaces.
In practically all of 
these systems, the GWA improves the quasiparticle energies relative to
the LDA eigenvalues.
The reason for the success of the GWA may be understood qualitatively 
by the fact that it is correct in some limiting cases as described in the
introduction. 
The GWA includes an important physical
ingredient in extended systems, namely screening or
polarization of the medium which is absent in the HFA. It is well known that
the neglect of screening leads to unphysical results in metals such as zero
density of states at the Fermi level and in 
semiconductors and insulators to too large band gaps. Even in atoms, the
inclusion of polarization effects lead to a significant improvement on the HF
eigenvalues. Since screening is a common feature in all electronic systems,
it is perhaps not surprising that the GWA works in a wide variety of
materials. It is also known empirically that second order perturbation theory
often takes into account most of the physical effects, in particular the
shifting of quasiparticle energies.

Despite of the success of the GWA, it has naturally some shortcomings.
One of these is related to satellite structure in the photoemission spectra.
Since the GWA describes a coupling of the electrons to one plasmon excitation,
represented by $W$, multiple plasmon excitations observed in the alkalis
are clearly beyond scope of the GWA. This problem is remedied by the cumulant
expansion theory described in section VII-C.
Apart from plasmon-related satellites, there are also satellite structure
originating from short-range interactions. This type of satellites 
appears in strongly correlated materials containing 3d or 4f orbitals.
The GWA is based on the RPA screening which takes into account the dominant
part of long-range (small momentum) screening. Short-range (large momentum)
intrasite interactions
of multiple holes usually present in strongly correlated systems
are therefore not well described by the GWA.
Here a T-matrix type approach may be appropriate as has been shown by model
calculations.
Apart from problems with satellite, there are also 
in some cases discrepancies in the quasiparticle energies. For example, the
band width in Na is $\approx$ 10 \% too large within the GWA compared with
experiment but these discrepancies are relatively small.

Another shortcoming is the absence of spin-dependence in the screened
interaction $W$ since the screening is purely Coulombic. 
The spin dependence enters only through the Green function.
One would therefore expect some problems when applying the GWA to magnetic
systems
where spin-spin correlations are important as indicated 
by $GW$ calculations on transition metal atoms. 
This area of research has not been explored extensively and it would require
inclusion of vertex corrections to take into account spin-spin correlations.
A T-matrix approach may be a first step in this direction.
 
Most $GW$ calculations performed so far are not self-consistent, i.e. the
Green function used to calculate the self-energy is not equal to the Green
function obtained from the Dyson equation with the very same self-energy.
Only very recently such self-consistent calculations were performed for the
electron gas. The results turn out to be worst than the straight $GW$
calculations (one iteration) and 
clearly show a cancellation between the effects of
self-consistency and vertex corrections.
In some way it is a blessing since fully self-consistent calculations are
numerically difficult to perform for real systems.
One interesting aspect of the self-consistent calculations
is that the total energies
are in very good agreement with the QMC results.
A recent finding shows that when the total energies are calculated within the
Luttinger-Ward functional and its extension
(Almbladh, von Barth, and van Leeuwen 1997)
using approximate $G$ and $W$ the results are
almost equally good, which circumvents the need for self-consistent
calculations (Hindgren 1997). 
This could be due to the variational property of the
functional and the fact that the self-consistent GWA is conserving in the
sense of Baym and Kadanoff. Calculating total energies
within the GWA
could be an alternative to the QMC method, particularly for extended systems.

Due to computational difficulties, applications of the GWA to large and
complex systems are still not feasible.   
Simplified $GW$ schemes, which are computationally efficient and yet maintain
the accuracy of the full calculations, are therefore very desirable.
Many schemes have been proposed but most of them are designed for
semiconductors. While they give reliable band gaps, details of the band
structure are not fully accounted for. 
Reliable schemes must probably
take into account non-locality as well as energy dependence
of the self-energy.
With efficient schemes, many interesting problems can then be tackled.
These include chemisorption at surfaces, 3d impurities in semiconductors,
interfaces, band off-sets in heterojunctions, and exotic materials such as
fullerenes in their diverse forms.
\section{Acknowledgment}

It is a pleasure to thank Professor Lars Hedin and Dr. Ulf von Barth
for their many valuable
comments and careful reading of the manuscript.

\parindent -8pt
\vskip0.4cm
{\bf REFERENCES}


Albrecht S, Onida G, and Reining L 1997
{\em Phys. Rev. B} {\bf 55} 10278-81

Almbladh C-O and Hedin L 1983
{\em Handbook on Synchrotron Radiation} {\bf 1} 686
ed. E. E. Koch (North-Holland)

Almbladh C-O and von Barth U 1985a 
{\em Phys. Rev. B} {\bf 31} 3231

Almbladh C-O and von Barth U 1985b
{\em Density Functional Methods in Physics}, Vol. 123 of 
{\em NATO Advanced Study Institute, Series B}, eds. Dreizler R M and da
Providencia J (Plenum, New York)

Almbladh C-O, von Barth U, and van Leeuwen R 1997, private communication

Alperin H A 1962 
{\em J. Phys. Soc. Jpn. Suppl. B} {\bf 17} 12

Andersen O K 1975 
{\em Phys. Rev. B} {\bf 12} 3060

Anisimov V I, Zaanen J, and Andersen O K 1991 
{\em Phys. Rev. B} {\bf 44} 943

Anisimov V I, Solovyev I V, Korotin M A, Czyzyk M T,
and Sawatzky G A 1993 
{\em Phys. Rev. B} {\bf 48} 16929

Anisimov V I, Aryasetiawan F, and Lichtenstein A I 1997
{\em J. Phys.: Condens. Matter} {\bf 9} 767-808

Arai M and Fujiwara T 1995
{\em Phys. Rev. B} {\bf 51} 1477-89

Arbman G and von Barth U 1975
{\em J. Phys. F} {\bf 5} 1155


Aryasetiawan F 1992a
{\em \ Phys. Rev. B} {\bf 46} 13051-64

Aryasetiawan F and von Barth U 1992b
{\em Physica Scripta} {\bf T45} 270-1

Aryasetiawan F and Gunnarsson O 1994a 
{\em Phys. Rev. B} {\bf 49} 16214-22

Aryasetiawan F and Gunnarsson O 1994b 
{\em \ Phys. Rev.B} {\bf 49} 7219

Aryasetiawan F and Gunnarsson O 1995
{\em \ Phys. Rev.  Lett.} {\bf 74} 3221-24

Aryasetiawan F and Karlsson K 1996 
{\em Phys. Rev. B} {\bf 54} 5353-7

Aryasetiawan F and Gunnarsson O 1996 
{\em Phys. Rev. B} {\bf 54} 17564-8

Aryasetiawan F, Hedin L, and Karlsson K 1996 
{\em Phys. Rev. Lett.} {\bf 77} 2268-71

Aryasetiawan F and Gunnarsson O 1997 (unpublished)

Ashcroft N W and Mermin N D 1976
{\em Solid State Physics} (Saunders College)

Aspnes D E 1976
{\em Phys. Rev. B} {\bf 14} 5331


Baldereschi A and Tosatti E 1979
{\em Solid State Commun.} {\bf 20} 131

Baldini G and Bosacchi B 1970
{\em Phys. Status Solidi} {\bf 38} 325

Bates D R 1947
{\em Proc. R. Soc. London Ser. A} {\bf 188} 350

Baym G and Kadanoff L P 1961
{\em Phys. Rev.} {\bf 124} 287

Baym G 1962
{\em Phys. Rev.} {\bf 127} 1391

Bechstedt F and Del Sole R 1988
{\em Phys. Rev. B} {\bf 38} 7710-6

Bechstedt F and Del Sole R 1990
{\em Solid State Commun.} {\bf 74} 41-4

Bechstedt F, Del Sole R, Cappellini G, and Reining L 1992
{\em Solid State Commun.} {\bf 84} 765-70

Bechstedt F, Tenelsen K, Adolph B, and Del Sole R 1997
{\em Phys. Rev. Lett.} {\bf 78} 1528-31

Becke A D 1988
{\em Phys. Rev. A} {\bf 38} 3098

Becke A D 1992
{\em J. Chem. Phys.} {\bf 96} 2155

Becke A D 1996
{\em J. Chem. Phys.} {\bf 104} 1040-6

Bergersen B, Kus F W, and Blomberg C 1973
{\em Can. J. Phys.} {\bf 51} 102-110

Berggren K -F and Sernelius B E 1981
{\em Phys. Rev. B} {\bf 24} 1971-86

Biermann L 1943
{\em Z. Astrophys.} {\bf 22} 157

Biermann L and L\"ubeck K 1948
{\em Z. Astrophys.} {\bf 25} 325


Blase X, Rubio A, Louie S G, and Cohen M L 1995
{\em Phys. Rev. B} {\bf 52} R2225-8

Bobbert P A and van Haeringen W 1994
{\em Phys. Rev. B} {\bf 49} 10326-31

Bohm D and Pines D 1953
{\em Phys. Rev.} {\bf 92} 609

Bonacic-Koutecky V, Fantucci P, and Koutecky J 1990a
{\em J. Chem. Phys.} {\bf 96} 3802

Bonacic-Koutecky V, Fantucci P, and Koutecky J 1990b
{\em Chem. Phys. Lett.} {\bf 166} 32

Bonacic-Koutecky V, Pittner J, Scheuch C, Guest M F, and Koutecky J 1992
{\em J. Chem. Phys.} {\bf 96} 7938

Boys S F 1950
{\em Proc. R. Soc. London A} {\bf 200} 542

%

Bylander D M and Kleinman L 1995a
{\em Phys. Rev. Lett.} {\bf 74} 3660

Bylander D M and Kleinman L 1995b
{\em Phys. Rev. B} {\bf 52} 14566


Calandra C and Manghi F 1992
{\em Phys. Rev. B} {\bf 45} 5819

Cardona M, Weinstein M, and Wolff G A 1965
{\em Phys. Rev.} {\bf 140} A633

Causa M and Zupan A 1994
{\em Chem. Phys. Lett.} {\bf 220} 145-53

Ceperley D M and Alder B J 1980
{\em Phys. Rev. Lett. }{\bf 45 }566

Charlesworth J P A, Godby R W and Needs R J 1993
{\em Phys. Rev. Lett.} {\bf 70} 1685-8

Cheetham A K and Hope D A O 1983
{\em Phys. Rev. B} {\bf 27} 6964

Cowan R D 1967 
{\em Phys. Rev. } {\bf 163} 54



Dal Corso A, Pasquarello A, Baldereschi A, and Car R 1996
{\em Phys. Rev. B }{\bf 53} 1180-5

Daling R and van Haeringen W 1989
{\em Phys. Rev. B }{\bf 40} 11659-65

Daling R, Unger P, Fulde P, and van Haeringen W 1991
{\em Phys. Rev. B }{\bf 43} 1851-4

de Groot H J, Bobbert P A, and van Haeringen W 1995 
{\em Phys. Rev. B }{\bf 52} 11000

Deisz J\ J, Eguiluz A, and Hanke W 1993 
{\em Phys. Rev.  Lett.}{\bf \ 71 }2793-96

Deisz J J and Eguiluz A 1997
{\em Phys. Rev. B} {\bf 55} 9195-9

Del Sole R, Reining L, and Godby R W 1994
{\em Phys. Rev. B }{\bf 49} 8024-8

Del Sole R and Girlanda R 1996
{\em Phys. Rev. B }{\bf 54} 14376-80

Dreizler R M and Gross E K U 1990
{\em Density Functional Theory} (Springer-Verlag, New York)

DuBois D F 1959a 
{\em Ann. Phys.} {\bf 7} 174

DuBois D F 1959b 
{\em Ann. Phys.} {\bf 8} 24

Dufek P, Blaha P, Sliwko V, and Schwarz K 1994
{\em Phys. Rev. B} {\bf 49} 10170-5

Dykstra C E 1988
{\em Ab Initio Calculation of the Structure and Properties of Molecules}
(Elsevier, Amsterdam)


Eberhardt W and Plummer E W 1980
{\em Phys. Rev. B} {\bf 21} 3245

Eddy C R, Moustakas T D, and Scanlon J 1993
{\em J. Appl. Phys.} {\bf 73} 448

Eguiluz A\ G, Heinrichsmeier M, Fleszar A, and Hanke W 1992 
{\em Phys. Rev. Lett. }{\bf 68 }1359-62

Engel G E, Farid B, Nex C M M, and March N H (1991)
{\em Phys. Rev. B} {\bf 44} 13356-73

Engel G E and Farid B (1992)
{\em Phys. Rev. B} {\bf 46} 15812-27

Engel G E and Farid B (1993)
{\em Phys. Rev. B} {\bf 47} 15931-4

Engel G E, Kwon Y, and Martin R M 1995
{\em Phys. Rev. B} {\bf 51} 13538-46


Erwin S C and Pickett W E 1991 
{\em Science} {\bf 254} 842


Farid B, Godby R W, and Needs R J 1990
{\em 20th International Conference on the Physics of Semiconductors}
editors Anastassakis E M and Joannopoulos J D, Vol. 3 1759-62
(World Scientific, Singapore)

Feldcamp L A, Stearns M B, and Shinozaki S S 1979
{\em Phys. Rev. B} {\bf 20} 1310

Fender B E F, Jacobson A J, and Wedgwood F A 1968
{\em J. Chem. Phys.} {\bf 48} 990

Fetter A\ L and Walecka J\ D 1971 
{\em Quantum Theory of Many-Particle Systems} (McGraw-Hill)


Fock V 1930
{\em Z. Phys.} {\bf 61} 126

Frota H O and Mahan G D 1992 
{\em Phys. Rev. B} {\bf 45} 6243

Fujimori A, Minami F, and Sugano S 1984 
{\em Phys. Rev. }B {\bf 29 }5225


Galitskii V M and Migdal A B 1958
{\em Sov. Phys. JETP} {\bf 7} 96

Galitskii V M 1958
{\em Sov. Phys. JETP} {\bf 7} 104

Gell-Mann M and Brueckner K 1957
{\em Phys.  Rev.} {\bf 106} 364

Godby R W, Schl\"{u}ter M, and Sham L J 1986
{\em Phys.  Rev. Lett.} {\bf 56} 2415-8

Godby R W, Schl\"{u}ter M, and Sham L J 1987a
{\em Phys.  Rev.} {\em B} {\bf 36} 6497-500

Godby R W, Schl\"{u}ter M, and Sham L J 1987b
{\em Phys.  Rev.} {\em B} {\bf 35} 4170-1

Godby R W, Schl\"{u}ter M, and Sham L J 1988 
{\em Phys.  Rev.} {\em B} {\bf 37} 10159-75

Godby R W and Needs R J 1989
{\em Phys.  Rev. Lett.} {\bf 62} 1169-72 

Godby R W and Sham L J 1994
{\em Phys.  Rev. B} {\bf 49} 1849-57


Grobman W\ D and Eastman D\ E 1972 
{\em Phys. Rev.  Lett. }{\bf 29 }1508

Gross E K U, Dobson J F, and Petersilka M 1996
{\em Density Functional Theory}, edited by Nalewajski R F,
(Springer)

Gu Z-q and Ching W Y 1994
{\em Phys.  Rev.} {\em B} {\bf 49} 10958-67

Guillot C, Ballu Y, Paign\'{e} J, Lecante J, Jain K P,
Thiry P, Pinchaux R, P\'{e}troff Y, and Falicov L M 1977
{\em Phys. Rev.  Lett.} {\bf 39} 1632
 
Gunnarsson O, Lundqvist B I, and Wilkins J W 1974
{\em Phys.  Rev.} {\em B} {\bf 10} 1319

Gunnarsson O and Jones R O 1980
{\em J. Chem. Phys.} {\bf 72} 5357


Gunnarsson O, Rainer D, and Zwicknagl G 1992
{\em Int. J. Mod. Phys. B} {\bf 6} 3993

Gunnarsson O, Meden V, and Sch\"onhammer K 1994
{\em Phys.  Rev.} {\em B} {\bf 50} 10462 

Gunnarsson O 1997 
{\em J. Phys.: Cond. Matt.} {\bf 9} 5635


Gygi F and Baldereschi A 1989
{\em Phys. Rev.  Lett. }{\bf 62 }2160


Hamada N, Hwang M, and Freeman A J 1990
{\em Phys. Rev. B} {\bf 41} 3620-6

Hanke W and Sham L J 1988 
{\em Phys. Rev. B} {\bf 38} 13361-70

Hanke W and Sham L J 1975 
{\em Phys. Rev. B} {\bf 12} 4501-11

Harris J and Jones R O 1978
{\em J. Chem. Phys.} {\bf 68} 3316

Hartree D R 1928
{\em Proc. Camb. Phil. Soc.} {\bf 24} 89, 111, 426

Hayashi E and Shimizu M 1969 
{\em J. Phys. Jpn. } {\bf 26} 1396

Hedin L 1965a 
{\em Phys. Rev}. {\bf 139} A796

Hedin L 1965b
{\em Arkiv f\"or Fysik} {\bf 30} 231-58

Hedin L and Johansson A 1969
{\em J. Phys. B} {\bf 2} 1336

Hedin L and Lundqvist S 1969 
{\em Solid State Physics vol. 23}, 
eds. H. Ehrenreich, F. Seitz, and D. Turnbull (Academic, New York).

Hedin L, Lundqvist B I, and Lundqvist S 1970
{\em J. Res. Natl. Bur. Stand. Sect. A} {\bf 74A} 417

Hedin L and Lundqvist B I 1971
{\em J. Phys. C} {\bf 4} 2064-83

Hedin L 1980
{\em Physica Scripta} {\bf 21} 477-80

Hedin L 1995 
{\em Int. J. Quantum Chem.} {\bf 56} 445-52

Himpsel F J, Knapp J A, and Eastman D E 1979 
{\em Phys. Rev. B} {\bf 19} 2919

Himpsel F J, van der Veen J F, and Eastman D E 1980 
{\em Phys. Rev. B}{\bf \ 22} 1967

Hindgren M 1997 (PhD thesis, University of Lund)

Hohenberg P and Kohn W 1964
{\em Phys. Rev.} {\bf 136} B864

Holm and von Barth 1997 (private communication)

Hong S and Mahan G D 1994
{\em Phys. Rev. B} {\bf 50} 8182-8


H\"ufner S, Wertheim G K, Smith N V, and Traum M M 1972
{\em Solid State Commun.} {\bf 11} 323

H\"ufner S and  Wertheim G K 1973
{\em Phys. Rev. B} {\bf 8} 4857

H\"ufner S, Osterwalder J, Riester T, and Hullinger F 1984
{\em Solid State Commun.} {\bf 52} 793

Hybertsen M S and Louie S G 1985a 
{\em Phys. Rev. Lett.} {\bf 55} 1418-21

Hybertsen M S and Louie S G 1985b
{\em Phys. Rev. B} {\bf 32} 7005-8

Hybertsen M\ S and Louie S\ G 1986 
{\em Phys. Rev. B} {\bf 34} 5390-413

Hybertsen M S and Louie S G 1987a
{\em Phys. Rev. B} {\bf 35} 5585-601, 5602-10


Hybertsen M S and Louie S G 1987b
{\em Comments Condens. Matter Phys.} {\bf 13} 223

Hybertsen M S and Louie S G 1988a
{\em Phys. Rev. B} {\bf 37} 2733-6

Hybertsen M S and Louie S G 1988b
{\em Phys. Rev. B} {\bf 38} 4033-44


Igarashi J 1983
{\em J. Phys. Soc. Jpn.} {\bf 52} 2827

Igarashi J 1985
{\em J. Phys. Soc. Jpn.} {\bf 54} 260

Igarashi J, Unger P, Hirai K, and Fulde P 1994
{\em Phys. Rev. B} {\bf 49} 16181

Inkson J C 1984 
{\em Many-body Theory of Solids}
(Plenum Press, New York, 1984) chap. 11.2

Ishii Y, Ohnishi S, and Sugano S 1986
{\em Phys. Rev. B} {\bf 33} 5271

Itchkawitz, B S, I -W Lyo, and E W Plummer 1985 
{\em Phys. Rev. B} {\bf 41} 8075


Jenkins S J, Srivastava G P, and Inkson J C 1993
{\em Phys. Rev. B} {\bf 48} 4388-97

Jensen, E, and E W  Plummer 1985 
{\em Phys. Rev. Lett.} {\bf 55} 1912

Johansson L S O, Uhrberg R I G, M\aa rtensson P and Hansson G V 1990
{\em Phys. Rev. B} {\bf 42} 1305

Jones R O and Gunnarsson O 1989
{\em Rev. Mod. Phys.} {\bf 61} 689-746


Kane E O 1971
{\em Phys. Rev. B} {\bf 4} 1910-7
 
Kanamori J 1963
{\em Prog. Theor. Phys.} {\bf 30} 275

Kanski J, Nilsson P O and Larsson C G 1980 
{\em Solid State Commun. }{\bf 35 }397

Kemeny P C and Shevchik 1975
{\em Solid State Commun.} {\bf 17} 255

Kohn W and Sham L J 1965 
{\em Phys. Rev.} {\bf 140} A1133

Kotani T 1995
{\em Phys. Rev. Lett.} {\bf 74} 2989

Kotani T and Akai H 1996
{\em Phys. Rev. B} {\bf 54} 16502

Kowalczyk, S P, Ley L, McFeely F R, Pollak R A, and Shirley D A 1973
{\em Phys. Rev. B} {\bf 8} 3583

Kress C, Fiedler M, and Bechstedt F 1994 in {\em Proceedings of the
Fourth International Conference on the Formation of Semiconductor
Interfaces} edited by B Lengeler, H L\"uth, W M\"onch and
J Pollmann (World Scientific, Singapore, 1994), p. 19


Lad R J and Henrich V E 1988
{\em Phys. Rev. B} {\bf 38} 10860

Landau L D 1956
{\em Zh. Eksperimen. i Teor. Fiz.} {\bf 30} 1058

Landau L D 1957
{\em Zh. Eksperimen. i Teor. Fiz.} {\bf 32} 59

Lang N D and Kohn W 1973 
{\em Phys. Rev. B }{\bf 7 }3541-50

Lang N D and Williams A R 1977
{\em Phys. Rev. B }{\bf 16} 2408

Langreth D C 1970
{\em Phys. Rev. B}{\bf 1 }471

Langreth D C and Mehl M J 1983
{\em Phys. Rev. B} {\bf 28} 1809

Lei T, Moustakas T D, Graham R J, He Y, and Berkowitz S J 1992a
{\em J. Appl. Phys.} {\bf 71} 4933

Lei T, Fanciulli M, Molnar R J,  Moustakas T D, Graham, R J, and Scanlon J
1992b
{\em Appl. Phys. Lett} {\bf 59} 944 

Levine Z H and Louie S G 
{\em Phys. Rev. B }{\bf 25 } 6310

Ley L, Kowalczyk S P, Pollak R A, and Shirley D A 1972 
{\em Phys. Rev. Lett. }{\bf 29} 1088

Lichtenstein A I, Zaanen J, and Anisimov V I 1995 
{\em Phys. Rev. B} {\bf 52} R5467

Liebsch A 1979
{\em Phys. Rev. Lett.} {\bf 43} 1431-4

Liebsch A 1981
{\em Phys. Rev. B} {\bf 23} 5203-12

Lindgren I 1971 
{\em Int. J. Quantum Chem.} {\bf 5} 411

Lindhard J 1954
{\em Kgl. Danske Videnskap. Selskap, Mat.-fys. Medd.} {\bf 28} 8

Lof R W, van Veenendaal M A, Koopmans B, Jonkman H T,
and Sawatzky G A 1992 
{\em Phys. Rev. Lett.} {\bf 68} 3924

Lundqvist B I 1967a 
{\em Phys. Kondens. Mater. }{\bf 6 }193

Lundqvist B I 1967b 
{\em Phys. Kondens. Mater. }{\bf 6 }206

Lundqvist B I 1968 
{\em Phys. Kondens. Mater.} {\bf 7} 117

Lundqvist B I and Samathiyakanit V 1969
{\em Phys. Kondens. Mater.} {\bf 9} 231

Lundqvist B I 1969
{\em Phys. Kondens. Mater.} {\bf 9} 236


Luttinger J M and Ward J C 1960
{\em Phys. Rev} {\bf 118} 1417-27

Lyo I -W and Plummer E W 1988 
{\em Phys. Rev. Lett.} {\bf 60} 1558


Madelung O 1984
{\em Crystal and Solid State Physics},
Landolt-B\"ornstein, New Series, Group 3, Vol. 17, Pt. a (Springer, Berlin)



Mahan G D and Sernelius B E 1989 
{\em Phys. Rev. Lett.} {\bf 62} 2718

Mahan G D 1990 
{\em Many-particle Physics} (Plenum Press, New York)

Mahan G D 1994 
{\em Comments Cond. Mat. Phys.} {\bf 16} 333

Manghi F and Calandra C 1994
{\em Phys. Rev. Lett.} {\bf 73} 3129-32

Manghi F, Bellini V, and Arcangeli C 1997
{\em Phys. Rev. B} {\bf 56} 7149-61

Martin P C and Schwinger J 1959
{\em Phys. Rev.} {\bf 115} 1342-73

M{\aa}rtensson H and Nilsson P O 1984
{\em Phys. Rev. B} {\bf 30} 3047

Massidda S, Continenza A, Posternak M and Baldereschi A 1995a 
{\em Phys. Rev. Lett.} {\bf 74} 2323-6

Massidda S, Resta R, Posternak M and Baldereschi A 1995b
{\em Phys. Rev. B} {\bf 52} R16977

Massidda S, Continenza A, Posternak M and Baldereschi A 1997
{\em Phys. Rev. B} {\bf 55} 13494-502

Mattheiss L F 1972a
{\em Phys. Rev. B} {\bf 5} 290

Mattheiss L F 1972b
{\em Phys. Rev. B} {\bf 5} 306

McFeely F R, Kowalczyk S P, Ley L, Cavell R G,
Pollak R A, and Shirley D A 1974 
{\em Phys. Rev. B }{\bf 9 }5268

Minnhagen P 1974
{\em J. Phys. C} {\bf 7} 3013


Moore C E 
{\em Atomic Energy Levels}, Natl. Bur. Stand. Ref. Data Ser.,
Natl. Bur. Stand. (U.S) Circ. No. 467 (U.S. GPO, Washington D C, 1949, 1952,
1958)

Moruzzi V L, Janak J F, and Williams A R 1978
{\em Calculated Electronic Properties of Metals} (Pergamon, New York)

Mott N F 1949 
{\em Phys. Soc. London Sect. A} {\bf 62} 416

M\"{u}ller W, Flesch J, and Meyer W 1984 
{\em J. Chem.  Phys.}{\bf \ 80} 3297-310



Northrup J E, Hybertsen M S, and Louie S G 1987
{\em Phys. Rev. Lett.} {\bf 59} 819

Northrup J E, Hybertsen M S, and Louie S G 1989 
{\em Phys. Rev. B} {\bf 39} 8198

Northrup J E 1993
{\em Phys. Rev. B} {\bf 47} 10032-5

Nozi\`{e}res P 1964
{\em Theory of Interacting Fermi Systems}, (Benjamin, New York)

Nozi\`{e}res P and de Dominicis C J 1969 
{\em Phys.  Rev.}{\bf \ 178} 1097

{\em Numerical Data and Functional Relationships in Science and Technology}
1982,
edited by Hellwege K-H and Madelung O, Landolt-B\"ornstein, New Series, Group
III, Vols. 17a and 22a (Springer, Berlin) 


Onida G, Reining L, Godby R W, Del Sole R and Andreoni W 1995
{\em Phys. Rev. Lett.} {\bf 75} 818-21

Ortuno M and Inkson J C 1979
{\em J. Phys. C} {\bf 12} 1065-71

Oschlies A, Godby R W, and Needs R J 1992
{\em Phys. Rev. B} {\bf 45} 13741-4

Oschlies A, Godby R W, and Needs R J 1995
{\em Phys. Rev. B} {\bf 51} 1527-35

Overhauser A W 1971
{\em Phys. Rev. B} {\bf 3} 1888

Overhauser A W 1985 
{\em Phys. Rev. Lett.} {\bf 55} 1916


Paisley M J, Sitar Z, Posthill J B, and Davis R F 1989
{\em J. Vac. Sci. Technol. A} {\bf 7} 701

Palummo M, Del Sole R, Reining L, Bechstedt F, and Cappellini G 1995
{\em Solid State Commun.} {\bf 95} 393-8

Penn D R 1962
{\em Phys. Rev.} {\bf 128} 2093

Penn D R 1979
{\em Phys. Rev. Lett.} {\bf 42} 921-4

Perdew J P and Zunger A 1981
{\em Phys. Rev. B} {\bf 23} 5048

Perdew J P and Levy M 1983
{\em Phys. Rev. Lett.} {\bf 51} 1884

Perdew J P, Burke K, and Ernzerhof M 1997
{\em Phys. Rev. Lett.} {\bf 77} 3865-8

Philipsen P H T and Baerends E J 1996
{\em Phys. Rev. B} {\bf 54} 5326-33

Pickett W E and Wang C S 1984
{\em Phys. Rev. B} {\bf 30} 4719-33

Pickett W E 1989
{\em Rev. Mod. Phys.} {\bf 62} 433

Pines D and Bohm D 1952
{\em Phys. Rev.} {\bf 85} 338

Pines D 1961
{\em Elementary Excitations in Solids} (W A Benjamin, New York)

Powell R J and Spicer W E 1970 
{\em Phys. Rev. B} {\bf 2} 2182


Quinn J J and Ferrell R A 1958 
{\em Phys. Rev.} {\bf 112} 812


Rauch W 1940
{\em Z. Phys.} {\bf 116} 652

Rice T M 1965
{\em Ann. Phys.} {\bf 31} 100

Rohlfing M, Kr\"{u}ger P and Pollmann J 1993 
{\em Phys. Rev. B }{\bf 48} 17791-805

Rohlfing M, Kr\"{u}ger P and Pollmann J 1995a
{\em Phys. Rev. Lett.} {\bf 19} 3489-92

Rohlfing M, Kr\"{u}ger P and Pollmann J 1995b
{\em Phys. Rev. B} {\bf 52} 1905-17

Rohlfing M, Kr\"{u}ger P and Pollmann J 1996
{\em Phys. Rev. B} {\bf 54} 13759-66

Rojas H N, Godby R W and Needs R J 1995 
{\em Phys. Rev. Lett.} {\bf 74} 1827-1830

Rubio A, Corkill J L, Cohen M L, Shirley E L, and Louie S G 1993 
{\em Phys. Rev. B} {\bf 48} 11810-6

Runge E and Gross E K U 1984
{\em Phys. Rev. Lett.} {\bf 52} 997


Saito S and Cohen M L 1988
{\em Phys. Rev. B} {\bf 38} 1123

Saito S, Zhang S B, Louie S\ G and Cohen M\ L 1989 
{\em Phys. Rev. B}{\bf \ 40} 3643-6

Satpathy S, Antropov V P, Andersen O K, Jepsen O, Gunnarsson O,
and Lichtenstein A I 1992  
{\em Phys. Rev. B} {\bf 46} 1773

Sawatzky G\ A and Allen J\ W 1984 
{\em Phys. Rev.  Lett. }{\bf 53} 2339

Schindlmayr A and Godby R W 1997 preprint

Sch\"onberger U and Aryasetiawan F 1995
{\em Phys. Rev. B} {\bf 52} 8788-93

Schwinger J 1951
{\em Proc. Natl. Acad. Sci. U. S.} {\bf 37} 452

Seidl A, G\"orling A, Vogl P, Majewski J A, and Levy M 1996
{\em Phys. Rev. B} {\bf 53} 3764-74

Sham L J and Kohn W 1966
{\em Phys. Rev. }{\bf 145} 561-7

Sham L J and Schl\"uter M 1983
{\em Phys. Rev. Lett.} {\bf 51} 1888

Sham L J and Schl\"uter M 1985
{\em Phys. Rev. B} {\bf 32} 3883

Shen Z X, Shih C K, Jepsen O, Spicer W E, Lindau I, and
Allen J W 1990 
{\em Phys. Rev. Lett.} {\bf 64} 2442

Shen Z X, List R S, Dessau D S, Wells B O, Jepsen O, 
Arko A J, Barttlet R, Shih C K, Parmigiani F, Huang J C, 
and Lindberg P A P 1991a
{\em Phys. Rev. B} {\bf 44} 3604

Shen Z X {\em et al} 1991b
{\em Solid State Commun.} {\bf 79} 623
 
Shirley E L, Zhu X, and Louie S G 1992
{\em Phys. Rev. Lett.} {\bf 69} 2955-8

Shirley E L and Martin R M 1993
{\em Phys. Rev. B} {\bf 47} 15404-27

Shirley E L and Louie S G 1993
{\em Phys. Rev. Lett.} {\bf 71} 133


Shirley E L 1996 
{\em Phys. Rev. B }{\bf 54} 7758-64

Shung K W -K and Mahan G D 1986
{\em Phys. Rev. Lett.} {\bf 57} 1076-79

Shung K W -K, Sernelius B E, and Mahan G D 1987
{\em Phys. Rev. B} {\bf 36} 4499

Shung K W -K and Mahan G D 1988 
{\em Phys. Rev. B} {\bf 38} 3856

Sinha S K 1969 
{\em Phys. Rev. B }{\bf 177} 1256

Sitar Z {\em et al} 1992
{\em J. Matter. Sci. Lett.} {\bf 11} 261

Slater J C 1930
{\em Phys. Rev.} {\bf 35} 210

Slater J C 1951a
{\em Phys. Rev.} {\bf 81} 385

Slater J C 1951b
{\em Phys. Rev.} {\bf 82} 538

Slater J C 1953
{\em Phys. Rev.} {\bf 91} 528

Slater J C 1974
{\em Quantum Theory of Molecules and Solids, Vol. IV,
The Self-consistent Field for Molecules and Solids}
(McGraw-Hill, New York)

Solovyev I V, Dederichs P H, and Anisimov V I (1994)
{\em Phys. Rev. B} {\bf 50} 16861-71

Solovyev I, Hamada N, and Terakura K 1996
{\em Phys. Rev. B} {\bf 53} 7158

Springer M, Svendsen P S, and von Barth U 1996
{\em Phys. Rev. B} {\bf 54} 17392-401

Steiner P, H\"ochst H, and H\"ufner S 1979 in
{\em Photoemission in Solids II}, edited by L. Ley and M. Cardona, Topics in
Applied Physics Vol. 27 (Springer-Verlag, Heidelberg)


Sterne P A and Inkson J C 1984
{\em J. Phys. C} {\bf 17} 1497-510

Strinati G, Mattausch H J, and Hanke W 1982
{\em Phys. Rev. B} {\bf 25} 2867-88

Sch\"onhammer and Gunnarsson 1989
{\em Phys. Rev. B} {\bf 37} 3128

Surh M P, Northrup J E, and Louie S G 1988 
{\em Phys. Rev. B} {\bf 38} 5976-80

Surh M P, Chacham H, and Louie S G 1995
{\em Phys. Rev. B} {\bf 51} 7464-70

Svane A and Gunnarsson O
1990 {\em Phys. Rev. Lett.} {\bf 65} 1148-51

Svendsen P S and von Barth U 1996
{\em Phys. Rev. B} {\bf 54} 17402-13

Szotek Z, Temmerman W M, and Winter H 1993 
{\em Phys. Rev. B} {\bf 47} 4029


Talman J D and Shadwick W F 1976
{\em Phys. Rev. A} {\bf 14} 36

Terakura K, Oguchi T, Williams A\ R and K\"{u}bler J 1984
{\em Phys. Rev. }B {\bf 30} 4734-47

Ting C S, Lee T K, and Quinn J J 1975 
{\em Phys. Rev. Lett.} {\bf 34} 870

Tjeng L H, Chen C\ T, Ghijsen J, Rudolf P and Sette F 1991
{\em Phys. Rev. Lett.} {\bf 67} 501

Troullier N and Martins J L 1992 
{\em Phys. Rev. B} {\bf 46} 1754

Tsui D C 1967
{\em Phys. Rev.} {\bf 164} 669

Tsui D C and Stark R W 1968
{\em J. Appl. Phys.} {\bf 39} 1056


Uhrberg R I G, Hansson G V, Nicholls J M, and Flodstr\"om S A 1981
{\em Phys. Rev. B } {\bf 24} 4684


van Elp J {\em et al} 1991
{\em Phys. Rev. B} {\bf 44} 1530

van Leeuwen R 1996
{\em Phys. Rev. Lett.} {\bf 76} 3610-3

Verdozzi C, Godby R W, and Holloway S 1995
{\em Phys. Rev. Lett.} {\bf 74} 2327-30

von Barth U and Hedin L (1972)
{\em J. Phys. C} {\bf 5} 1629-42


von Barth U and Holm B 1996 
{\em Phys. Rev. B} {\bf 54} 8411-9

von der Linden W and Horsch P 1988 
{\em Phys. Rev. B} {\bf 37} 8351


Walter J P and Cohen M L 1972
{\em Phys. Rev. B} {\bf 5} 3101

Wang C S and Pickett W E 1983
{\em \ Phys. Rev. Lett.} {\bf 51} 597-600

Wang C, Pollack S, Cameron D, and Kappes M M 1990
{\em J. Chem. Phys} {\bf 93} 3787

Ward J C 1950
{\em Phys. Rev.} {\bf 78} 182

Watson R E, Herbst J F, Hodges L, Lundqvist B I, and Wilkins J W 1976
{\em Phys. Rev. B} {\bf 13} 1463

Whited R C, Flaten C J, and Walker W C 1973
{\em Solid State Commun.} {\bf 13} 1903

Wick G C 1950
{\em Phys. Rev.} {\bf 80} 268



Zakharov  O, Rubio A, Blase X, Cohen M L, and Louie S G 1994
{\em Phys. Rev. B} {\bf 50} 10780-7

Zhang S B, Tom\`{a}nek D, Louie S G, Cohen M L, and Hybertsen M S 1988
{\em Solid State Commun.} {\bf 66} 585-8


Zhang S B, Wei S-H, and Zunger A 1995
{\em Phys. Rev. B} {\bf 52} 13975


Zhu X and Louie S G 1991
{\em Phys. Rev. B} {\bf 43} 14142-56

Zunger A, Perdew J P, and Oliver G L 1980 
{\em Solid State Commun.} {\bf 34} 933 

Zunger A and Lindefelt U 1983
{\em Solid State Commun.} {\bf 45} 343

\end{document}